\documentclass[10pt,journal,cspaper,compsoc,twoside]{IEEEtran}

\usepackage{fancyhdr}

\usepackage{amsmath, amsthm}
\usepackage{graphicx,epsfig,color,endnotes,alltt}
\usepackage{subfigure}
\usepackage{listings}

\usepackage{url}
\usepackage{balance}
\usepackage{algorithm}
\usepackage[noend]{algorithmic}
\usepackage{subfigure}
\usepackage{fancyvrb}
\usepackage{bbding}
\usepackage[normalem]{ulem}
\usepackage{array}
\usepackage{color}
\usepackage{multirow}
\usepackage[nocompress]{cite}
\definecolor{Magenta}{rgb}{1,0.5,0}

\newcommand{\tabincell}[2]{\begin{tabular}{@{}#1@{}}#2\end{tabular}}

\ifCLASSOPTIONcompsoc
\else
\fi
\ifCLASSINFOpdf
\else
\fi

\begin{document}
\title{Rank-Aware Dynamic Migrations and Adaptive Demotions for DRAM Power Management}

\author{Yanchao~Lu, Donghong~Wu, Bingsheng~He, Xueyan Tang, Jianliang~Xu and Minyi~Guo%
\IEEEcompsocitemizethanks{\IEEEcompsocthanksitem Bingsheng He and
Xueyan Tang are with Nanyang Technological University, Singapore.
Corresponding author: Bingsheng He, bshe@ntu.edu.sg.
\IEEEcompsocthanksitem Yanchao Lu, Donghong Wu and Minyi Guo are
with Shanghai Jiao Tong University. The work of Yanchao Lu and
Donghong Wu is done when they were visiting students in Nanyang
Technological University, Singapore. \IEEEcompsocthanksitem
Jianliang Xu is with Hong Kong Baptist
University.}%
\thanks{}}

\markboth{}{}
\IEEEcompsoctitleabstractindextext{%
\begin{abstract}
Modern DRAM architectures allow a number of low-power states on individual
memory \emph{ranks} for advanced power management. Many previous studies have
taken advantage of demotions on low-power states for energy saving. However,
most of the demotion schemes are statically performed on a limited number of
pre-selected low-power states, and are suboptimal for different workloads and
memory architectures. Even worse, the idle periods are often too short for
effective power state transitions, especially for memory intensive applications.
Wrong decisions on power state transition incur significant energy and delay
penalties. In this paper, we propose a novel memory system design named RAMZzz
with rank-aware energy saving optimizations including dynamic page migrations
and adaptive demotions. Specifically, we group the pages with similar access
locality into the same rank with dynamic page migrations. Ranks have their
hotness: hot ranks are kept busy for high utilization and cold ranks can have
more lengthy idle periods for power state transitions. We further develop
adaptive state demotions by considering all low-power states for each rank and a
prediction model to estimate the power-down timeout among states. We
experimentally compare our algorithm with other energy saving policies with
cycle-accurate simulation. Experiments with benchmark workloads show that RAMZzz
achieves significant improvement on energy-delay$^2$ and energy consumption over
other energy saving techniques.
\end{abstract}
}

\maketitle

\IEEEdisplaynotcompsoctitleabstractindextext
\IEEEpeerreviewmaketitle

\section{Introduction}\label{sec:intro}

\IEEEPARstart{E}{nergy} consumption has become a major factor for the design and
implementation of computer systems. Inside many computing systems, main memory
(or DRAM) is a critical component for the performance and energy consumption. As
processors have moved to multi-/many-core era, more applications run
simultaneously with their working sets in the main memory. The hunger for main
memory of larger capacity makes the amount of energy consumed by main memory
approaching or even surpassing that consumed by processors in many servers~\cite
{Hoelzle:2009:DCI:1643608,Lefurgy:2003:EMC:957964.957972}. For example, it has
been reported that main memory contributes to as much as 40--46\% of total
energy consumption in server
applications~\cite{Meisner:2009:PES:1508244.1508269,s.ware:architecting,
Lefurgy:2003:EMC:957964.957972}. For these reasons, this paper studies the
energy saving techniques of main memory.

Current main memory architectures allow power management on individual memory
ranks. Individual ranks at different power states consume different amounts of
energy. There have been various energy-saving techniques on exploiting the power
management capability of main
memory~\cite{Huang:2003:DIP:1247340.1247345,Huang:2005:IEE:1077603.1077696,Delaluz:2002:SDE:513918.514095,DBEnergy}.
The common theme of those research studies is to exploit the transition of
individual memory ranks to low-power states (i.e., \emph{demotion}) for energy
saving. Fan et al. concluded that immediate transitions to the low-power state
save the most energy consumption for most single-application
workloads~\cite{Fan:2001:MCP:383082.383118}. However, the decision can be wrong
for more memory intensive workloads such as multi-programmed executions. Huang
et al.~\cite{Huang:2005:IEE:1077603.1077696} has shown that only sufficiently
long idle periods can be exploited for energy saving because state transitions
themselves take non-negligible amount of time and energy. Essentially, the
amount of energy saving relies on the distributions of idle periods and the
effectiveness of how power management techniques exploit the idle periods.
Existing techniques are suboptimal in the following aspects: (1) they do not
effectively extend the idle period, either with static page
placement~\cite{Diniz:2007:LPC:1250662.1250699,Fan:2001:MCP:383082.383118} or
with heuristics-based page
migrations~\cite{Huang:2003:DIP:1247340.1247345,Huang:2005:IEE:1077603.1077696};
(2) the prediction on the power-down timeout (\emph{the amount of time
spent since the beginning of an idle period before transferring to a low-power
state}) for a state transition is limited and static, either with
heuristics~\cite{Huang:2003:DIP:1247340.1247345,Huang:2005:IEE:1077603.1077696}
or regression-based model~\cite{Fan:2001:MCP:383082.383118}; (3) most of the
demotion schemes are statically performed on a limited number of pre-selected
low-power states (e.g., Huang et al.~\cite{Huang:2005:IEE:1077603.1077696}
selects two low-power states only, out of five in DDR3). The static demotion
scheme is suboptimal for different workloads and different memory architectures.

To address the aforementioned issues, we propose a novel memory
design named RAMZzz with rank-aware power management techniques
including dynamic page migrations and adaptive demotions. Instead of
having static page placement, we develop dynamic page migration
mechanisms to exploit the access locality changes in the workload.
Pages are placed into different ranks according to their access
locality so that the pages in the same rank have roughly the same
hotness. As a result, ranks are categorized into hot and cold ones.
The hot rank is highly utilized and has very short idle periods. In
contrast, the cold rank has a relatively small number of long idle
periods, which is good for power state transitions for energy
saving.

Instead of adopting static demotion schemes, we develop adaptive
demotions to exploit the power management capabilities of all
low-power states for individual ranks. The decisions are guided by a
prediction model to estimate the idle period distribution. The
prediction model combines the historical page access frequency and
historical idle period distribution, and is specifically designed
with the consideration of page migrations among ranks. Based on the
prediction model, RAMZzz is able to optimize for different goals
such as energy saving and energy-delay$^2$ (ED$^2$). In this
paper, we focus on the optimization goal of minimizing ED$^2$ (or
energy consumption) of the memory system while keeping the program
performance penalty within a given budget. The budget is a
pre-defined performance slowdown relative to the maximum performance
without any power management (e.g., 10\% performance loss).

We evaluate our design using detailed simulations of different workloads
including SPEC 2006 and PARSEC~\cite{bienia11benchmarking}. We evaluate RAMZzz
in comparison with representative power saving
policies~\cite{Lebeck:2000:PAP:378993.379007, Huang:2005:IEE:1077603.1077696,
Diniz:2007:LPC:1250662.1250699} and an ideal oracle approach. Our experiments
with the optimization goal of ED$^2$ (for a maximum acceptable
performance degradation of 4\%) on three different DRAM architectures show that
(1) both page migrations and adaptive demotions well adapt to the workload.
Page migrations achieve an average ED$^2$ improvement of 17.1--21.8\% over
schemes without page migrations, and adaptive demotions achieve an average
ED$^2$ improvement of 22.4--36.4\% over static demotions; (2) with both page
migrations and adaptive demotions, RAMZzz achieves an average ED$^2$ improvement
of 63.0--64.2\% over the basic approach without power management, and achieves
only 3.7--5.7\% on average larger ED$^2$ than the ideal oracle approach. The
experiments with the optimization goal of energy consumption have demonstrated
similar results.

{\bf Organization.} The rest of the paper is organized as follows.
We introduce the background on basic power management of DRAM and review related
work in Section~\ref{sec:motivation}. Section~\ref{sec:overview} gives an
overview of RAMZzz design, followed by detailed implementations in
Section~\ref{sec:implementation}. The experimental results are presented in
Section~\ref{sec:evaluation}. We conclude this paper in Section~\ref{sec:conclusion}.

\section{Background and Related Work}\label{sec:motivation}

\subsection{DRAM Power Management} \label{subsec:dram_power}

In this paper, we use the terminology of DDR-series memory architectures (e.g.,
DDR2 and DDR3 etc) to describe our approach. We will evaluate RAMZzz on
different DDR-series memory architectures in the experiments. DDR is usually
packaged as modules, or DIMMs. Each DIMM contains multiple ranks. In power
management, a rank is the smallest physical unit that we can control. Individual
ranks can service memory requests independently and also operate at different
\emph{power states}. The power consumption of a memory rank can be divided into
two main categories: active power and background power.
Active power consists of the power that is required to activate the banks and
service memory reads and writes. Background power is the power consumption
without any DRAM accesses. Background power is a major component in the total
DRAM power consumption, and tends to be more significant in the
future~\cite{Huang:2005:IEE:1077603.1077696, Zheng:2010:PPT:1850266.1850273}.
For example, Huang et al.~\cite{Huang:2005:IEE:1077603.1077696} found that the
background power contributes to 52\% of the total DRAM power in their
evaluation. Memory capacity and bandwidth will become larger and is usually
provisioned with peak usage, which causes severe under-utilization~\cite{Malladi:2012:TED:2337159.2337164}.
Therefore, we focus on reducing the background power consumption.

Different power states have different power consumptions. Entering a
low-power state when a rank is idle reduces the background power consumption. To
exit from a low-power state, the disabled hardware components need to be
reactivated and the rank needs to be restored to the active state. State
transitions among different power states cause latency and energy penalties.

\begin{table}
\centering
\caption{Power states for three typical DRAM architectures.}
\label{tb:DRAMState}
\vspace{-2ex}
{\footnotesize
\begin{tabular}{|c|c|c|}
\hline
{\bf Power State} & {\bf Normalized Power} & {\bf
\tabincell{c}{Resynchronization \\ Time (ns)}} \\
\hline
\hline
\multicolumn{3}{|l|}{DDR3 DRx4 at 1333 MHz~\cite{ddr3spec}} \\
\hline
ACT & 1.0 & 0 \\
\hline
ACT\_PDN & 0.612 & 6 \\
\hline
PRE\_PDN\_FAST & 0.520 & 18 \\
\hline
PRE\_PDN\_SLOW & 0.299 & 24 \\
\hline
SR\_FAST & 0.170 & 768 \\
\hline
SR\_SLOW & 0.104 & 6768 \\
\hline
\hline
\multicolumn{3}{|l|}{DDR2 DRx8 at 800 MHz~\cite{ddr2spec}} \\
\hline
ACT & 1.0 & 0 \\
\hline
ACT\_PDN\_FAST & 0.619 & 5 \\
\hline
ACT\_PDN\_SLOW & 0.325 & 18 \\
\hline
PRE\_PDN & 0.237 & 25 \\
\hline
SR & 0.178 & 500 \\
\hline
\hline
\multicolumn{3}{|l|}{LPDDR2 DRx16 at 800 MHz~\cite{lpddr2spec}} \\
\hline
ACT & 1.0 & 0 \\
\hline
ACT\_PDN & 0.523 & 8 \\
\hline
PRE\_PDN & 0.303 & 26 \\
\hline
SR & 0.194 & 100 \\
\hline
\end{tabular}
}
\vspace{-4ex}
\end{table}

Depending on which hardware components are disabled, modern memory
architectures support a number of power states with complicated
transitions~\cite{ddr2spec, ddr3spec}. Each state is characterized
with its power consumption and the time that it takes to transition
back to the active state (resynchronization time). Typically, the
lower power consumption the low-power state has, the higher the
resynchronization time is. Table~\ref{tb:DRAMState} summarizes the
major power state transitions of three typical DRAM architectures:
DDR3, DDR2 and LPDDR2. We do not consider some advanced power
management modes in LPDDR2, like Deep Power-down (DPD) and Partial Array Self Refresh (PASR),
because data retention cannot be kept when the LPDDR2 enters those
states. For each state, we show its dynamic power consumption
(normalized to that of ACT) and the resynchronization times back to
ACT. The power consumption values are calculated with DRAM System
Power Calculator~\cite{calc}. The resynchronization times are
obtained from DRAM manufacturers' data sheets~\cite{ddr3spec,
ddr2spec, lpddr2spec}.

From Table~\ref{tb:DRAMState}, we have the following observations on state
demotions on different memory architectures.

First, on a specific memory architecture, power states have quite
different latency and energy penalties as well as different power
consumptions. Take DDR3 as an example. Pre-charge power-down with
fast exit state (PRE\_PDN\_FAST) consumes 52\% of the power of
active idle state (ACT), with relatively small latency as well as
energy penalties. In contrast, self-refresh with fast exit state
(SR\_FAST) consumes only 17\% of the power of ACT, with much higher
latency and energy penalties. The resynchronization time of SR\_FAST
is over an order of magnitude higher than that of PRE\_PDN\_FAST.

Second, different memory architectures have their own specifications on power
states as well as power state energy consumption and resynchronization time.
First, different memory architectures may have different sets of power states.
For example, DDR3 has a special low-power state, i.e., self-refresh with slow
exit state (SR\_SLOW), whereas DDR2 and LPDDR2 do not have any equivalent state.
SR\_SLOW has a very high resynchronization time and consumes only 10\% of the
power of ACT. Second, the energy consumption or the resynchronization time of
the same power state can vary for different memory architectures. Take
self-refresh states (SR) as an example. While SR consumes a similar normalized
power consumption for the three architectures (about 17--19\%), the
resynchronization time varies significantly. The resynchronization times on
DDR3, DDR2, LPDDR2 are 768ns (SR\_FAST), 500ns (SR) and 100ns (SR), respectively.

The above-mentioned observations have significant implications to
DRAM power management design.

First, the above-mentioned observations clearly show the deficiency
of the static demotion
schemes~\cite{Huang:2003:DIP:1247340.1247345,Huang:2005:IEE:1077603.1077696,Delaluz:2002:SDE:513918.514095,Fan:2001:MCP:383082.383118}.
The static demotion schemes are performed on the pre-selected low-power
states (even for all ranks in the same architecture, and for
different memory architectures). On a specific memory
architecture, the static decision loses the opportunities for
demoting to the most energy-effective low-power state for different
idle period lengths. Moreover, since the latency and energy
consumption penalties and power consumption of a low-power state
vary with different memory architectures, the static decision loses
the opportunities for adapting to different memory architectures.

Second, because the latency and energy penalty for switching from
deeper low-power states is substantially higher than the penalty of
switching from shallower states, entering deep power-down states for
short idle times could in fact hurt energy efficiency because the
power savings might not be able to offset the high latency penalty
of switching back to the active state. Thus, the effective use of deeper
low-power state is contingent on having long idle periods on a rank.
That naturally leads to two problems for reducing background power
consumption: 1) how to create longer idle periods without modifying
the application, and 2) how to make correct decisions on state
transitions.

The design of RAMZzz are guided by the aforementioned two
implications. It embraces dynamic migrations and adaptive demotions,
adapting to different workloads and different memory architectures.

\subsection{Related Work}\label{sec:related}

We briefly review the related work on energy saving with power states and with
other hardware and software approaches.

Saving energy by transiting memory power states has attracted many research
efforts, covering memory controller design, compilers and operating systems.

Different power state transition approaches have been developed for DRAM
systems. Hur et al.~\cite{hur:a} developed adaptive history-based scheduling in
the memory controller. Based on page migration, Huang et
al.~\cite{Huang:2005:IEE:1077603.1077696} stored frequently-accessed pages into
hot ranks and left infrequently-used and unmapped pages on cold ranks. Their
decisions on page migrations are based on heuristics. Lebeck et
al.~\cite{Lebeck:2000:PAP:378993.379007} studied different page allocation
strategies. Their approach does not have any analytical model to guide the
decision, or utilize both recency and frequency to capture rank hotness. Diniz
et al.~\cite{Diniz:2007:LPC:1250662.1250699} limited the energy consumption by
adjusting the power states of DRAM. Our prediction model offers a novel way of
power management on guiding page migrations and power state transitions. Fan et
al.~\cite{Fan:2001:MCP:383082.383118} developed an analytic model on estimating
the idle time of DRAM chips using an exponential distribution. Their model does
not consider page migrations. Kshitij et al.~\cite{Sudan:6212453} used a similar
page migration mechanism between cold and hot ranks, but always set cold ranks
with a pre-selected low-power state. Instead of relying on the presumed
knowledge of distribution, our prediction model combines the historical
information on idle period distribution and page access locality. More
importantly, compared with all previous studies that pre-define a number of
fixed states for all ranks~\cite{hur:a, Huang:2005:IEE:1077603.1077696,
Lebeck:2000:PAP:378993.379007, Diniz:2007:LPC:1250662.1250699,
Fan:2001:MCP:383082.383118, Sudan:6212453}, this paper develops adaptive
demotions to exploit the energy-saving capabilities of all power states, and the
adaptation is on the granularity of individual ranks for different memory
architectures.

DRAM power state transitions have been implemented in operating
systems and compilers. Delaluz et
al.~\cite{Delaluz:2002:SDE:513918.514095} present an operating
system based solution letting the scheduler decide the power state
transitions. This approach requires the interfaces of exposing and
controlling the power states. Huang et
al.~\cite{Huang:2003:DIP:1247340.1247345} proposed power-aware
virtual memory systems. For energy efficient compilations, Delaluz
et al.~\cite{Delaluz:2000:ECO:354880.354900} proposed compiler
optimizations for memory energy consumption of array allocations.
They further combined the hardware-directed approach and
compiler-directed approaches~\cite{Delaluz:2001:DEM:580550.876438}
for more energy saving.

There are other approaches for reducing the DRAM power
consumption. We review three representative categories. The first
category is to reduce the active power consumption. Zheng et
al.~\cite{Zheng:2008:MAD:1521747.1521797} suggested the subdivision
of a conventional DRAM rank into mini-ranks comprising of a subset
of DRAM devices to improve DRAM energy efficiency. Anh et
al.~\cite{Ahn:2009:FSP:1654059.1654102} proposed Virtual Memory
Devices (VMDs) comprising of a small number of DRAM chips. Decoupled
DIMMs~\cite{Zheng:2009:DDB:1555754.1555788} proposed the DRAM
devices at a lower frequency than the memory channel to reduce DRAM
power. The second category is to reduce the power consumption of
power state transitions. Bi et al.~\cite{bi:delay-hiding} took
advantage of the I/O handling routines in the OS kernel to hide the
delay incurred by memory power state transitions. Balis et
al.~\cite{Cooper-Balis:2010:FAP:1849300.1849320} proposed finer
grained memory state transition. The third category is to adjust the
voltage and frequency of DRAM. Memory voltage and frequency scaling
(DVFS) is a recent approach to reduce DRAM energy
consumption~\cite{David:2011:MPM:1998582.1998590,
Deng:2011:MAL:1950365.1950392}. Those approaches are complementary
to the state transition-based energy saving approaches.

Recently, different architectural designs of DRAM
systems~\cite{Ahn:2009:FSP:1654059.1654102, kim:HPCA2010,
Zheng:2010:PPT:1850266.1850273, Udipi:2010:RDD:1815961.1815983} are explored on
multi-core processors for performance, energy, reliability and other issues.
Cache-centric optimizations (either
cache-conscious~\cite{He:2006:CAX:1175893.1176184} or
cache-oblivious~\cite{He:2008:CDL:1366102.1366105,DBLP:conf/cidr/HeL07}) reduce
memory access and create more opportunities for energy saving. Besides
optimizations targeting at general DRAM systems, some researchers have also
proposed energy saving techniques for specific applications such as
databases~\cite{DBEnergy, Kumar:2011:MEM:2016802.2016864} and video
processing~\cite{Kumar:2011:MEM:2016802.2016864}.

A preliminary version of RAMZzz has been presented in a previous
paper~\cite{Wu:2012:RRD:2388996.2389040}. This paper improves the
previous paper in many aspects, with two major improvements. First,
we have enhanced RAMZzz with adaptive demotions, which further
increases the effectiveness of state demotions on individual ranks.
Second, we have evaluated the effectiveness of RAMZzz on different
memory architectures, and demonstrated the self-tuning feature of
RAMZzz for different workloads and different memory architectures.

\section{Design Overview}\label{sec:overview}

In this section, we give an overview of the design rationales and workflow of
RAMZzz.

\subsection{Motivations}

Our goal is to reduce the background power of DRAM. Due to the inherent power
management mechanisms of DRAM, there are three obstacles in the effectiveness of
reducing the background power.

First, due to the latency and power penalty of transiting from
low-power state to active state, it requires a minimum length
threshold for an idle period that is worthwhile to make the state
transition. Furthermore, the threshold value varies with the amount
of energy and delay penalties of different state transitions. Since
there is a length threshold for an idle period, an energy saving
technique needs to determine whether an idle period on a rank is
longer than threshold or not. Ideally, if the idle period is longer
than the threshold value, the rank should jump to the low-power
state at the beginning of the idle period; otherwise, we should keep
the rank in the active state. However, it is not easy to predict the
length of each idle period, due to dynamic memory references.

Second, the state transition-based power saving approaches cannot take full
advantage of idle periods, especially for memory intensive workloads. In memory
intensive workloads, the number of idle periods is large, and many of the idle
periods are too short to be exploited for power saving. It is desirable to
reshape the page references to different ranks so that the idle periods become
longer and the number of idle periods is minimized.

Third, static demotion schemes cannot adapt to different workloads and different
memory architectures. With page migrations, we further need adaptation for power
management on individual ranks (differentiating the rank hotness).

\subsection{Workflow of RAMZzz}

\begin{figure}
  \centering
  \includegraphics[width=0.45\textwidth]{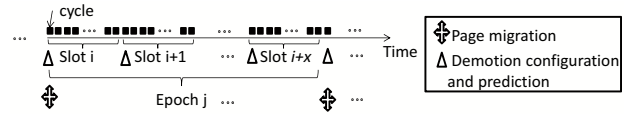}
  \vspace{-2ex}
  \caption{Overview of RAMZzz.}\label{fig:RAMZzzOverall}
  \vspace{-4ex}
\end{figure}

We propose a novel memory design RAMZzz with dynamic migrations and
adaptive demotions to address the aforementioned obstacles. We
develop a dynamic page placement policy that is likely to create
longer idle periods. The policy takes advantage of recency and
frequency of pages stored in the ranks, and ranks are categorized
into hot and cold ones. The hot ranks tend to have very short idle
periods, and the cold ranks with relatively long idle periods. Page
migrations are periodically performed to maintain the rank hotness
(the period is defined as \emph{epoch}). With dynamic page
migrations, short idle periods are consolidated into longer ones and
the number of idle periods is reduced on the cold ranks. On the
other hand, the configuration for adaptive demotions is determined
periodically (the period is called \emph{slot}). For each slot, a
\emph{demotion configuration} (i.e., the power-down timeouts for all
power states) is used to guide the demotion within the slot.

We further develop an analytical model to periodically estimate the idle period
distribution of one slot. Our analytical model is based on the locality of
memory pages and the idle period distribution of the previous slot. Given an
optimization goal (such as minimizing energy consumption or minimizing ED$^2$),
we use the prediction model to determine the demotion configuration for the new
slot. Since the prediction has much lower overhead than the page migration, a
slot is designed to be smaller than an epoch. In our design, an epoch consists of
multiple slots. Figure~\ref{fig:RAMZzzOverall} illustrates the relationship
between slot and epoch. RAMZzz performs demotion configuration and prediction at the beginning of
each slot and performs page migration at the beginning of each epoch.

The overall workflow of RAMZzz is designed as shown in
Algorithm~\ref{alg:workflow}. RAMZzz maintains the performance model by updating
the data structures used in the prediction model (Section~\ref{subsec:model}).
As the idle period length increases, actions of the adaptive demotion scheme may
be triggered. At the beginning of each epoch, RAMZzz decides the page migration
schedule and starts to migrate the pages to the destination ranks
(Section~\ref{subsec:migration}). At the beginning of each slot, RAMZzz performs
prediction and determines the demotion configuration for the new slot
(Section~\ref{subsec:demotion}). The next section will describe the design and
implementation details of each component.

\begin{algorithm}
  \caption{Workflow of RAMZzz}
\begin{footnotesize}
\begin{algorithmic}[1]\label{alg:workflow}
    \IF{any memory reference to rank $r$}
    \IF{rank $r$ is in the low-power state}
    \STATE Set $r$ to be ACT;
    \ENDIF
    \STATE Maintains the prediction model;
    /*Section~\ref{subsec:model}*/
    \ELSE
    \STATE Update the current idle period of rank $r$;
    \STATE Perform demotions (if necessary) according to the demotion
    configuration of rank $r$; /*Section~\ref{subsec:demotion}*/
    \ENDIF
    \IF{the current cycle is the beginning of an epoch}
    \STATE Run page migration algorithm and schedule page
    migrations; /*Section~\ref{subsec:migration}*/
    \ENDIF
    \IF{the current cycle is the beginning of a slot}
    \STATE Determine the demotion configuration for the new slot; /*Section~\ref{subsec:demotion}*/
    \ENDIF
\end{algorithmic}
\end{footnotesize}
\end{algorithm}

\section{Design and Implementation Details}\label{sec:implementation}

After giving an overview on RAMZzz, we describe the details for the following
components in rank-aware power management: dynamic page migration, prediction
model and adaptive demotions. Finally, we discuss some other implementation
issues in integrating RAMZzz into memory systems.

\subsection{Dynamic Page Migration}\label{subsec:migration}

When an epoch starts, we first group the pages according to their locality and
each group maps to a rank in the DRAM. Next, pages are migrated according to the
mapping from groups to ranks.

{\bf Rank-aware page grouping.} We place the pages with similar hotness into the
same rank. We adopt the main memory management policy named
MQ~\cite{Zhou:2001:MRA:647055.715773}. We briefly describe the idea of MQ, and
refer the readers to the original paper for more details. MQ has $M$ LRU queues
numbered from 0 to $M$-1. We assume $M=16$ following previous
studies~\cite{Zhou:2001:MRA:647055.715773, Ramos:2011:PPH:1995896.1995911}. Each
queue stores the page descriptor including the page ID, a frequency counter and
a logical expiration time. The queue with a larger ID stores the page
descriptors of those most frequently used pages. On the first access, the page
descriptor is placed to the head of queue zero, with initialization on its
expiration time. A page descriptor in Queue $i$ is promoted to Queue $i+1$ when
its frequency counter reaches $2^{i+1}$. On the other hand, if a page in Queue
$i$ is not accessed recently based on the expiration time, its page descriptor
will be demoted to Queue $i-1$. We use a modified MQ structure to group
physical memory pages~\cite{Ramos:2011:PPH:1995896.1995911}. The updates to the
MQ structure are performed by the memory controller, which is designed to be off
the critical path of memory accesses. More implementation details are described
in Appendix A of the supplementary file.

An observation in MQ is that MQ has clustered the pages with similar access
locality into the same queue. Moreover, unlike LRU, MQ considers both frequency
and recency in page accesses (we study how the locations of pages in the
MQ queues correlate with their access patterns in Appendix F of the
supplementary file). As a result, we have a simple yet effective approach
to place the pages in the ranks. Suppose each rank has a distinct hotness value.
We assign the rank that a page is placed in a manner such that: given any two
pages $p$ and $p'$ with the descriptors in Queues $q$ and $q'$, $p$ and $p'$ are
stored in ranks $r$ and $r'$ ($r$ is hotter than $r'$) if and only if $q>q'$ or
if $q=q'$ and $p$ is ahead of $p'$ in the queue. That means, the pages whose
descriptors are stored in a higher queue in MQ are stored in hotter ranks.
Within the same queue in MQ, the more recently accessed pages are stored in
hotter ranks. Algorithm~\ref{alg:page} shows the process of grouping the pages
into $R$ sets, and each set of pages is stored in a memory rank. Each rank has a
capacity of $C$ pages.

\begin{algorithm}
\caption{Obtain $R$ page groups in the increasing hotness}
\begin{footnotesize}
\begin{algorithmic}[1]\label{alg:page}
    \STATE initiate $R$ empty sets, $S_0$, $S_1$, ..., $S_{R-1}$;
    \STATE $\mathit{curSet}=0$;
    \FOR{Queue $i=M-1$, $M-2$, ...,  $0$ in MQ}
    \FOR{Page $p$ from head to tail in Queue $i$}
    \STATE Add $p$ to $S_\mathit{curSet}$;
    \IF{$|S_\mathit{curSet}|=C$}
    \STATE $\mathit{curSet}++$;
    \ENDIF
    \ENDFOR
    \ENDFOR
\end{algorithmic}
\end{footnotesize}
\end{algorithm}

\begin{figure}
  \centering
  \includegraphics[width=0.7\linewidth]{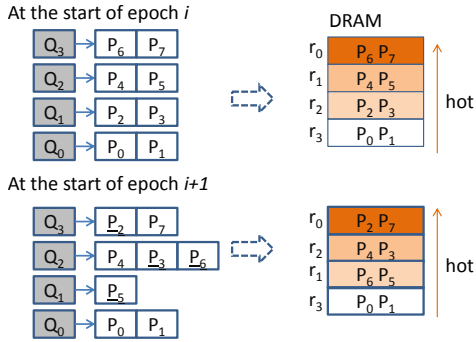}
  \vspace{-1ex}
  \caption{An example of page placement on ranks.}\label{fig:MQ}
  \vspace{-1ex}
\end{figure}

Figure~\ref{fig:MQ} illustrates an example of page placement onto
the ranks. There are four ranks in DRAM, and each rank can hold two
pages. At epoch $i$, we run Algorithm~\ref{alg:page} on the MQ
structures, and obtain the page placement on the right. For example,
$P_6$ and $P_7$ belong to $Q_3$, which are the hottest pages, and
they are placed into the hottest rank (here $r_0$). At epoch $i+1$,
there are some changes in the MQ (the underlined page descriptors) and the
update page placement is shown on the right.

{\bf Page migrations.} %
To update page placement at each epoch, we first need to determine
the mappings from groups to ranks, i.e., which rank stores which set
(or group) of pages determined in Algorithm~\ref{alg:page}.
According to the current page placement among ranks, different
mappings from groups to ranks can result in different amounts of
page migrations, leading to different amounts of penalty in energy
and latency. We should find a mapping to minimize page migrations.

\begin{figure}
  \centering
  \includegraphics[width=0.8\linewidth]{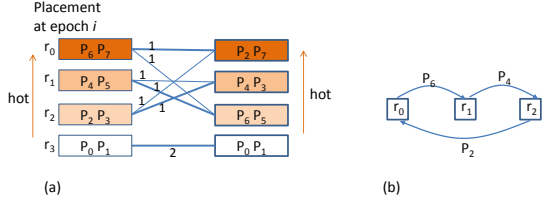}
  \vspace{-1ex}
  \caption{An example of page migrations: (a) calculate the maximum matching on the bipartite graph;
  (b) calculate Eulerian cycle for page migrations.}\label{fig:migration}
  \vspace{-4ex}
\end{figure}

We formulate this problem as finding a maximum weighted matching on a balanced bipartite
graph. The bipartite graph is defined as $G$ whose partition has the parts $U$
and $V$. %
Here, $U$ and $V$ are defined as the page placement among ranks in
the previous epoch and the page groups obtained with
Algorithm~\ref{alg:page} in the current epoch respectively. An edge
between $r_i$ and $S_j$ has a weight equaling to the number of pages
that exist in both rank $r_i$ and $S_j$. Since $|U| =|V|$, that is,
the two subsets have equal cardinality, $G$ is a balanced bipartite
graph. We find the maximum weighted matching of such a balanced
bipartite graph with the classic Hopcroft-Karp algorithm. The
maximum weighted matching means the maximum number of pages that are
common in both sides, and equivalently the
matching minimizes the number of page migrations. %
Figure~\ref{fig:migration}(a) illustrates the calculation of the maximum
matching for the bipartite graph for the example in Figure~\ref{fig:MQ}. In this
example, there are multiple possible matchings with the same maximum matching
weight. The thick edges represent one of such maximum matchings.

After the page mappings to individual ranks are determined, we know which pages
should be migrated from one rank to another. Then, we need to schedule the page
migrations in a manner to minimize the runtime overhead. Inspired by the
Eulerian cycle in graph theory, we develop a novel approach to perform multiple
page migrations in parallel. We consider a labeled directed graph $G_m$ where
each node represents a distinct rank. An edge from node $r_i$ to node $r_j$ is
labeled with a page descriptor, representing the pages to be migrated from rank
$r_i$ to rank $r_j$.

Each strongly connected component of $G_m$ has Eulerian cycles.
According to graph theory, a directed graph has a Eulerian cycle if and only if
every vertex has equal in degree and out degree, and all of its vertices with
nonzero degree belong to a single strongly connected component. By definition,
each strongly connected component of $G_m$ satisfies both properties, and thus
we can find Eulerian cycles in $G_m$. The page migration follows the Eulerian
cycle. We divide the Eulerian cycle into multiple segments so that each segment
is a simple path or cycle. Then, the page migrations in each segment can be
performed concurrently. Figure~\ref{fig:migration}(b) illustrates one example of
Eulerian cycle according to the maximum matching on the left. The three
migrations form a Eulerian cycle, and they are performed in one segment.

To facilitate concurrent page migrations according to the Eulerian cycle,
each rank is equipped with one extra row-buffer for storing the incoming page.
When migrating a page, a rank first writes the outgoing page to the buffer of
the target rank, and then reads the incoming page from its buffer. We provide
more implementation details and overhead analysis in Appendix A of the
supplementary file.

\subsection{Prediction Model}\label{subsec:model}

When a new slot starts, we run a prediction model against each rank.
The model predicts the idle period distribution. Our estimation
should be adapted to the potential changes in the page locality as
well as the set of pages in each rank.

We use the histogram to represent the idle period distribution.
Suppose the slot size is $T$ cycles, and the histogram has $T$
buckets. We denote the histogram to be $\mathit{Hist[i]}$, $i=0$, 1,
..., $T$. The histogram means there are $\mathit{Hist[i]}$ number of idle
periods with the length of $i$ cycles each. One issue is the storage
overhead of the histogram. A basic approach is to store the
histogram into an array, and each bucket is represented as a 32-bit
integer. However, the storage overhead of this basic approach is too
high. Consider a slot size of $10^8$ cycles in our experiments. The
basic approach consumes around 400MB per rank. In practice, the
histogram is usually very sparse, and there are at most $\sqrt{T}$
idle periods longer than $\sqrt{T}$ cycles. Thus, we develop a
simple approach to store the short and the long idle periods
separately. In particular, we maintain two small arrays: the
histogram counters for the short idle periods no longer than
$\sqrt{T}$ cycles, and another array of $\sqrt{T}$ integers to store
the actual lengths of the long idle periods that are longer than
$\sqrt{T}$ cycles. This simple approach reduces the storage overhead
to $2\sqrt{T}$ integers. It takes only 80KB per rank to support a
slot size of $10^8$ cycles. We calculate the histogram for idle
periods longer than $\sqrt{T}$ cycles with just one scan on the
array.

Our estimation specifically consider page migrations. If the new slot is
\emph{not} the beginning of an epoch, there is no page migration and
we use the actual histogram in the previous slot,
$\mathit{Hist}'[i]$, to be the prediction of the current slot, i.e.,
$\mathit{Hist}[i]=\mathit{Hist}'[i]$ ($0\le i \le T$). Otherwise, we
need to combine the access locality of the migrated pages with the
historical histogram.

Our estimation after page migration works as follows. We model the references to
the same page conforming to a Poisson distribution.
Suppose a page $i$ is accessed with $f$ times in a slot. Under the Poisson
distribution, the probability of having one access to page $i$ within a cycle is
$p_i=\frac{g\cdot f}{T}$, where $g$ is the memory access latency. In our
implementation, we take advantage of the frequency counter and the expiration
time in the MQ structure (as described in the previous section) to approximate
$p_i$. This already offers a sufficiently accurate approximation in practice.
Given a rank consisting of $N$ pages (pages 0, 1, ..., $N-1$), the probability
of an idle cycle in the rank is $Q=(1-p_0)\cdot(1-p_1)... \cdot(1-p_{N-1})$.
Based on $Q$, we can estimate the probability of forming an idle period with
length of $k$ cycles (followed by a busy cycle in $(k+1)^{th}$ cycle). That is,
the probability of having an idle period of $k$ cycles is $W_k=Q^k\cdot (1-Q)$.

We denote the old values of those probability values in the previous
epoch to be $W_k'$ ($k$=0, 1, 2, ..., $T$). After page migrations,
we calculate $W_k$ ($k$=0, 1, 2, ..., $T$) according to the updated pages in
the rank. Given the actual histogram in the previous slot,
$\mathit{Hist}'[i]$, we can estimate the histogram of the new
slot with the ratio $W_i/W_i'$, that is, $\mathit{Hist}^+[i]
=W_i/W_i'\cdot\mathit{Hist}'[i]$. Finally, we normalize the
histogram so that the histogram represents the total time length of
a slot. Denote
$s'$=$\sum_{i=0}^{T}(\mathit{Hist}^+[i] \cdot (i+g))$.
We normalize the histogram with the value of $\frac{T}{s'}$, i.e.,
$\mathit{Hist}[i] =\mathit{Hist}^+[i]\times \frac{T}{s'}$. We use
$\mathit{Hist}[i]$ as the prediction on the idle period distribution
for the new slot.

Based on the prediction model, we will estimate the power-down timeout
for the new slot in the next sub-section.

\subsection{Adaptive Demotions}\label{subsec:demotion}

With the predicted idle period distribution, there are opportunities
to avoid the state transitions upon those short idle periods, and to
have instant state transitions for long idle periods. For example,
if we know all the idle periods are expected to be very long, we can
set the power-down timeout to be zero, thus performing instant state
transitions. Thus, we have developed a simple approach to reduce the
total penalty of state transitions. The basic idea is, for each
low-power state, we use one power-down timeout to determine the state
transition within the entire slot. Suppose a DDR-series memory
architecture has $M$ low-power states, denoted as $S_1,\ldots,S_M$
in the descending order of their power consumptions. For each
low-power state $S_i$, RAMZzz performs the state transition to $S_i$
after an idle period threshold $\Delta_i$. If the idle period is
shorter than $\Delta_i$, RAMZzz does not make the state transition
to $S_i$.

Since we need to exploit all power states in order to adapt to
different workloads and different memory architectures, a naive
approach is to consider all the possible state transitions. However,
the demotion configuration of the naive approach is too complex to
derive. Instead of considering all state transitions, we view
multiple state transitions as a chain of state transitions from
higher-power states to lower-power states. We will show that our
adaptive demotion scheme can identify the unnecessary power states
in a chain of states, and thus further simplify the demotion scheme.
We define the demotion configuration to be a vector of demotion
times $\vec{\Delta} = (\Delta_1,\ldots,\Delta_M)$ where $\Delta_i$
represents the power-down timeout of low-power state $S_i$, $i = 1,
\cdots, M$. In the chain, when the idle period length is longer than
$\Delta_i$, we perform states transition from $S_\mathit{i-1}$ to
$S_\mathit{i}$.

Given the estimated histogram on idle periods, we estimate the demotion
configuration of each rank for a given optimization goal. We use energy
consumption as the optimization goal to illustrate our algorithm design on
estimating the demotion configuration. One can similarly extend it to other
goals such as ED$^2$. Since the choice on different power-down timeouts does not
affect the energy consumption of memory reads and writes, our metric can be
simplified as the total energy consumption of background power and the state
transition penalty.

We analyze the energy consumption on different demotions over an idle
period. Suppose the idle period length is $t$ cycles, and the power
consumption of active state $ACT$ and a low-power state $S_i$ are
$P_\mathit{ACT}$ and $P_\mathit{S_i}$ ($i = 1, \cdots , M$),
respectively. Given a demotion configuration $\vec{\Delta}$, if $t
\le \Delta_1$, there is no state transition to low-power states.
Otherwise, denote $I(t)$ to be the maximum $i$ such that $\Delta_i < t$
($i = 1, \cdots, M$). In the chain, there are at most $I(t)$ state
transitions, from $S_1$ to $S_\mathit{I(t)}$. At the end of the idle period, a
memory access comes and the rank transits from low-power state $S_\mathit{I(t)}$
back to ACT. Thus, the energy consumption of the idle period can be
calculated as $\mathcal{B}(\vec{\Delta}, t)$ in
Eq.~(\ref{eq:multi_energy}).
{\footnotesize
\begin{equation}
\label{eq:multi_energy}
\begin{array}{rcl}
\mathcal{B}(\vec{\Delta}, t) & = & P_\mathit{ACT} \cdot \Delta_1 +
\sum_{j=1}^{\mathit{I(t)}-1}(P_{\mathit{S_j}} \cdot (\Delta_\mathit{j+1} -
\Delta_j))
\\ [1ex]
& & + P_\mathit{S_\mathit{I(t)}} \cdot (t - \Delta_\mathit{I(t)}) +
E_\mathit{S_\mathit{I(t)}}
\end{array}
\end{equation}
}%
\noindent where $E_\mathit{S_\mathit{I(t)}}$ is resynchronization energy
penalty from low-power state $S_\mathit{I(t)}$ back to ACT.

Given the histogram $\mathit{Hist}[t]$ ($t=0, 1, \cdots, T$), each
$\mathit{Hist}[t]$ means there are $\mathit{Hist}[t]$ idle periods with length
$t$ cycles. We can calculate the total energy consumption for all the idle
periods, as $E(\vec{\Delta})$ in Eq.~(\ref{eq:total_energy}).
{\footnotesize
\begin{equation}
\label{eq:total_energy}
E(\vec{\Delta}) = \sum_{t=0}^{\Delta_1}(P_\mathit{ACT} \cdot t \cdot \mathit{Hist}[t]) + \sum_{t=\Delta_1 + 1}^{T}(\mathcal{B}(\vec{\Delta}, t) \cdot \mathit{Hist}[t])
\end{equation}
}%
\noindent RAMZzz also allows users to specify a delay budget to
limit the delay penalty incurred by state resynchronization. We
calculate the total resynchronization delay as $D(\vec{\Delta})$ in
Eq.~(\ref{eq:total_delay}).
{\footnotesize
\begin{equation}
\label{eq:total_delay} D(\vec{\Delta}) =
\sum_{t=\Delta_1 + 1}^{T}(R_\mathit{S_\mathit{I(t)}} \cdot \mathit{Hist}[t])
\end{equation}
}%
\noindent where $R_\mathit{S_\mathit{I(t)}}$ is resynchronization delay from
low-power state $S_\mathit{I(t)}$ back to ACT. Our goal is to determine the
suitable demotion configuration $\vec{\Delta}$ so that $E(\vec{\Delta})$ is
minimized. If a delay budget is given, we choose the $\vec{\Delta}$ value that
minimizes $E(\vec{\Delta})$ with the constraint that the total delay
$D(\vec{\Delta})$ is no larger than the given delay budget.

We note that $E(\vec{\Delta})$ is neither concave nor monotonic.
Therefore, we have to iterate all the possible values for $\Delta_i$=0, 1, ...,
$T$ ($i = 1, \cdots, M$), and find the best combination of $\Delta_i$ ($i = 1,
\cdots, M$). The complexity of this naive approach of increases exponentially
with the number of low-power states in the DRAM architecture. In the following,
we develop an efficient greedy algorithm to find a reasonably good demotion
configuration (illustrated in Algorithm~\ref{alg:online_alg}).

We start by assuming that only one low-power state is used in the
entire slot, and select the best suitable low-power state and its
power-down timeout which leads to a smallest estimated $E(\vec{\Delta})$
among all $M$ low-power states. Then, we keep the estimated demotion
time of the selected low-power state unchanged, and select a new
low-power state and its power-down timeout from the rest $M-1$ low-power
states, which results in a smallest estimated $E(\vec{\Delta})$. We
repeat this process to add one more new low-power state into the previous
selected subset of low-power states together with its power-down timeout
in each step. Algorithm~\ref{alg:online_alg} has much lower
computational complexity than the naive approach.

\begin{algorithm}[t]
\caption{The greedy algorithm to find the suitable demotion
configuration $\vec{\Delta}$}
\begin{footnotesize}
\begin{algorithmic}[1]\label{alg:online_alg}
    \item[\textbf{Input:}] ~~\\
    All low-power states set $\vec{S} = (S_1, \ldots, S_M)$,
    and associated power consumptions set $\vec{P} = (
    P_\mathit{S_1}, \ldots, P_\mathit{S_M})$;
    \item[\textbf{Initialization:}] ~~\\
    $\vec{\Delta} = \phi$, $\vec{S}_\mathit{select} = \phi$;
    \WHILE{$|\vec{S}_\mathit{select}| \neq M $}
    \FORALL{$S_i$ $\in$ $\vec{S}$}
    \STATE Add $S_i$ into $\vec{S}_\mathit{select}$;
    \FOR{each possible $\Delta_i$ value}
    \STATE Calculate $E(\vec{\Delta})$ using
    Eq.~(\ref{eq:total_energy}) with selected low-power states subset
    $\vec{S}_\mathit{select}$;
    \ENDFOR
    \STATE Find the suitable $\Delta_i$ that has the best
    $E(\vec{\Delta})$;
    \STATE Remove $S_i$ from $\vec{S}_\mathit{select}$;
    \ENDFOR
    \STATE Find the low-power state $S_k$ that has a best $E(\vec{\Delta})$;
    \STATE Add $\Delta_k$ into $\vec{\Delta}$;
    \STATE Remove $S_k$ from $\vec{S}$;
    \ENDWHILE
    \item[\textbf{Output:}] ~~\\
    power-down timeout set $\vec{\Delta}$
\end{algorithmic}
\end{footnotesize}
\end{algorithm}

Algorithm~\ref{alg:online_alg} has a low runtime overhead in most cases.
First, it does not need to iterate through all values from 0 to $T$ ($T$ is the
slot size). Instead, it only searches those values with non-zero frequencies in
the predicted histogram. This number is far smaller than $T$ in practice.
Second, as more low-power states are selected during the process (one state per
step), the search space for rest low-power states is further reduced since the
power-down timeout of $S_i$ is bounded by that of $S_\mathit{i-1}$ and
$S_\mathit{i+1}$, i.e., $\Delta_\mathit{i-1} \leq \Delta_i \leq
\Delta_\mathit{i+1}$. Moreover, we further optimize
Algorithm~\ref{alg:online_alg} in two ways. First, we adopt the branch-bound
optimization in order to further reduce the search space (That is, we try
possible values from the highest to the lowest until the program performance
penalty violates the given budget). Second, we use an exponential search
approach by iterating in the form of $2^i$ ($0 \leq i \leq log_2 T$) for each
power-down timeout. On the current architectures, the greedy algorithm has a low
runtime overhead and provides near-optimal demotion configurations, as shall be
shown in our evaluation (Section~\ref{sec:evaluation}).

The adaptive demotion scheme is applied on each rank at the beginning of a slot.
The demotion configurations can be different among different ranks and at
different slots. This is a distinct feature of adaptive demotion, in comparison
with the previous work on static demotion
schemes~\cite{Huang:2003:DIP:1247340.1247345,Huang:2005:IEE:1077603.1077696,Delaluz:2002:SDE:513918.514095,Fan:2001:MCP:383082.383118}.

\subsection{Other Implementation Issues}\label{subsec:other_issues}

RAMZzz can be implemented with a combination of modest hardware and
software supports. First, RAMZzz adds a few new components to the
memory controller and operating system. Following the previous
study~\cite{Ramos:2011:PPH:1995896.1995911}, RAMZzz extends a
programmable controller~\cite{Xilinx} by adding its own new
components. Four new modules including MQ, Migration, Remap and
Demotion are added into the memory controller for implementing the
functionality of page grouping, page migration, page remapping and
power state control in RAMZzz, respectively. Other functionalities
including page grouping and the prediction model are offloaded to
the OS (like previous
studies~\cite{Ramos:2011:PPH:1995896.1995911,Huang:2003:DIP:1247340.1247345}).
Second, we note that the structure complexity and storage overhead
of RAMZzz are similar to the previous proposals, e.g.,
~\cite{Huang:2005:IEE:1077603.1077696,
Deng:2011:MAL:1950365.1950392, Diniz:2007:LPC:1250662.1250699,
Fan:2001:MCP:383082.383118, Sudan:2010:MID:1736020.1736045}. For
example, our design has small DRAM space requirement (less than 2\%
of the total amount of DRAM). We have included both performance and
energy penalties of these new modules in our simulation and
evaluation, and demonstrated that their overheads are acceptable in
current architectures in Section~\ref{sec:evaluation}.

In Appendix A of the supplementary file, we provide more discussions
on implementation details, including energy/performance/storage
overhead analysis and optimizations of RAMZzz.

\section{Evaluation}\label{sec:evaluation}
In this section, we evaluate our design using ED$^2$ and energy
consumption as metrics. We have conducted a number of experiments. We compare
the behavior of RAMZzz and its alternatives in order to show the effectiveness
of RAMZzz on different memory architectures, and the impact of individual
techniques. We focus on the DRAM component. For the space interests, we present
the results with ED$^2$ as the optimization goal in this paper, and leave the
results with the total energy consumption as the optimization metric to Appendix
C of the supplementary file. We also study the impact of RAMZzz on full
system energy savings in Appendix D of the supplementary file.

\subsection{Methodology} \label{subsec:meth}

Our evaluation is based on trace-driven simulations. In the first
step, we use cycle-accurate simulators to collect memory access
traces (last-level cache misses and writebacks) from running
benchmark workloads. In the second step, we replay the traces using
our detailed memory system simulator. Our simulation models all the
relevant aspects of the OS, memory controller, and memory devices,
including page replacements, memory channel and bank contention,
memory device power and timing, and row buffer management. The
memory controller exploits the page interleaving mechanism. More
implementation details can be found in Appendix A of the
supplementary file.

We evaluate workloads from SPEC 2006 and
PARSEC~\cite{bienia11benchmarking}. We use two different approaches to collect
their memory traces: one with PTLSim~\cite{ptlsim} simulator and the other with
Sniper~\cite{carlson2011etloafsaapms} simulator. On the one hand, the memory
footprints of SPEC 2006 are usually smaller than those of PARSEC. On the other
hand, PARSEC cannot run on PTLSim, which we use to collect the memory trace from
SPEC 2006. Also, PTLSim can offer more control on the hardware configurations
for sensitivity studies.

{\bf SPEC 2006 Workloads.} We use PTLSim~\cite{ptlsim} to collect memory access
traces of SPEC 2006 workloads. The main architectural characteristics of
the simulated machine are listed in Table~\ref{tb:ptlsim_config}. We model and
conduct the evaluation with an in-order processor following previous
studies~\cite{Ramos:2011:PPH:1995896.1995911,Deng:2011:MAL:1950365.1950392}.
More complex and recent processors are studied with Sniper-based simulations.
We evaluate our techniques with three different memory architectures, as shown
in Table~\ref{tb:DRAMState} (Section~\ref{subsec:dram_power}). Those memory
architectures are used in different computing systems. We simulate different
capacities (1GB, 2GB and 4GB) and different numbers of ranks (4, 8, 12 and 16)
for the memory system. All the ranks have the same configurations (DRAM
parameters) and capacities. By default, we assume a 2GB DRAM with 8 ranks. We
pick these small memory sizes to match the footprint of the workloads'
simulation points. We calculate the memory power consumption following Micron's
System Power Calculator~\cite{calc}, with the power and delay illustrated in
Table~\ref{tb:DRAMState}. The energy and performance overheads caused by new MC
and OS modules (e.g., remapping, migration and demotion) are derived from our
analysis in Appendix A of the supplementary file, which are consistent with
those of other
authors~\cite{Ramos:2011:PPH:1995896.1995911,Guo:2010:RCA:1815961.1816012,Sudan:2010:MID:1736020.1736045}.

\begin{table}[t]
  \centering
  \caption{Architectural configurations of PTLSim. The default setting is
  highlighted.}\label{tb:ptlsim_config}
  \vspace{-2ex}
  {\footnotesize
    \begin{tabular}{|p{3.1cm}|p{4.6cm}|}
    \hline
    Component & Features \\
    \hline
    CPU & In-order cores running at 2.66GHz \\
    Cores & 4 \\
    TLB & 64 entries \\
    L1 I/D cache (per core) & 48KB \\
    L2/L3 cache (shared) & 256KB/4MB \\
    Cache line/OS page size & 64B/4KB \\
    \hline
    DRAM & DDR3-1333, DDR2-800, LPDDR2-800 \\
    Channels & 4 \\
    Ranks & 4, {\bf 8}, 12, 16\\
    Capacity (GB) & 1, {\bf 2}, 4\\
    Delay and Power & see Table~\ref{tb:DRAMState} \\
    \hline
    \end{tabular}
  }
  \vspace{-2ex}
\end{table}

We have used 19 applications from SPEC 2006 with the \emph{ref} inputs. These
workloads have widely different memory memory access rates, footprints and
localities. Due to space limitations, we do not present the results for single
applications; instead, we report their geometric mean ({\bf GM}), and also four
particular applications with different memory intensiveness. They are omnetpp,
cactusADM, mcf and lbm (denoted as S1, S2, S3 and S4, respectively). To assess
our algorithm under the context of multi-core CPUs, we study mixed workloads of
four different applications from SPEC 2006 (Table~\ref{tb:mix}). The four
workloads start at the same time. The mixed workloads form multi-programmed
executions on a four-core CPU, ordered by the average number of memory accesses
($\mathit{Mean}$). The standard deviation and mean values are calculated based
on memory access statistics per $5\times 10^8$ CPU cycles. For each workload, we
select the simulation period of $15 \times 10^9$ cycles in the original PTLSim
simulation, which represents a stable and sufficiently long execution behavior.

\begin{table}[t]
  \centering
  \caption{Mixed workload: memory footprint (FP), memory accesses statistics per
  $5\times 10^8$ cycles ($\mathit{Mean}$ and
  $\frac{\mathit{Stdev}}{\mathit{Mean}}$).} \label{tb:mix}
  \vspace{-2ex}
  {\footnotesize
    \begin{tabular}{|c|p{0.63cm}|p{0.53cm}|p{0.53cm}|c|}
    \hline
    Name & FP (MB) & $\mathit{Mean}$ ($10^6$) &
    $\frac{\mathit{Stdev}}{\mathit{Mean}}$ & Applications \\
    \hline
    M1&661.3&0.6&1.02 & gromacs, gobmk, hmmer, bzip \\
    \hline
    M2&1477.4&1.7&1.11 &bzip, soplex, sjeng, cactusADM \\
    \hline
    M3&626.6&2.9&0.59 &soplex, sjeng, gcc, zeusmp \\
    \hline
    M4&537.8&3.5&0.47 &zeusmp, gcc, leslie3d, omnetpp \\
    \hline
    M5&1082.9&4.4&0.71 &gcc, leslie3d, calculix, gemsFDTD \\
    \hline
    M6&988.2&7.8&0.40 &libquantum, milc, mcf, lbm \\\hline
    \end{tabular}
  }
 \vspace{-4ex}
\end{table}

{\bf PARSEC Workloads.} Since current PTLSim cannot support PARSEC
benchmarks, we use another
simulator--Sniper~\cite{carlson2011etloafsaapms} to collect memory
access traces of PARSEC. We also note that, some workloads in PARSEC
like \emph{dedup, facesim, canneal} cannot run successfully. In this
study, we focus on the results for four applications including
blackscholes, bodytrack, ferret and streamcluster. By default, we
use the simulated CPU architecture as shown in
Table~\ref{tb:sniper_config} (Intel's Gainestown CPUs), which
simulates a four-core processor running at 2.66 GHz based on the
Intel's Nehalem micro architecture. By default, we simulate a
four-core CPU, and 2GB DRAM with 8 ranks. The memory architecture
has the same power consumption and performance configurations as the
PTLSim-based simulations. Also, we conduct simulation studies with
main-stream servers with a large number of cores and large memory
capacity in Section~\ref{subsec:large_sys}. Each PARSEC workload
runs with four threads, and each thread is assigned to one core. We
use the \emph{sim-medium} inputs for PARSEC workloads, and perform
the measurement on the specified Region-of-Interest (ROI) of PARSEC
workloads~\cite{bienia11benchmarking}.

\begin{table}[t]
  \centering
  \caption{Architectural configurations of Sniper. The default setting is
  highlighted.}\label{tb:sniper_config}
  \vspace{-2ex}
  {\footnotesize
    \begin{tabular}{|p{3.1cm}|p{4.6cm}|}
    \hline
    Component & Features \\
    \hline
    CPU & Out-order cores running at 2.66GHz \\
    Cores & {\bf 4}, 8, 16 \\
    DTLB/ITLB & 64/128 entries \\
    L1 I/D cache (per core) & 32KB/32KB\\
    L2 cache (per core) & 256KB \\
    L3 cache (shared) & 8MB \\
    Cache line/OS page size & 64B/4KB \\
    \hline
    DRAM & DDR3-1333 \\
    Channels & 4 \\
    Ranks & {\bf 8}, 16\\
    Capacity (GB) & {\bf 2}, 32, 64\\
    Delay and Power & see Table~\ref{tb:DRAMState} \\
    \hline
    \end{tabular}
  }
  \vspace{-4ex}
\end{table}

{\bf Comparisons.} In our previous study~\cite{Wu:2012:RRD:2388996.2389040}, we
have already shown that the preliminary version of RAMZzz significantly
outperforms other power management
techniques~\cite{Lebeck:2000:PAP:378993.379007,Fan:2001:MCP:383082.383118,
Huang:2005:IEE:1077603.1077696} in terms of both ED$^2$ and energy consumption.
Due to the space limitation, we focus on evaluating the impact of individual
techniques developed in this paper. In particular, we consider two RAMZzz
variants namely RZ--SP and RZ--SD. They are the same as RAMZzz except that
RZ--SP uses the static page management scheme without page migrations, whereas
RZ--SD uses the static demotion scheme. The static demotion scheme simply
transits a rank to a pre-selected low-power state according to the prediction
model.

In addition to RAMZzz variants, we also simulate the following
techniques for comparison. \emph{All the metrics reported in this
paper are normalized to those of BASE.}
\begin{itemize}
 \item {\bf No Power Management (BASE)}: no power management technique is used,
 and ranks are kept active even when they are idle.
 \item {\bf Ideal Oracle Approach (ORACLE)}: ORACLE is the same as RAMZzz,
 except the power-down timeout in ORACLE is determined with the future information,
 instead of history. Specifically, at the beginning of each slot, we perform an
 offline profiling on the current slot, and get the real histogram of idle
 periods. Based on the histogram, we calculate the optimal power-down timeout.
\end{itemize}

RAMZzz allows users to specify the slot and epoch sizes and delay
budgets. By default, the slot size is $10^8$ cycles and an epoch
consists of ten slots ($10^9$ cycles), and delay budget is set to be
4\% of the slot size. We evaluate the impacts of these parameters in
Appendix E of the supplementary file.

\begin{figure}
  \centering
  \includegraphics[width=0.8\linewidth]{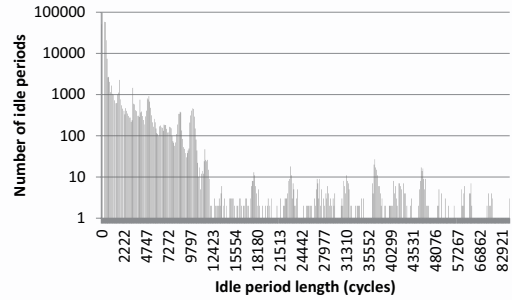}
  \vspace{-2ex}
  \caption{The histogram of idle periods with M2 on Rank 0.} \label{fig:hist}
  \vspace{-4ex}
\end{figure}

{\bf Idle Period Distribution.} We study the distribution of idle periods.
Figure~\ref{fig:hist} shows the histogram of idle period lengths of the
collected traces on Rank 0 on DDR3 under BASE approach. Many idle periods are
too short to be exploited for state transitions, e.g., shorter than the
threshold idle period length for demoting to SR\_FAST (2500 cycles on DDR3). We
observed similar results on other ranks.

\subsection{Results on SPEC 2006 Workloads}
\label{subsec:evaled2}

\begin{figure*}[htb]
\centering
\subfigure[Results on ED$^2$ for DDR3]{
\label{fig:ddr3_ed2_overall}
\begin{minipage}[b]{0.31\linewidth}
\centering
\includegraphics[width=0.75\linewidth]{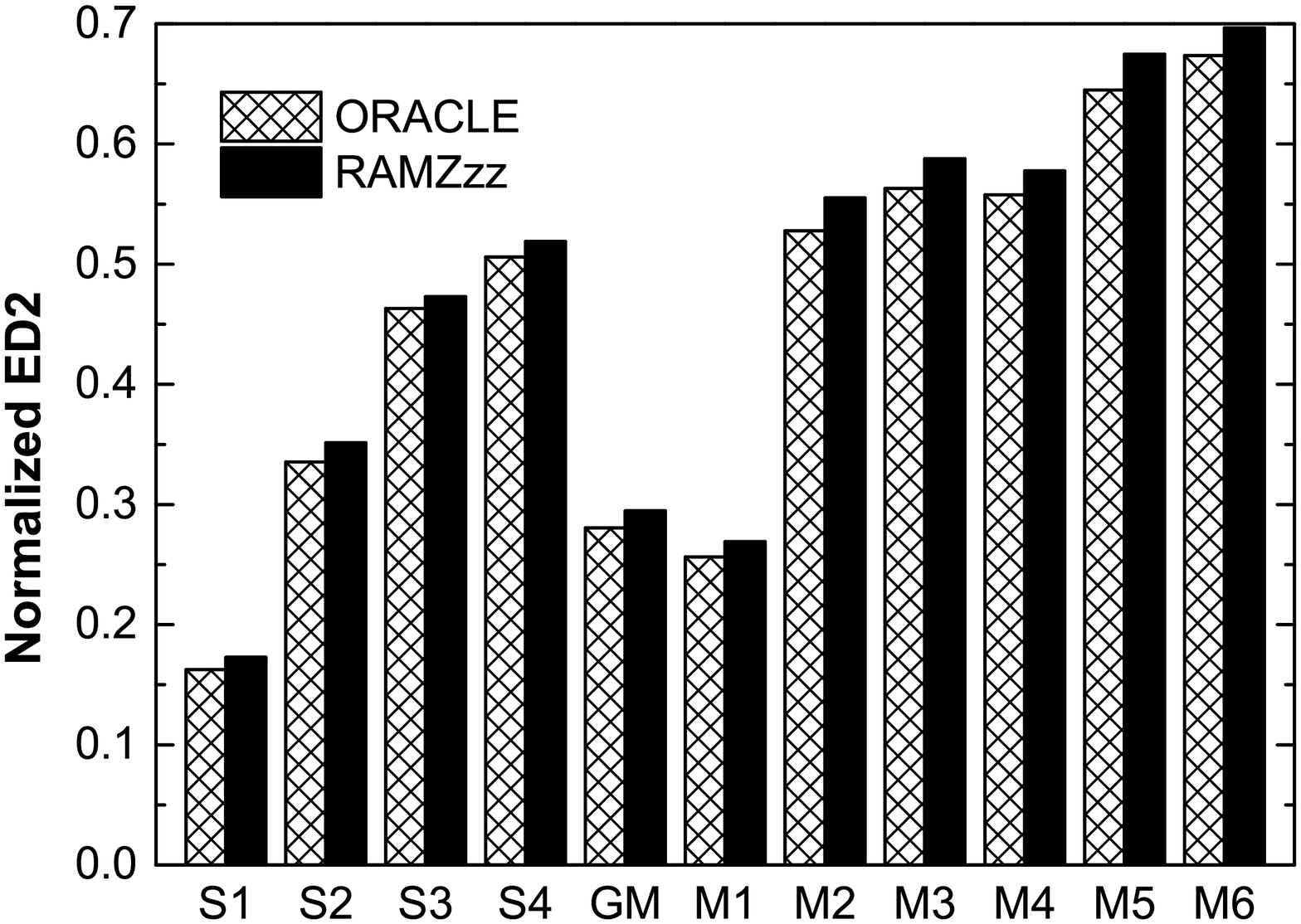}
\end{minipage}}
\hspace{0.1cm}
\subfigure[Results on ED$^2$ for DDR2]{
\label{fig:ddr2_ed2_overall}
\begin{minipage}[b]{0.31\linewidth}
\centering
\includegraphics[width=0.75\linewidth]{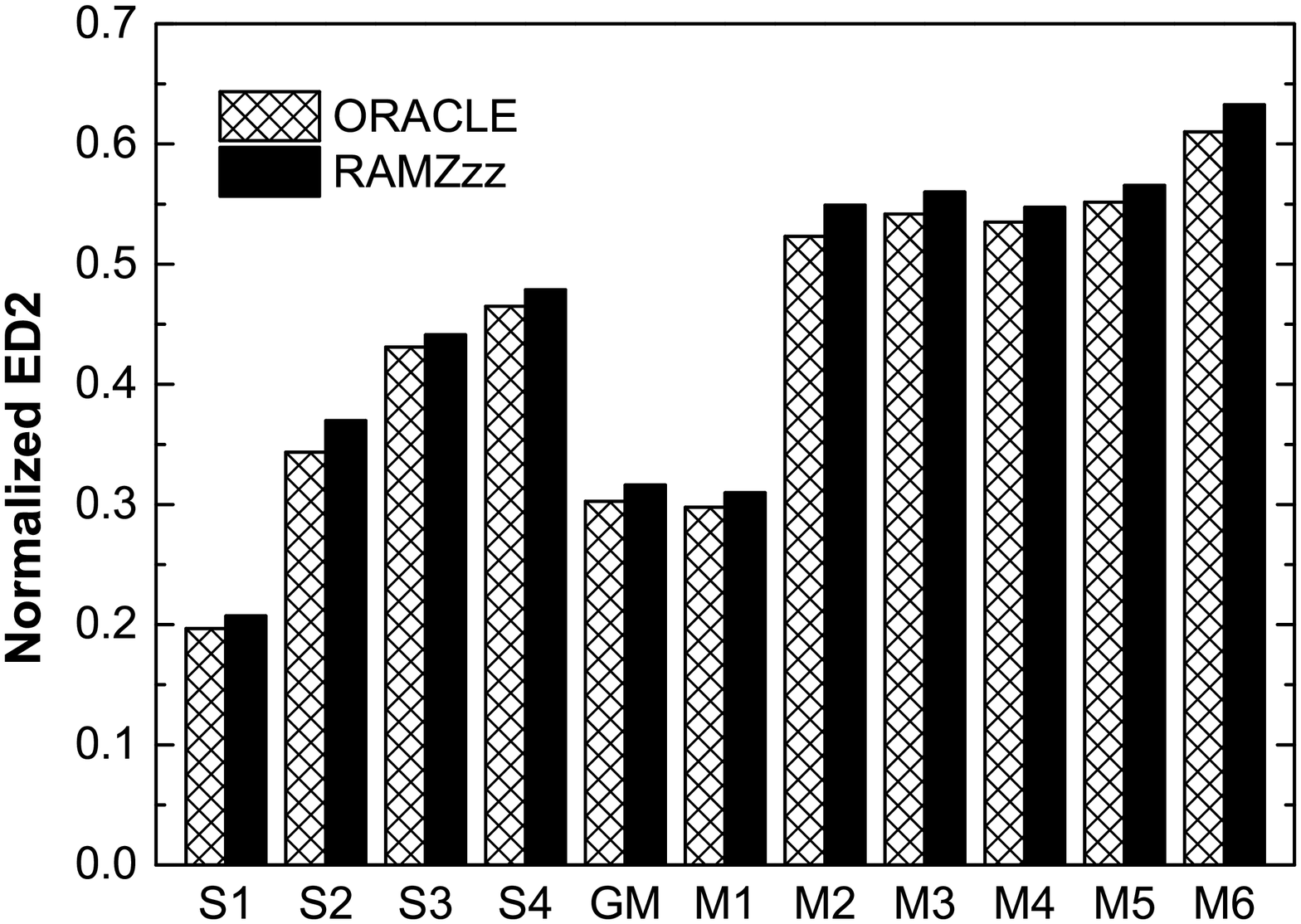}
\end{minipage}}
\hspace{0.1cm}
\subfigure[Results on ED$^2$ for LPDDR2]{
\label{fig:lpddr2_ed2_overall}
\begin{minipage}[b]{0.31\linewidth}
\centering
\includegraphics[width=0.75\linewidth]{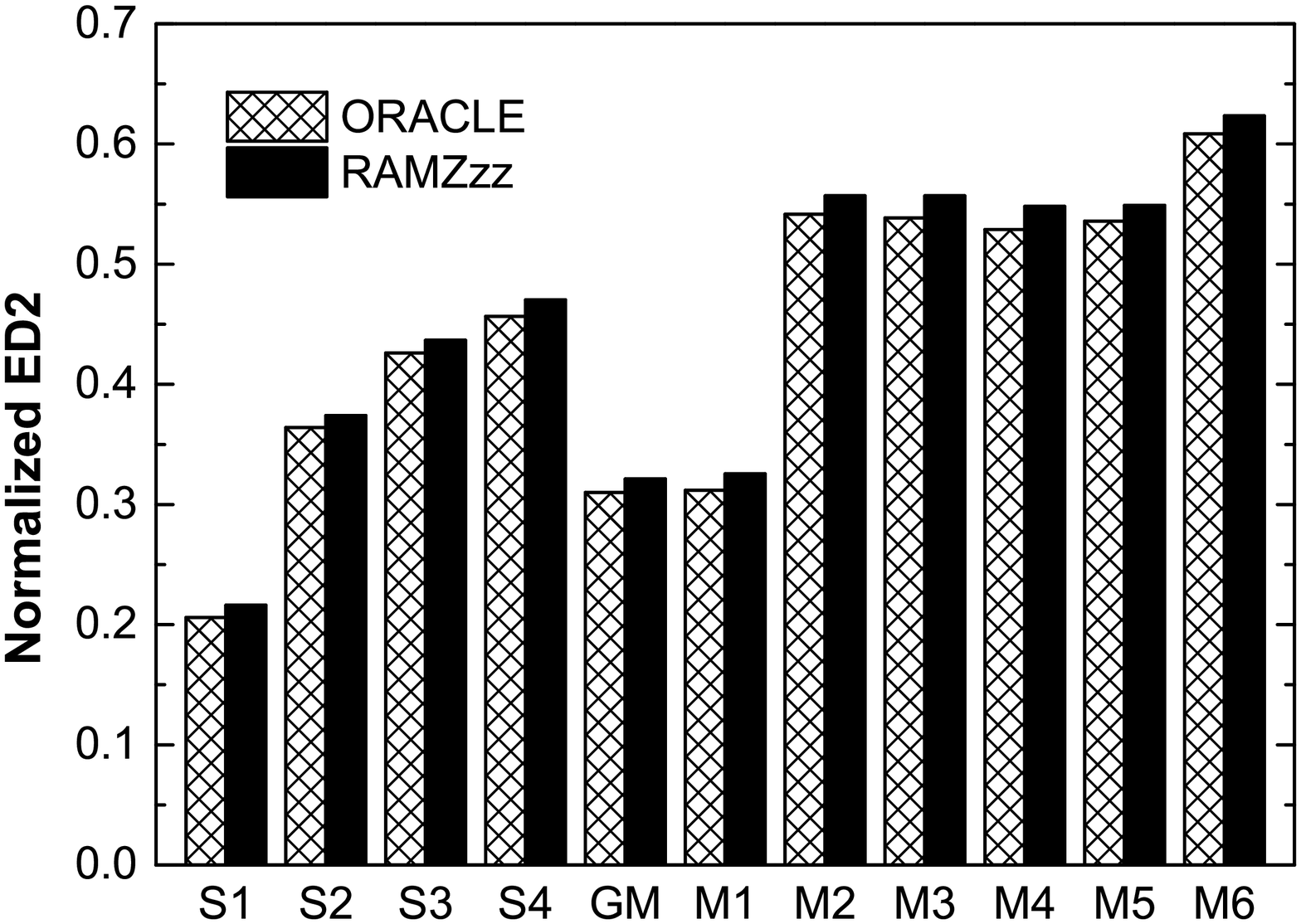}
\end{minipage}}
\caption{Comparing ED$^2$ of RAMZzz and ORACLE with the optimization goal of
ED$^2$ on three memory architectures.}
\label{fig:dram_ed2_overall}
\vspace{-1ex}
\end{figure*}

\begin{figure*}[htb]
\centering
\subfigure[The breakdown of time for DDR3]{
\label{fig:ddr3_ed2_ramzzz_bd}
\begin{minipage}[b]{0.31\linewidth}
\centering
\includegraphics[width=0.90\linewidth]{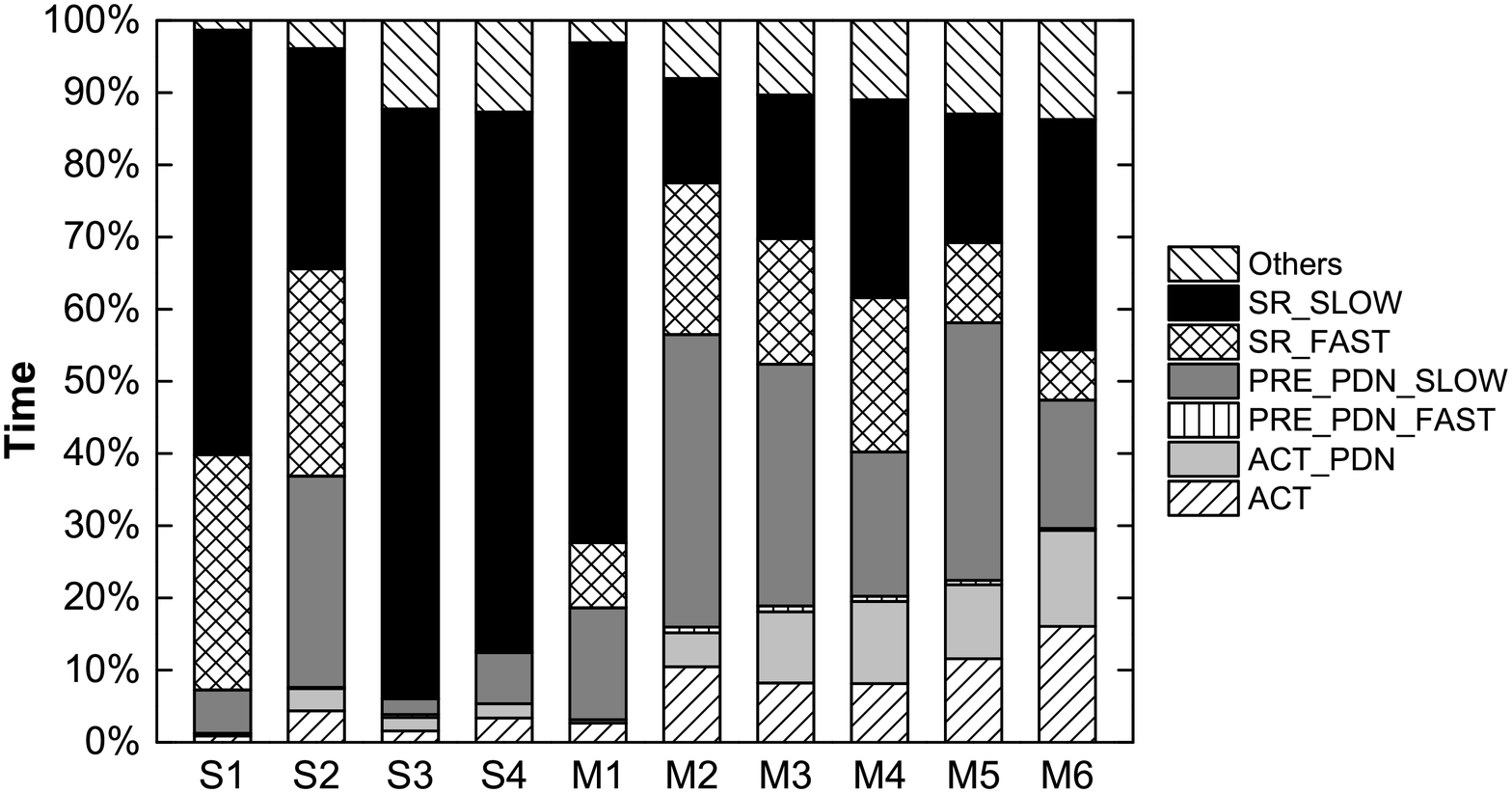}
\end{minipage}}
\hspace{0.1cm}
\subfigure[The breakdown of time for DDR2]{
\label{fig:ddr2_ed2_ramzzz_bd}
\begin{minipage}[b]{0.31\linewidth}
\centering
\includegraphics[width=0.90\linewidth]{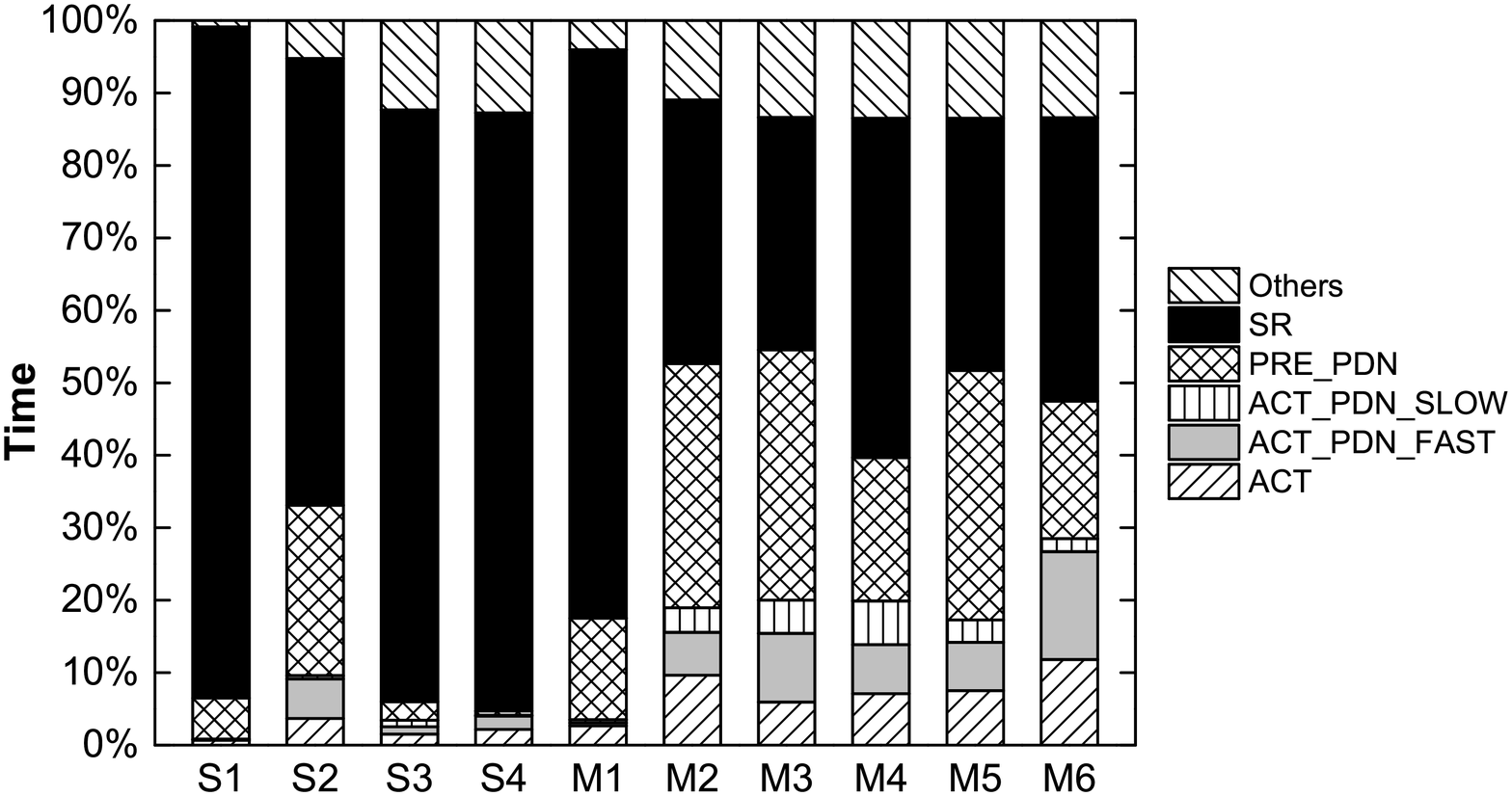}
\end{minipage}}
\hspace{0.1cm}
\subfigure[The breakdown of time for LPDDR2]{
\label{fig:lpddr2_ed2_ramzzz_bd}
\begin{minipage}[b]{0.31\linewidth}
\centering
\includegraphics[width=0.83\linewidth]{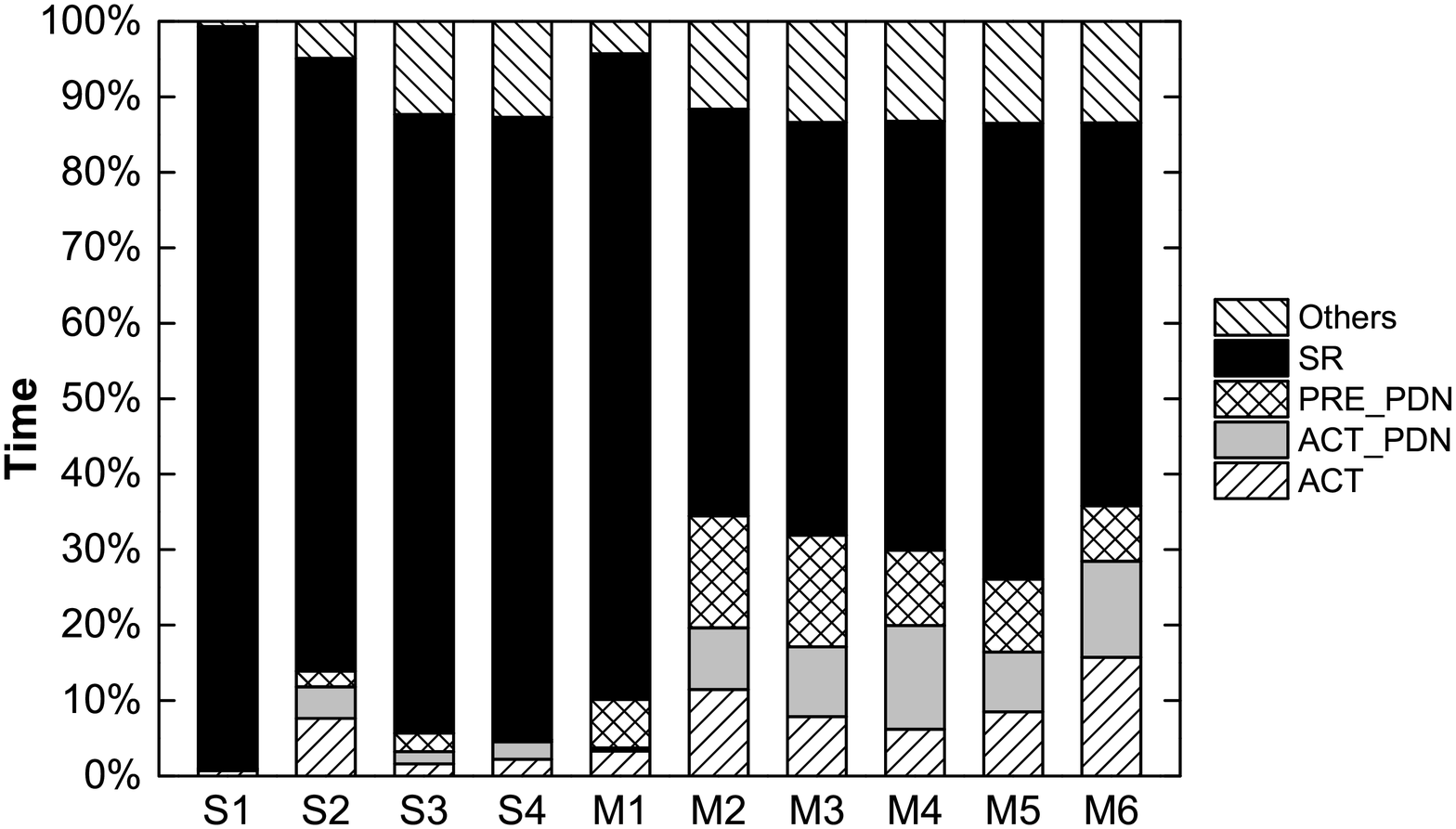}
\end{minipage}}
\caption{The breakdown of time stayed in different power states for RAMZzz with
the optimization goal of ED$^2$.}
\label{fig:dram_ed2_ramzzz_bd}
\vspace{-3ex}
\end{figure*}

We first compare the algorithms with the optimization goal of ED$^2$ on
SPEC 2006 workloads, because ED$^2$ is a widely used metric for energy
efficiency.

We study the overall impact of RAMZzz in comparison with BASE and ORACLE. The
comparison with BASE shows the overall effectiveness of energy saving techniques
of RAMZzz, and the comparison with ORACLE shows the effectiveness of our
prediction model. Figure~\ref{fig:dram_ed2_overall} presents normalized ED$^2$
results for RAMZzz and ORACLE approaches on three different DRAM architectures
(more randomly-mixed workloads, which are chosen from SPEC 2006, are evaluated
in Appendix B of the supplementary file). If the normalized ED$^2$ of an approach
is smaller than 1.0, the approach is more energy efficient than BASE.

Thanks to the rank-aware power management, RAMZzz is significantly more
energy-efficient than BASE. Compared with BASE, the reduction on ED$^2$ is
64.2\%, 63.3\% and 63.0\% on average on DDR3, DDR2 and LPDDR2, respectively. The
reduction is more significant on the workloads of single applications (e.g.,
S1--S4) than the mixed workloads. There are two main reasons. First, since the
single-application workload has a smaller memory footprint, the page migration
has a smaller overhead and the number of cold ranks is larger. The number of
page migrations becomes very small after the first few epochs. In contrast, the
execution process of the workloads with a large memory footprint (such as M5 and
M6) consistently has a fair amount of page migrations at all epochs.
Secondly, on single-application workloads, there are more opportunities for
saving background power using lower-power states (such as SR\_FAST and SR\_SLOW
in DDR3, SR in DDR2 and LPDDR2). Figure~\ref{fig:dram_ed2_ramzzz_bd} shows the
breakdown of time stayed in different power states for RAMZzz on DDR3, DDR2 and
LPDDR2. In Figure~\ref{fig:dram_ed2_ramzzz_bd}, each power state represents the
percentage of time when ranks are in this state during the total simulation
period. And \emph{Others} represents the percentage of time that includes DRAM
operations, page remapping delay, page migration delay and resynchronization
delay. As the workload becomes more memory-intensive, the portion of time that a
rank is in lower-power states becomes less significant, indicating that many
idle periods are too short and they are not worthwhile to perform state
transitions into lower-power states (even with page migration).
For the less memory-intensive workloads like S1--4 and M1, lower-power states
have very significant portions in the total simulation time, indicating
significant energy saving compared with BASE.

\begin{figure*}[htb]
\begin{center}
\begin{minipage}[b]{0.23\linewidth}
  \centering
  \includegraphics[width=0.75\linewidth]{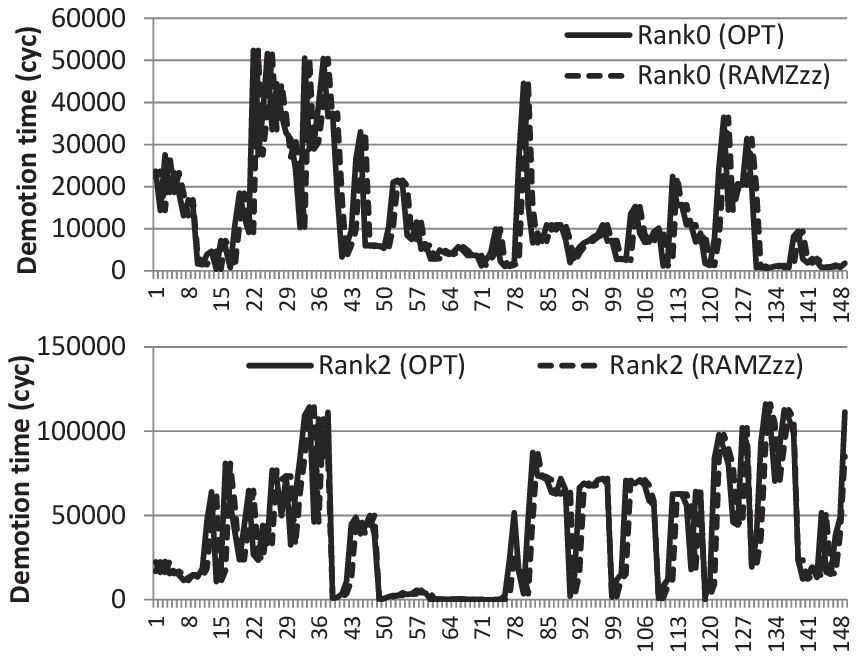}
  \caption{Power-down timeout comparison}\label{fig:demotiontime}
\end{minipage}
\hspace{0.1cm}
\begin{minipage}[b]{0.23\linewidth}
  \centering
  \includegraphics[width=1.0\textwidth]{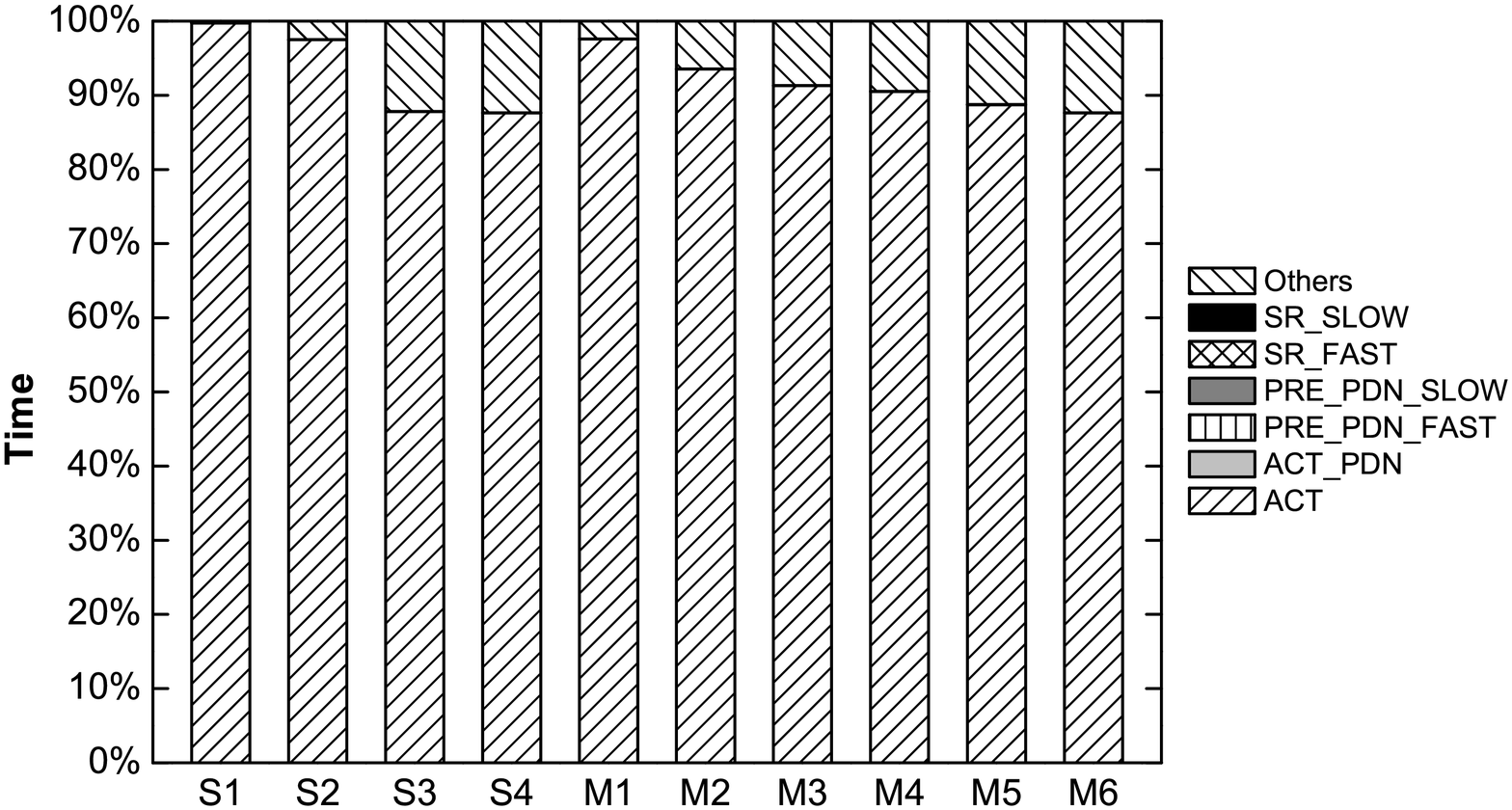}
  \caption{The breakdown of time for BASE.}\label{fig:ddr3_bd_base}
\end{minipage}
\hspace{0.1cm}
\begin{minipage}[b]{0.23\linewidth}
  \centering
  \includegraphics[width=1.0\linewidth]{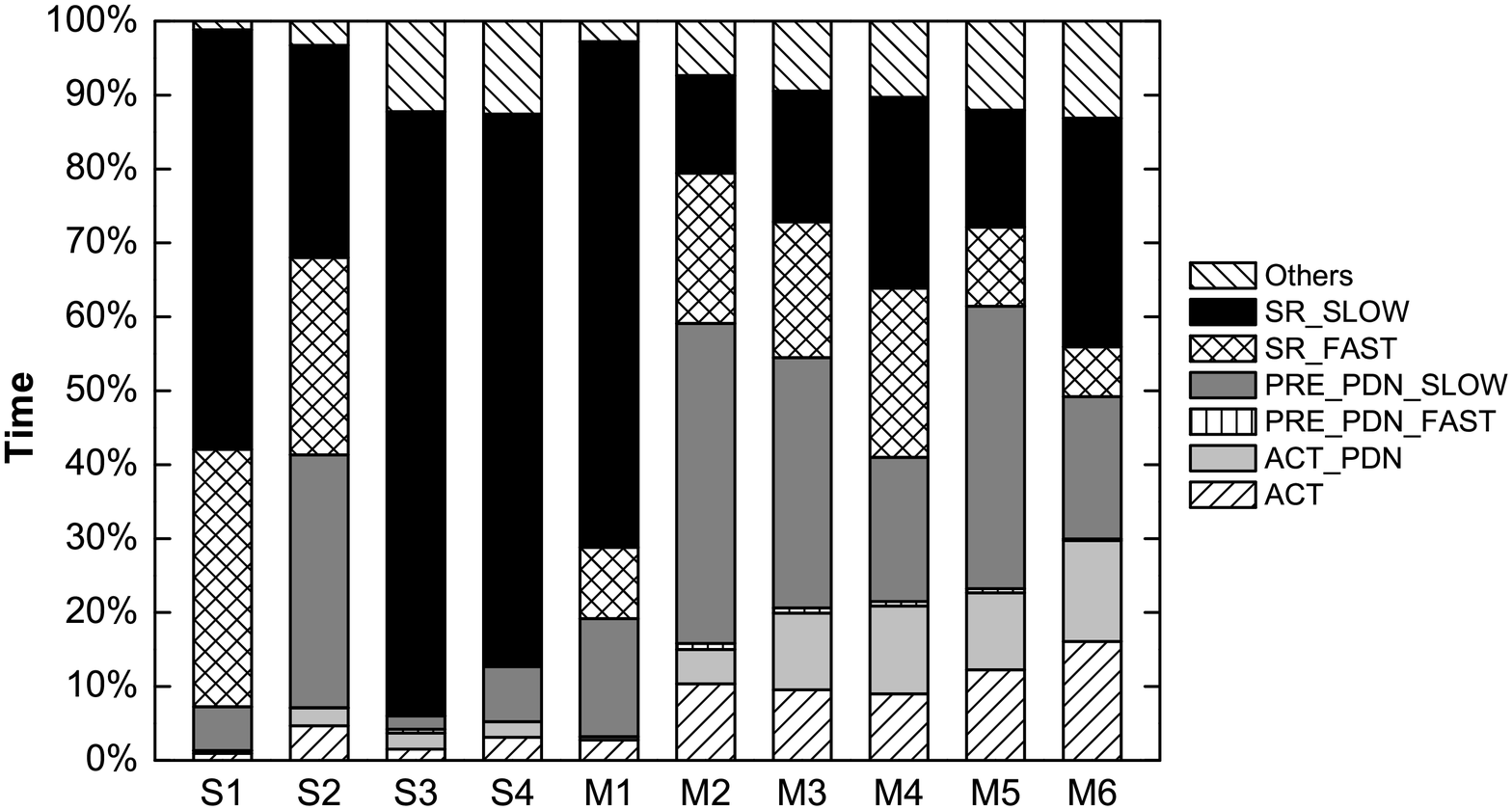}
  \caption{The breakdown of time for ORACLE.}\label{fig:ddr3_bd_oracle}
\end{minipage}
\hspace{0.1cm}
\begin{minipage}[b]{0.23\linewidth}
  \centering
  \includegraphics[width=0.75\textwidth]{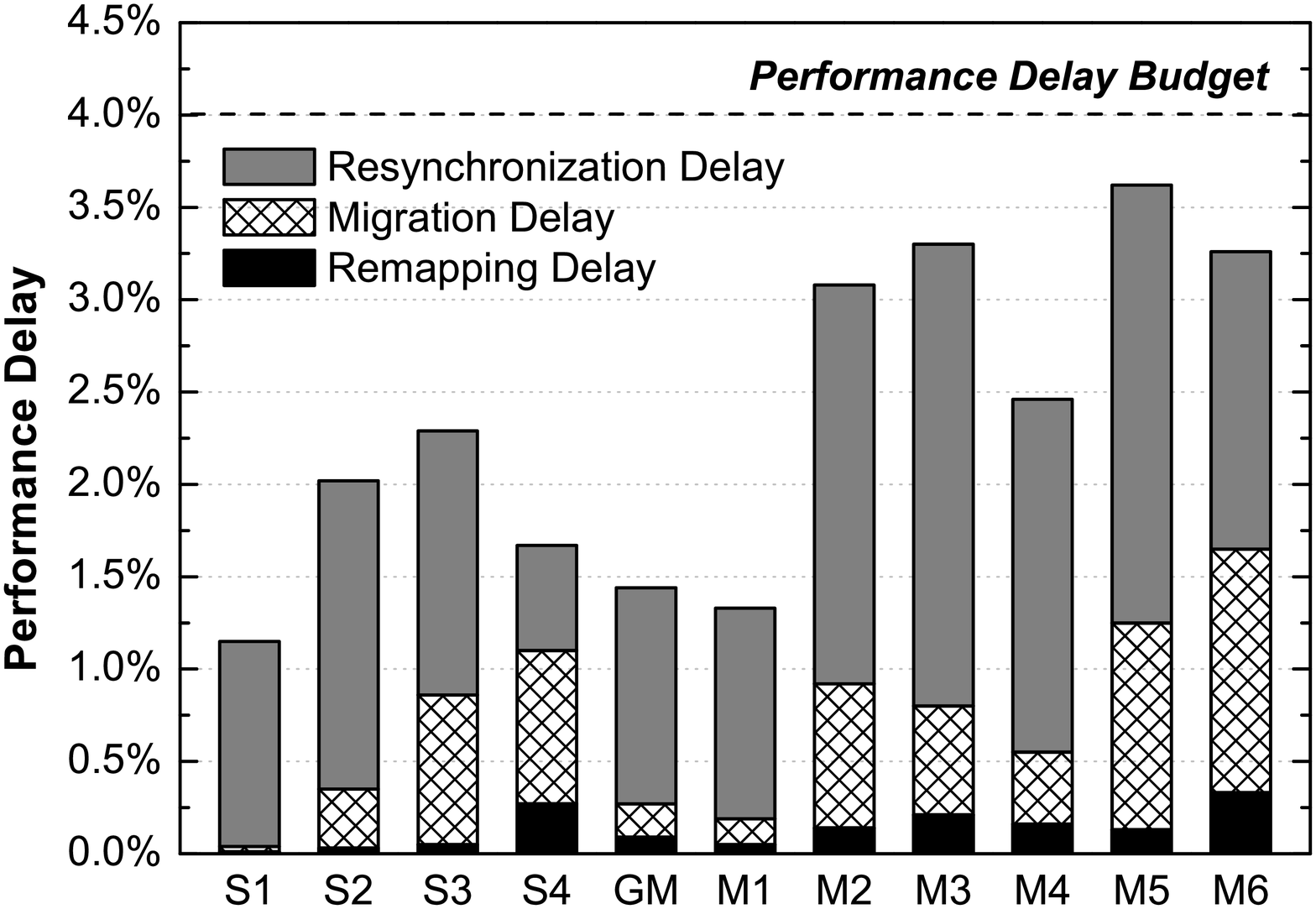}
  \caption{The breakdown of delay for RAMZzz.}\label{fig:ddr3_ramzzz_delay}
\end{minipage}
\vspace{-2ex}
\end{center}
\end{figure*}

It can also be seen from Figure~\ref{fig:dram_ed2_overall} that
RAMZzz achieves a very close ED$^2$ to ORACLE on all workloads and
memory architectures. RAMZzz achieves 5.7\%, 4.4\% and 3.7\% on
average larger ED$^2$ than ORACLE on DDR3, DDR2 and LPDDR2,
respectively. This good result is because our histogram-based
prediction model is able to accurately estimate the suitable
power-down timeout for the sake of minimizing ED$^2$.
Figure~\ref{fig:demotiontime} compares RAMZzz's estimated power-down
timeouts to SR\_FAST with ORACLE on ranks 0 and 2 of executing M4 on
DDR3. Our estimation is very close to the optimal value on the two
ranks. We observe similar results for different ranks and different
workloads and also for the power-down timeouts of other low-power
states and other DRAM architectures. We also find that our
model has high accuracy in predicting rank idle period distribution
(detailed results are presented in Appendix G of the supplementary
file).

We have further made the following observations on the result of breakdown in
Figure~\ref{fig:dram_ed2_ramzzz_bd}. First, on a specific memory architecture,
the portion of time for different low-power states varies significantly across
different workloads. Different workloads have different choices on the most
energy-effective low-power state. For most single-application workloads, RAMZzz makes the
decision to demote into SR\_SLOW on DDR3 in most idle periods, whereas the
decision of demotion is to SR\_FAST or PRE\_PDN\_SLOW for the mixed workloads.
Second, on different DRAM architectures, the portion of time for different
low-power states varies significantly, even for the same workload.
SR on LPDDR2 has a much higher significance in all workloads than on DDR3 and
DDR2. That is because, as we have seen in Table~\ref{tb:DRAMState}, SR on LPDDR2
consumes a similar normalized power consumption but a relative smaller
resynchronization time when compared with the other two DRAM architectures.
These two observations have actually demonstrated the effectiveness of adaptive
demotions of RAMZzz for different workloads and different memory architectures.
We will experimentally study the impact of adaptive demotions in
Section~\ref{subsubsec:evaldemotions}.

Figures~\ref{fig:ddr3_bd_base} and~\ref{fig:ddr3_bd_oracle}
show the breakdown of time stayed in different power states for BASE and ORACLE
on DDR3, respectively. Compared with Figure~\ref{fig:ddr3_ed2_ramzzz_bd}, RAMZzz
has a very similar power state distribution to ORACLE on all workloads, which
again demonstrates the effectiveness of our estimation. Compared to BASE, both
RAMZzz and ORACLE significantly reduce the percentage of time when ranks are in
the \emph{ACT} state by the adaptive use of all available low-power states. We
observe similar results for other workloads and DRAM architectures.

\begin{figure*}[htb]
\begin{center}
\begin{minipage}[b]{0.23\linewidth}
  \centering
  \includegraphics[width=0.75\linewidth]{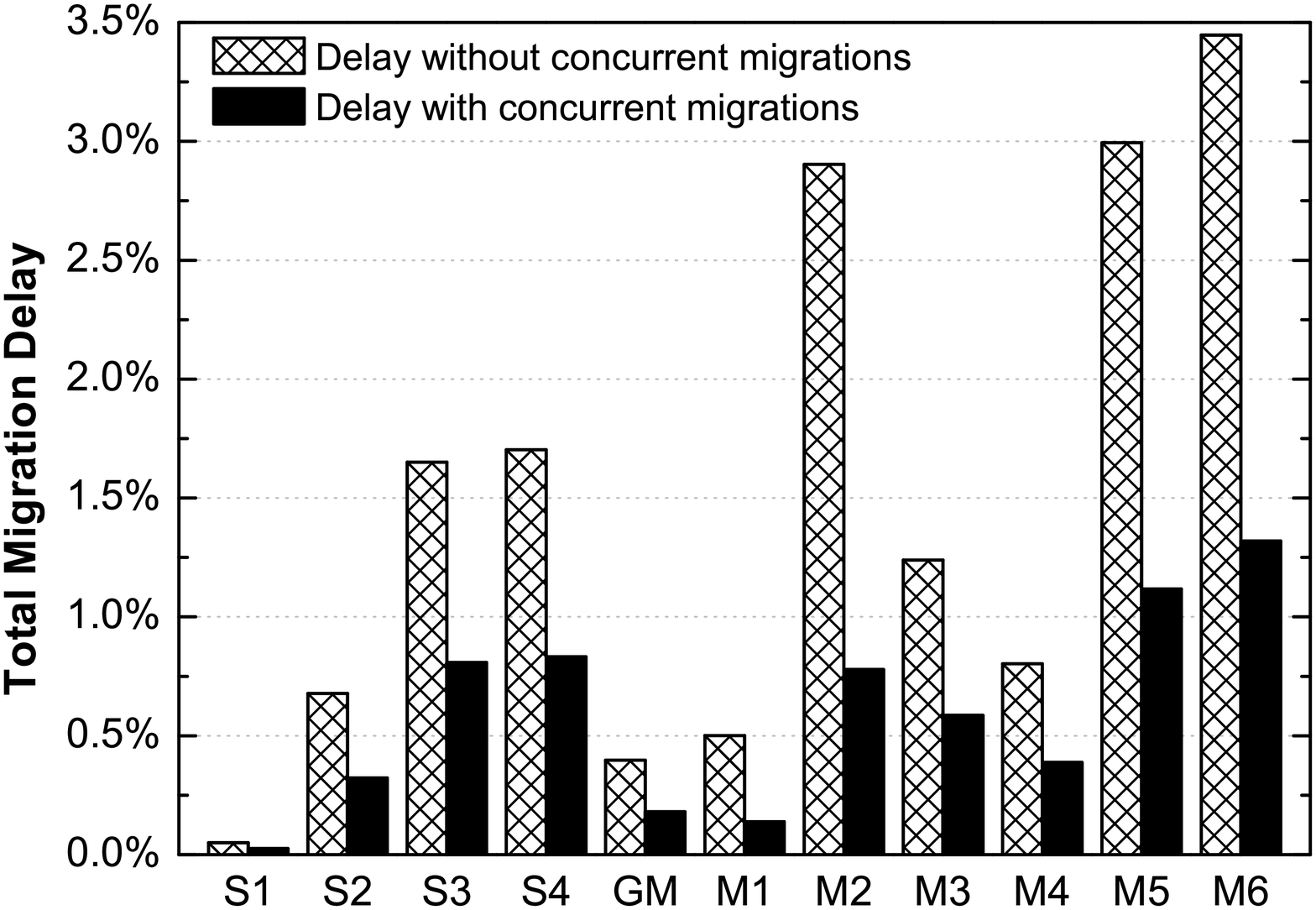}
  \caption{The optimization of page migration
  delay.}\label{fig:ddr3_ed2_migration_ed2}
\end{minipage}
\hspace{0.1cm}
\begin{minipage}[b]{0.23\linewidth}
  \centering
  \includegraphics[width=0.75\linewidth]{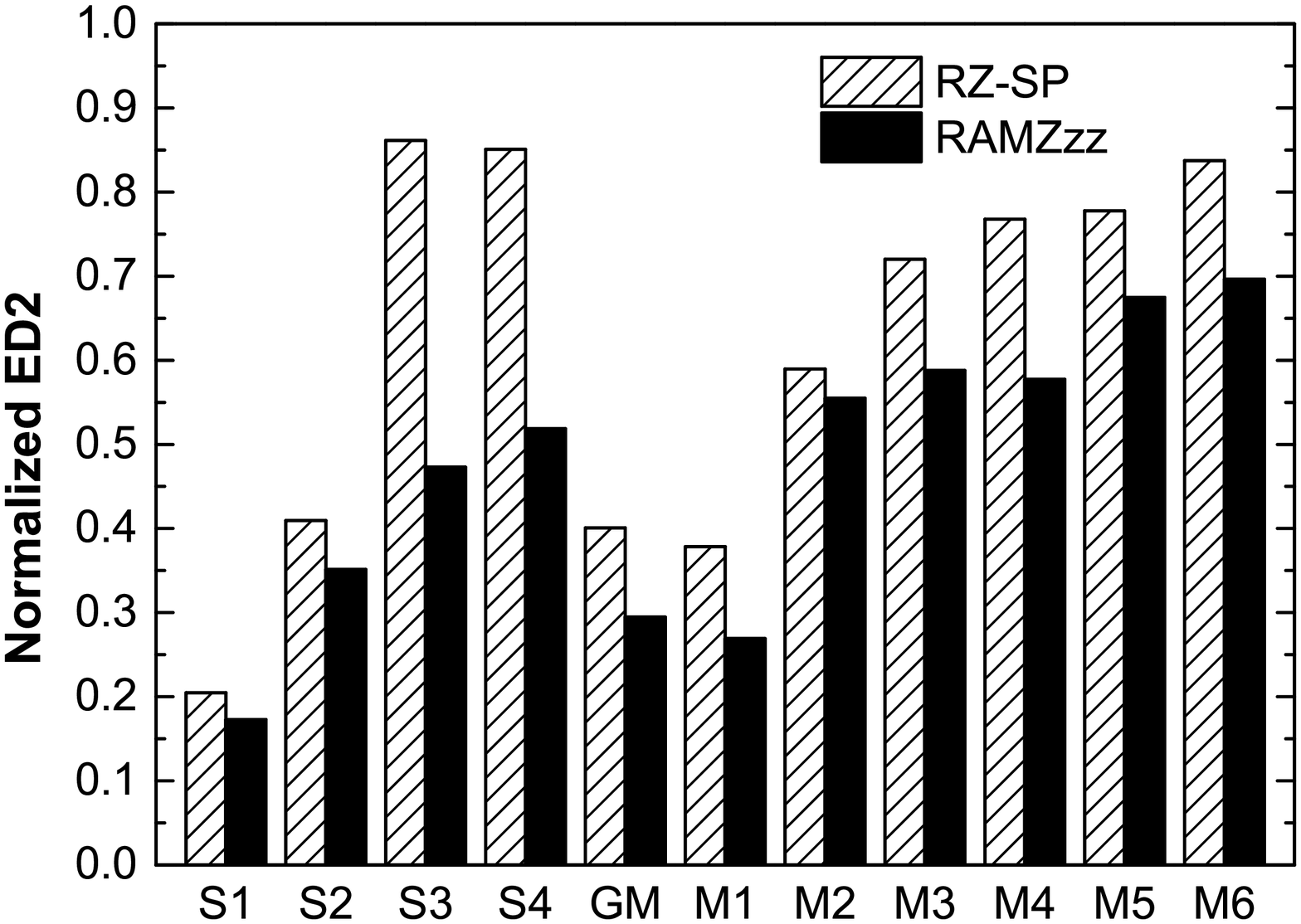}
  \caption{Comparing ED$^2$ of RAMZzz and RZ--SP.}\label{fig:study_on_migration}
\end{minipage}
\hspace{0.1cm}
\begin{minipage}[b]{0.23\linewidth}
  \centering
  \includegraphics[width=1.0\linewidth]{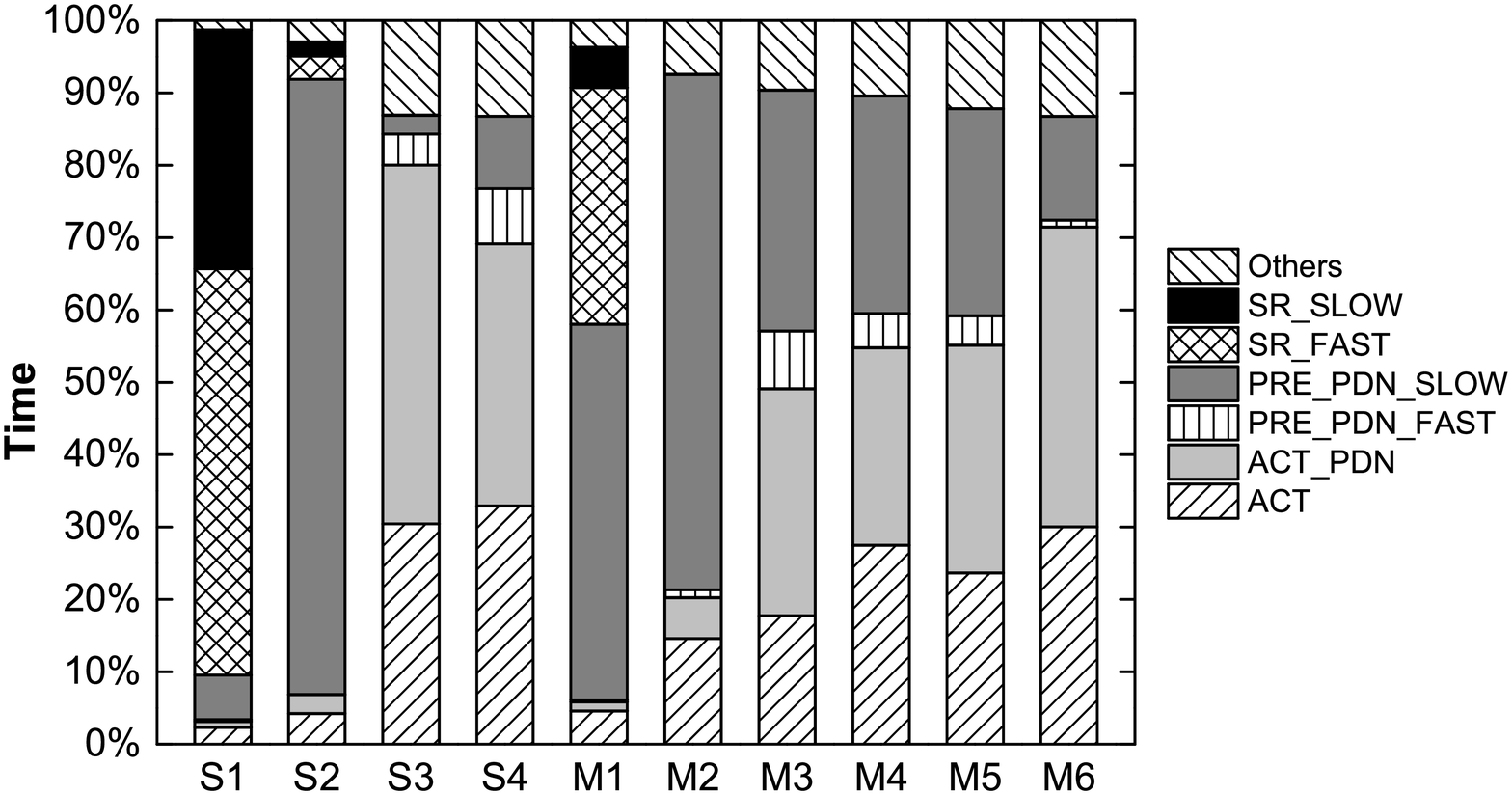}
  \caption{The breakdown of time for RZ--SP.}\label{fig:ddr3_ed2_rz(n,a)_bd}
\end{minipage}
\hspace{0.1cm}
\begin{minipage}[b]{0.23\linewidth}
  \centering
  \includegraphics[width=1.0\linewidth]{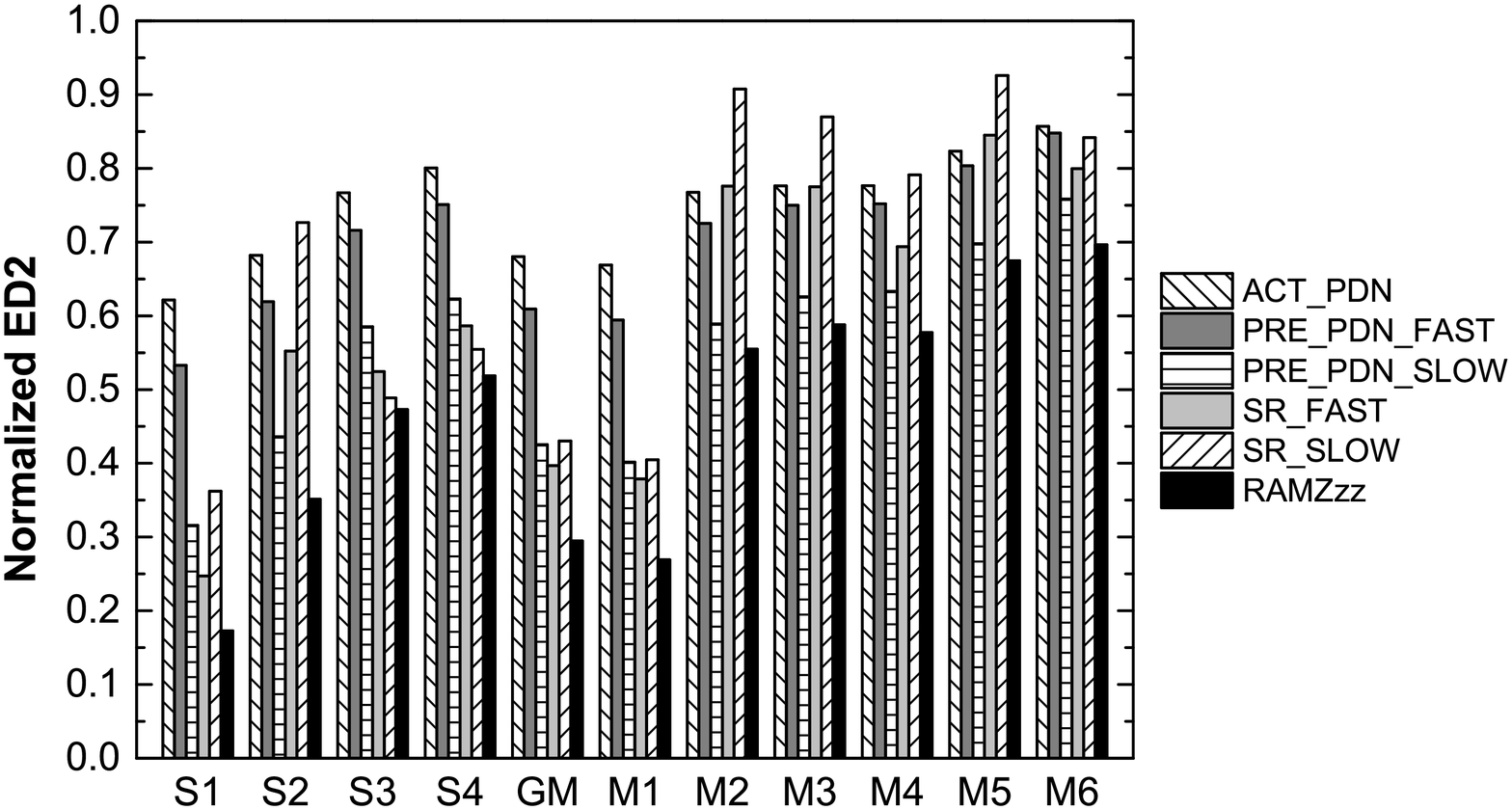}
  \caption{Comparing ED$^2$ of RAMZzz and RZ--SD.}\label{fig:study_on_demotion_ramzzz}
\end{minipage}
\vspace{-4ex}
\end{center}
\end{figure*}

Next, we study the performance delay in detail.
Figure~\ref{fig:ddr3_ramzzz_delay} shows the breakdown of performance delay
for RAMZzz on DDR3. We divide the delay into three parts: resynchronization
delay (caused by state transitions), migration delay (caused by page migrations) and
remapping delay (caused by \emph{Remapping Table} lookup and address remapping).
The performance delay of RAMZzz is well controlled under the pre-defined penalty
budget (i.e., 4\% in this experiment). The results demonstrate that our model is
able to limit the performance delay within the pre-defined threshold. The
resynchronization delay contributes the largest portion of performance delay on most
workloads. Due to concurrent migrations, the migration delay is kept within an
acceptable range (i.e., less than 1.5\% on all workloads).  The remapping delay
only accounts for a very small portion of the performance delay on all workloads. This
observation is consistent with our analysis in Appendix A of the supplementary
file. The remapping operation is performed when a request is added to the MC
queues and does not extend the critical path in the common case because queuing
delays at the MC are substantial.

As seen from Figure~\ref{fig:ddr3_ramzzz_delay}, the migration delay is
higher on the workloads with a large memory footprint (such as S3, M5 and M6).
To further study the migration delay, Figure~\ref{fig:ddr3_ed2_migration_ed2}
presents the total migration delay of RAMZzz with/without our graph-based
optimizations on DDR3. Thanks to our graph-based optimizations (as described in
Section~\ref{subsec:migration}), the total migration delay is significantly
decreased, with the reduction of 50.0\% to 74.4\%. Concurrent migrations
prevent significant performance degradation in all workloads.

Finally, we discuss the overhead of calculating the migration information
(Eulerian cycle) and the demotion configuration. We find that the number of
those values with non-zero frequencies in the predicted histogram is far smaller
than the slot size ($10^8$) in our evaluation. Thus, the search space of
Algorithm~\ref{alg:online_alg} is acceptable at runtime. The average time for
calculating of the demotion configuration is around several milliseconds on
current architectures. Such calculation is performed only once per slot (around
40 ms by default). Moreover, the calculation of the migration information is
performed only once per epoch (around 400 ms by default). Thus, their overheads
are low on current architectures. The results are consistent with previous
studies~\cite{Sudan:6212453,Sudan:2010:MID:1736020.1736045,Ramos:2011:PPH:1995896.1995911,Deng:2011:MAL:1950365.1950392}.

\subsection{Individual Impacts}

We now study the individual impact of dynamic migrations and adaptive demotions
on RAMZzz with the optimization goal of ED$^2$ on SPEC 2006 workloads. Due to
the space limitation, we present the figures for DDR3 memory architecture only
and comment on other architectures without figures when appropriate.

\subsubsection{Studies on Dynamic Migrations} \label{subsubsec:evalmigration}

We study the impact of dynamic migrations, comparing RAMZzz and RZ--SP (RZ--SP
uses the adaptive demotion scheme with no page migration).
Figure~\ref{fig:study_on_migration} presents ED$^2$ results for RAMZzz and RZ--SP on DDR3.

RAMZzz has much lower ED$^2$ than RZ--SP, with an average reduction of 21.8\%.
The reduction depends on the memory footprint and memory access intensiveness.
The reduction is more significant on memory-intensive workloads (such as M4) or
workloads with small memory footprint (such as S3 and S4). If a workload has a
small memory footprint, the page migration has a small overhead on both the
delay and the energy consumption, and the portion of cold ranks is higher. If
the memory access of a workload is more intensive, many idle periods are too
short and RZ--SP has less opportunities for saving background power (even with
our proposed adaptive demotion scheme). On those two kinds of workloads, page
migration is important for the effectiveness of power management. In contrast,
when the memory access is less intensive or has a large memory footprint, RZ--SP
is quite competitive to RAMZzz. The example workloads include S1 and M2.

Figure~\ref{fig:ddr3_ed2_rz(n,a)_bd} shows the breakdown of time stayed in
different power states for RZ--SP on DDR3. Comparing with
Figure~\ref{fig:ddr3_ed2_ramzzz_bd}, the portion of time of those lower-power
states with a higher resynchronization time are much smaller. RZ--SP demotes the
ranks into PRE\_PDN\_FAST and PRE\_PDN\_SLOW states in most times, whereas
RAMZzz demotes into even lower-power states, i.e., SR\_FAST and SR\_SLOW. That
is because dynamic page migration is able to create longer idle periods. For
example, the percentage of SR\_SLOW is almost zero in RZ--SP for memory
intensive workloads, such as S3, S4 and M2--4, while SR\_SLOW has a significant
portion in RAMZzz for those workloads. Since page migrations are disabled
in RZ--SP, the total delay of RZ--SP is slightly smaller than that of RAMZzz,
less than 2.5\% for all workloads. We show that RAMZzz is still more
energy-efficient than RZ--SP on full system ED$^2$ (or energy consumption) in
Appendix D of the supplementary file.

To summarize the impact of dynamic page migrations, we observe RAMZzz has much
lower ED$^2$ than RZ--SP on three DRAM architectures, with an average reduction
of 21.8\%, 15.8\% and 17.1\%, and a range of 5.2--45.1\%, 2.8--39.6\% and
1.7--41.1\% on DDR3, DDR2 and LPDDR2 respectively. The reduction is more
significant on memory-intensive workloads or workloads with small memory
footprint on DDR2 and LPDDR2.

\subsubsection{Studies on Adaptive Demotions} \label{subsubsec:evaldemotions}

In this section, we study the impact of adaptive demotions, that is to compare
the performance of RAMZzz and RZ--SD (RZ--SD uses the dynamic page migration
without the adaptive demotion).

Figure~\ref{fig:study_on_demotion_ramzzz} presents the comparison of ED$^2$ for
RAMZzz and RZ--SD on DDR3. We compare the performance of RAMZzz with every
possible RZ--SD approach on all workloads. That is, we use every available
low-power state as the pre-selected low-power state in the RZ--SD approach.
Since DDR3 has five low-power states, we have five RZ--SD approaches where each
approach is denoted as the name of pre-selected low-power state
Figure~\ref{fig:study_on_demotion_ramzzz} (such as \emph{SR\_FAST} represents
the RZ--SD approach which uses SR\_FAST as the pre-selected low-power state).

We observe that RAMZzz outperforms all RZ--SD approaches on all workloads, with
the reduction from 26.4\% to 51.1\% (36.4\% on average). Moreover, different
workloads have different choices on the most energy-efficient RZ--SD approach,
indicating that the static demotion scheme can not adapt to different workloads.
The efficiency of the static demotion scheme is closely related to the decision
on the pre-selected low-power state, justifying the necessity of adaptive
demotions. The total delay of RZ--SD is close to that of RAMZzz, less than 3\%
for all workloads. We observe that RAMZzz is also more efficient than RZ--SD on
full system ED$^2$ (or energy consumption) in Appendix D of the supplementary
file.

Finally, we study the impact of the number of available low-power states. In
Table~\ref{tb:adaptive}, we change the number of available low-power states
used on DDR3, DDR2 and LPDDR2 on M1 from 1 to 5, 1 to 4 and 1 to 3,
respectively. We add a low-power state with smaller power consumption
when increasing the number of available low-power states. As the number of
available low-power states increasing, the normalized ED$^2$ becomes smaller.
The improvement in normalized ED$^2$ by increasing the number of available
low-power states from 1 to the maximum is 59.8\%, 54.1\% and 45.2\% on DDR3,
DDR2 and LPDDR2, respectively. This further proves the self-adapting feature brought
by our proposed adaptive demotion scheme.

\begin{table}
 \centering
 \caption{Comparing ED$^2$ of RAMZzz with different number of low-power states
 on three DRAM architectures on M1.}
 \vspace{-2ex}
 {\footnotesize
    \label{tb:adaptive}
    \begin{tabular}{|c|c|c|c|c|c|c|}
    \hline
    {\bf \tabincell{c}{Number of low \\ -power states}} & {\bf 1} &
    {\bf 2} & {\bf 3} & {\bf 4} & {\bf 5} \\
    \hline
    {\bf DDR3} & 0.67 & 0.59 & 0.40 & 0.31 & 0.27 \\
    \hline
    {\bf DDR2} & 0.68 & 0.43 & 0.35 & 0.30 & N/A \\
    \hline
    {\bf LPDDR2} & 0.59 & 0.41 & 0.33 & N/A & N/A \\
    \hline
    \end{tabular}
 }
 \vspace{-4ex}
\end{table}

To summarize the impact of adaptive demotions, we observe RAMZzz has much lower
ED$^2$ than RZ--SD on three DRAM architectures, and with the reduction of
26.4--51.1\% (36.4\% on average), 12.0--48.7\% (25.0\% on average) and
5.0--41.9\% (22.4\% on average) on DDR3, DDR2 and LPDDR2, respectively.

\begin{figure*}[htb]
\begin{center}
\begin{minipage}[b]{0.31\linewidth}
  \centering
  \includegraphics[width=0.80\linewidth]{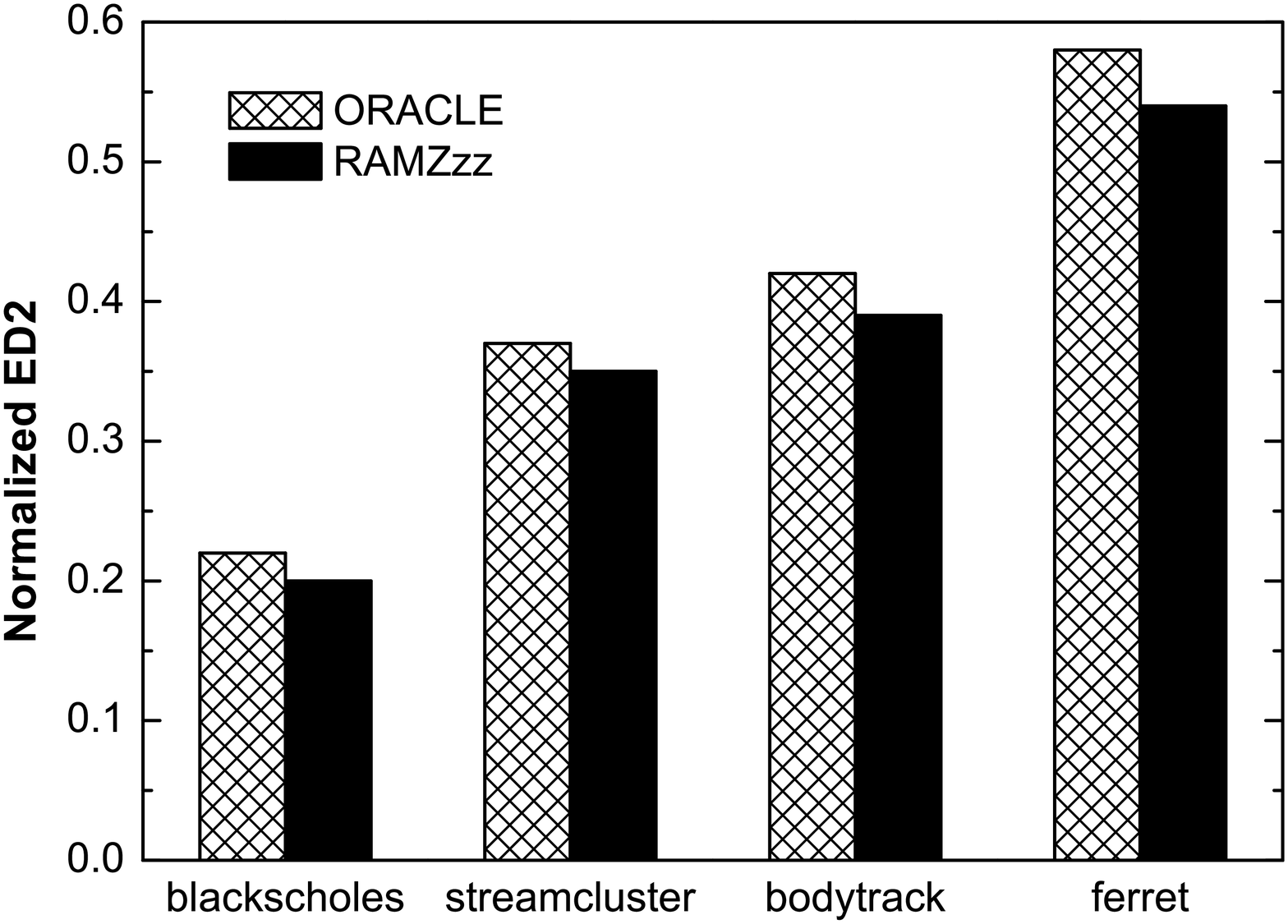}
  \caption{The overall results of PARSEC workloads.}\label{fig:parsec_2gb_ddr3}
\end{minipage}
\hspace{0.1cm}
\begin{minipage}[b]{0.31\linewidth}
  \centering
  \includegraphics[width=0.70\linewidth]{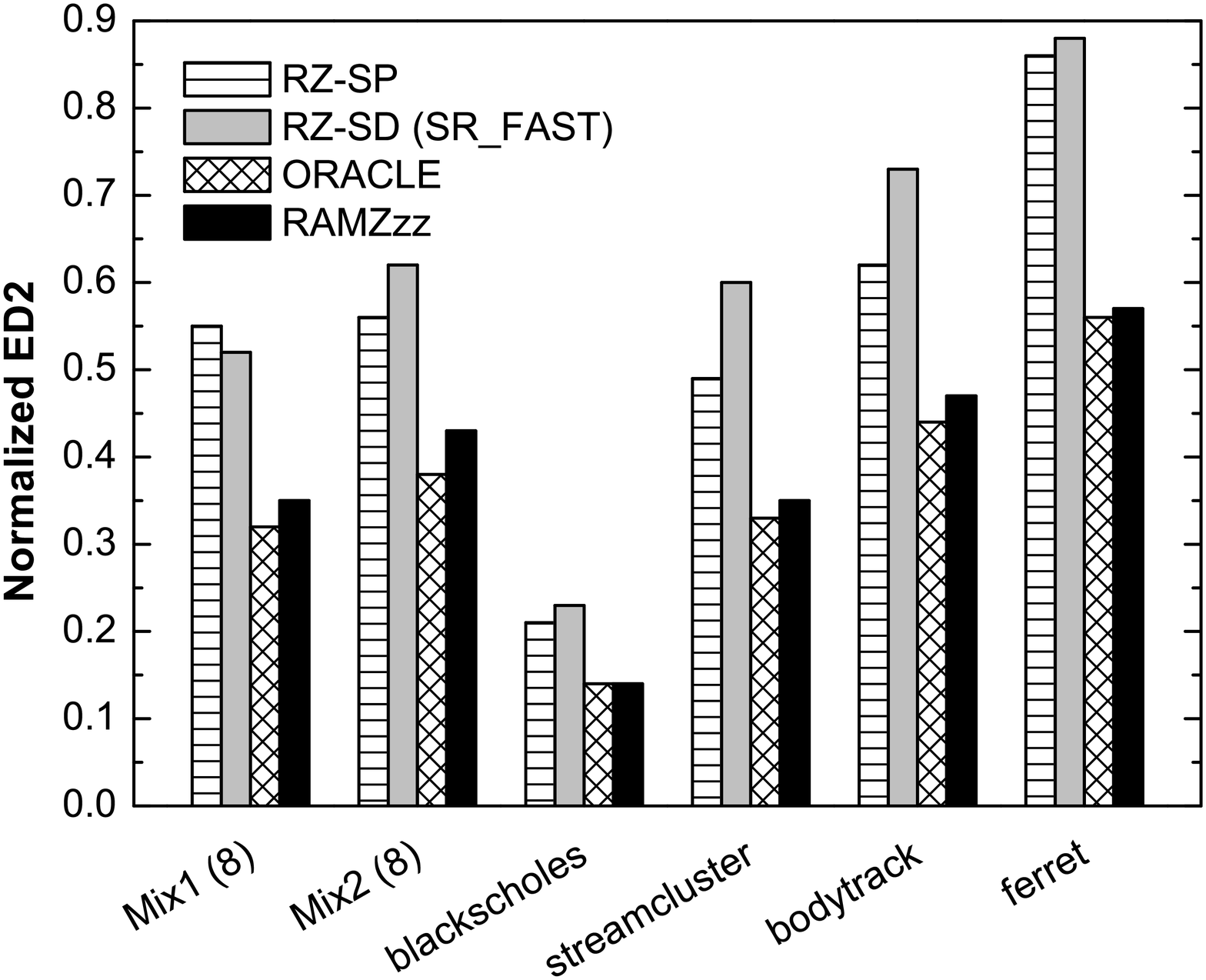}
  \caption{Results on a 8-core with 32GB DDR3 system.}\label{fig:32gb_ddr3}
\end{minipage}
\hspace{0.1cm}
\begin{minipage}[b]{0.31\linewidth}
  \centering
  \includegraphics[width=0.70\linewidth]{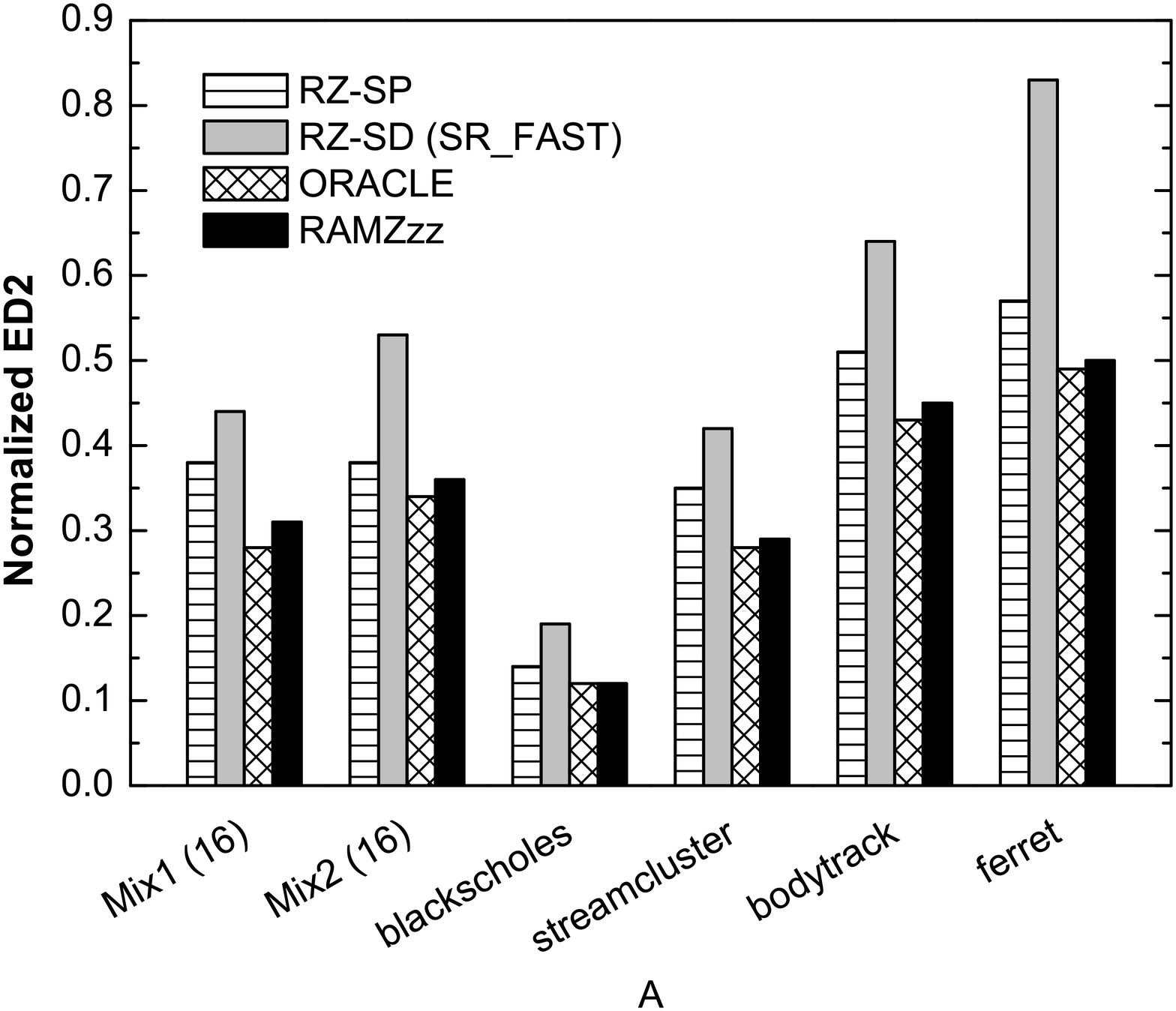}
  \caption{Results on a 16-core with 64GB DDR3 system.}\label{fig:64gb_ddr3}
\end{minipage}
\vspace{-4ex}
\end{center}
\end{figure*}

\subsection{Results on PARSEC workloads}

Figure~\ref{fig:parsec_2gb_ddr3} shows the normalized ED$^2$ results of
RAMZzz and ORACLE approaches on DDR3 architecture using PARSEC workloads. We use
the default experimental setting (e.g., the delay budget is 4\%). RAMZzz is
also significantly more energy-efficient than BASE on PARSEC workloads. We
observe similar results to those on the SPEC 2006 workloads. For example, the
reduction is more significant for the workloads with less intensive memory
accesses (such as blackscholes). RAMZzz achieves a very close ED$^2$ to ORACLE
on PARSEC workloads (as shown in Figure~\ref{fig:parsec_2gb_ddr3}).

\subsection{Results on More Powerful Computer Systems} \label{subsec:large_sys}

We also perform the simulation studies on more powerful computer systems
with a larger number of CPU cores and large memory capacity. We use Sniper to
collect memory access traces of multi-threaded/multi-programmed workloads from
SPEC 2006 and PARSEC benchmarks. With the default simulated CPU architecture, we
run mixed workloads of 8 (and 16) applications from SPEC 2006 and four PARSEC
workloads (i.e., blackscholes, bodytrack, ferret and streamcluster) executed
with 8 (and 16) threads on a 8-core (and 16-core) processor with 32 (and 64) GB
DDR3 with 8 (and 16) ranks memory system.

Figures~\ref{fig:32gb_ddr3} and~\ref{fig:64gb_ddr3} present the comparison
of normalized ED$^2$ of BASE, ORACLE, RAMZzz, RZ--SP and RZ--SD (SR\_FAST is
used as the pre-selected low-power state) with the optimization goal of ED$^2$
and with delay budget of 4\%. We make the following observations. First, RAMZzz
has very significant ED$^2$ reduction compared with BASE. The reduction in
ED$^2$ is 60.7\% and 67.2\% on average on the 8-core and 16-core systems,
respectively.
Second, RAMZzz achieves only 6.0\% higher ED$^2$ on average than ORACLE on all
workloads and systems. For the impact of dynamic page migrations, RAMZzz has an
average ED$^2$ reduction of 45.4\% and 61.2\% over RZ--SP on the 8-core and
16-core systems, respectively.
For adaptive demotions, RAMZzz has an average ED$^2$ reduction of 40.5\% and
49.2\% over RZ--SD on the 8-core and 16-core systems, respectively.

\section{Conclusion}\label{sec:conclusion}

In this paper, we have proposed a novel memory design RAMZzz to reduce the DRAM
energy consumption. It embraces two rank-aware power saving techniques to
address the major obstacles in state transition-based power saving approaches:
dynamic page migrations and adaptive demotions. A cost model is developed to
guide the optimizations for different workloads and different memory
architectures. We evaluate RAMZzz with SPEC 2006 and PAESEC benchmarks in
comparison with other power saving techniques on three main memory architectures
including DDR3, DDR2 and LPDDR2. Our simulation results demonstrate significant
improvement in ED$^2$ and energy consumption over other power saving techniques.
Moreover, RAMZzz performs very close to the ideal oracle approach for different
workloads and memory architectures.

\appendices

\section{Detailed Implementation Issues} \label{sec:other_issues}

\begin{figure}[htb]
  \centering
  \includegraphics[width=0.4\textwidth]{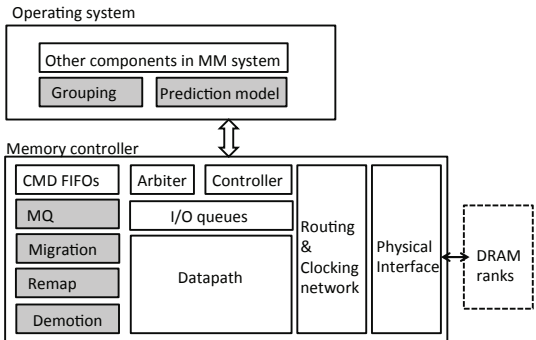}
  \vspace{-2ex}
  \caption{Memory controller and operating system with RAMZzz's new
  modules highlighted.}\label{fig:MC}
  \vspace{-2ex}
\end{figure}

RAMZzz adds a few new components to the memory controller and operating system.
Following the previous study~\cite{Ramos:2011:PPH:1995896.1995911}, RAMZzz
extends a programmable controller~\cite{Xilinx} by adding its own new components
(shaded in Figure~\ref{fig:MC}). Other functionalities including page grouping
and the prediction model are offloaded to operating systems (like previous
studies~\cite{Ramos:2011:PPH:1995896.1995911, Huang:2003:DIP:1247340.1247345}).

\textbf{Memory Controller Structure.} The memory controller (MC) receives
read/write requests from the cache controller via the CMD FIFOs. The Arbiter
dequeues requests from those FIFO queues, and the controller converts those
requests into the necessary instructions and sequences required to communicate
with the memory. The Datapath module handles the flow of reads and writes
between the memory devices. The physical interface converts the controller
instructions into the actual timing relationships and signals required for
accessing the memory device.

We assume the MC exploits page interleaving. Page interleaving exploits
higher data locality but also makes accesses to multiple banks less uniform, which may
cause row-buffer conflicts in some cases. However, it is better for grouping and
migrating physical pages based on the frequency and recency of accesses, as
described in Section 4 (the hot ranks are more likely to have very short idle
periods, and the cold ranks have relatively long idle periods). A further
optimization is to use the permutation-based page interleaving
scheme~\cite{Zhang:2000:PPI:360128.360134}, which retains high data locality and
reduces row-buffer conflicts. Other interleaving schemes, such as cache line
interleaving (which maps consecutive cache lines to different memory banks), may
not effectively exploit the data locality. In such situation, a possible
approach is to make migration decisions based on other locality indicators
(e.g., row buffer misses as exploited by Wang et
al.~\cite{Wang:2013:EHM:2523721.2523737}) rather than LLC misses/writebacks.
We leave the optimization and evaluation on other interleaving schemes as our
future work. Memory requests are handled on a FCFS basis. There are more
sophisticated access scheduling optimizations like finite queue length, critical-word-first optimization, and prioritizing reads. Those
techniques are orthogonal to our study, and we do not include them into the
simulations.

Four new modules including MQ, Migration, Remap and Demotion are added into the
memory controller for implementing the functionality of page grouping, page
migration, page remapping and power state control in RAMZzz, respectively.
All the logics of the new modules are performed by the memory controller, and
are designed off the critical path of memory accesses, giving the priority to
the memory accesses from applications. We add a flag bit to indicate whether
this request is from applications or new modules. The total on-chip storage of
new MC modules in our design is 112KB (as described in the following).

\textbf{MQ Module.} To avoid performance degradation, MQ
module contains the small on-chip cache (64KB with 4K entries) to
store the MQ structure and a separate queue (10KB) for the updates to the
MQ structure. To find the MQ entry of a physical page, MC uses
hashing with the corresponding page number. Misses in the entry
cache produce requests to DRAM. MQ module's logic snoops the CMD
FIFO queue, creating one update per new request. The updates to the
MQ structure are performed by the MC off the critical path of memory
accesses (via the aforementioned flag bit). The update queue is
implemented as a small circular buffer, where a new update precludes
any currently queued update to the same entry. In our design, each
physical page descriptor in the MQ queues takes 124 bits. Each
descriptor contains the corresponding page number (22 bits), the
reference counter (14 bits), the queue number in MQ (4 bits), the
last-access time (27 bits), the pointers to other descriptors (54
bits), and the reserved bit for flags (3 bits). The space overhead
of our design is low. For the 2GB DRAM, the total space taken by the
descriptors is about 8MB (only 0.4\% of the total DRAM space).

\textbf{Migration Module.} The Migration module contains the queue of scheduled
migrations. The migrations are enqueued in a manner such that concurrent
migrations of a Eulerian cycle are put in consecutive positions. At the beginning of each
epoch, the OS accesses the current MQ structure to perform grouping and
calculate the Eulerian cycle. Then, the OS updates the queue of scheduled
migrations (10KB) which is stored in the Migration module. Page migrations start
from the beginning of an epoch, and is scheduled once there are idle periods. Priority is
given to longer segments because they involve more pages. Memory requests are
buffered until the migration is concluded. To facilitate concurrent page
migrations according to the Eulerian cycle, each rank is equipped with one extra
row-buffer for storing the incoming page. When migrating a page, a rank first
writes the outgoing page to the buffer of the target rank, and then
reads the incoming page from its buffer.

\textbf{Remap Module.} Similar to the previous
design~\cite{Ramos:2011:PPH:1995896.1995911}, we introduce a new
layer of translation between physical addresses assigned by the OS
(and stored in the OS page table) and those used by the MC to access
DRAM devices. Specifically, the MC maintains the \emph{Remapping Table}, a
hash table for translating physical page addresses coming from the
LLC to actual remapped physical page addresses. The OS can access
the \emph{Remapping Table} as well. After the migration is completed
at the beginning of an epoch, the \emph{Remapping Table} is updated
accordingly. Periodically or when the table fills up (at which point
the MC interrupts the CPU), the OS commits the new translations to
its page table and invalidates the corresponding TLB entries. For
example, if the OS uses a hashed inverted page table, e.g.,
UltraSparc and PowerPC architectures, it considerably simplifies the
commit operation. Then, the OS sets a flag in a memory-mapped
register in the MC to make sure that the MC prevents from migrating
pages during the commit process, and clears the \emph{Remapping
Table}.

When a memory request (with physical address assigned by the
OS) arrives at the MC, it searches the address in the
\emph{Remapping Table}. On a hit, the new physical page address is
used by the MC to issue the appropriate commands to retrieve the
data from its new location. Otherwise, the original address is used.
In terms of access latency, the remapping operation happens when a
request is added to the MC queues and does not extend the critical
path in the common case because queuing delays at the MC are
substantial. For memory-intensive workloads, memory requests usually
wait in the MC queues for a long time before being serviced. The
above translation can begin when the request is queued and the delay
for translation can be easily hidden behind the long waiting time. The
notion of introducing \emph{Remapping Table} for the MC has been
widely used in the
past~\cite{Ramos:2011:PPH:1995896.1995911,Sudan:2010:MID:1736020.1736045,Sudan:6212453}.

The Remap module maintains the \emph{Remapping Table} (28KB
with 4K entries) and the logic to remap target addresses. At the end
of migration, the Migration module submits the migration information
to the Remap module, which creates new mappings in the
\emph{Remapping Table}. The Remap module snoops the CMD queue to
check if it is necessary to remap its entries. We assume each
\emph{Remapping Table} lookup and each remapping take 1 memory
cycle. However, these operations only delay a memory request if it
finds the CMD queue empty (which is not the common case). Note that
the migration and remapping of a segment blocks the accesses to only
the pages involved, and concurrent accesses to other pages are still
possible.

\textbf{Demotion Module.} The Demotion module performs the demotion to control
the power state of each rank according to its demotion configuration. The
demotion configuration of each rank is updated by the OS at the beginning of a
slot.

\textbf{OS Modules.} Two major new components Grouping and Prediction Model are
added to the memory management sub-system in operating system. The Grouping
module performs grouping and calculates the Eulerian cycle according to the MQ
structure at the beginning of an epoch. At the beginning of each epoch, the OS
accesses the current MQ structure to perform grouping and calculate the Eulerian
cycle. Then, the OS updates the queue of scheduled migrations which is stored in
the Migration module. The Prediction Model module runs the prediction model and
obtains the demotion configuration for memory controller at the beginning of each
slot.

Note that the structure complexity and storage overhead of RAMZzz are similar to
the previous proposals, e.g., ~\cite{Huang:2005:IEE:1077603.1077696,
Deng:2011:MAL:1950365.1950392, Diniz:2007:LPC:1250662.1250699,
Fan:2001:MCP:383082.383118, Sudan:2010:MID:1736020.1736045}. For example, our
design has small DRAM space requirement (less than 2\% of the total amount of DRAM).

\section{Experimental Results on Mixed Workloads}

\begin{figure}
  \centering
  \includegraphics[width=0.8\linewidth]{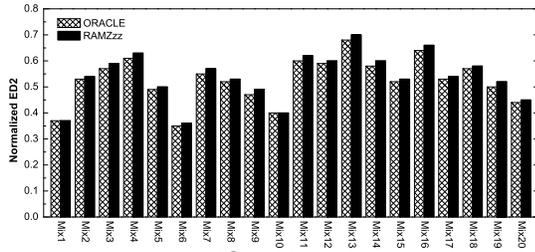}
  \vspace{-2ex}
  \caption{Comparing ED$^2$ of RAMZzz and ORACLE with the optimization goal of
ED$^2$ on DDR3.}\label{fig:ddr3_mix20}
  \vspace{-3ex}
\end{figure}

Figure~\ref{fig:ddr3_mix20} presents the experimental results of RAMZzz
for twenty additional mixed workloads with the optimization goal of ED$^2$ on
DDR3. The memory trace of each mixed workload is collected by executing four
randomly chosen SPEC 2006 applications concurrently on PTLSim. RAMZzz has much
lower ED$2$ than BASE on all these workloads, with an average reduction of
46.2\%, and a range of 30.5--63.9\%. RAMZzz also achieves very close ED$^2$ to
ORACLE on all twenty workloads.

\section{Results on Energy-Oriented Optimizations of SPEC 2006 Workloads}
\label{sec:energy_overall}

\begin{figure*}[htb]
\centering
\subfigure[Results on energy consumption for DDR3]{
\label{fig:ddr3_energy_overall}
\begin{minipage}[b]{0.31\linewidth}
\centering
\includegraphics[width=0.90\linewidth]{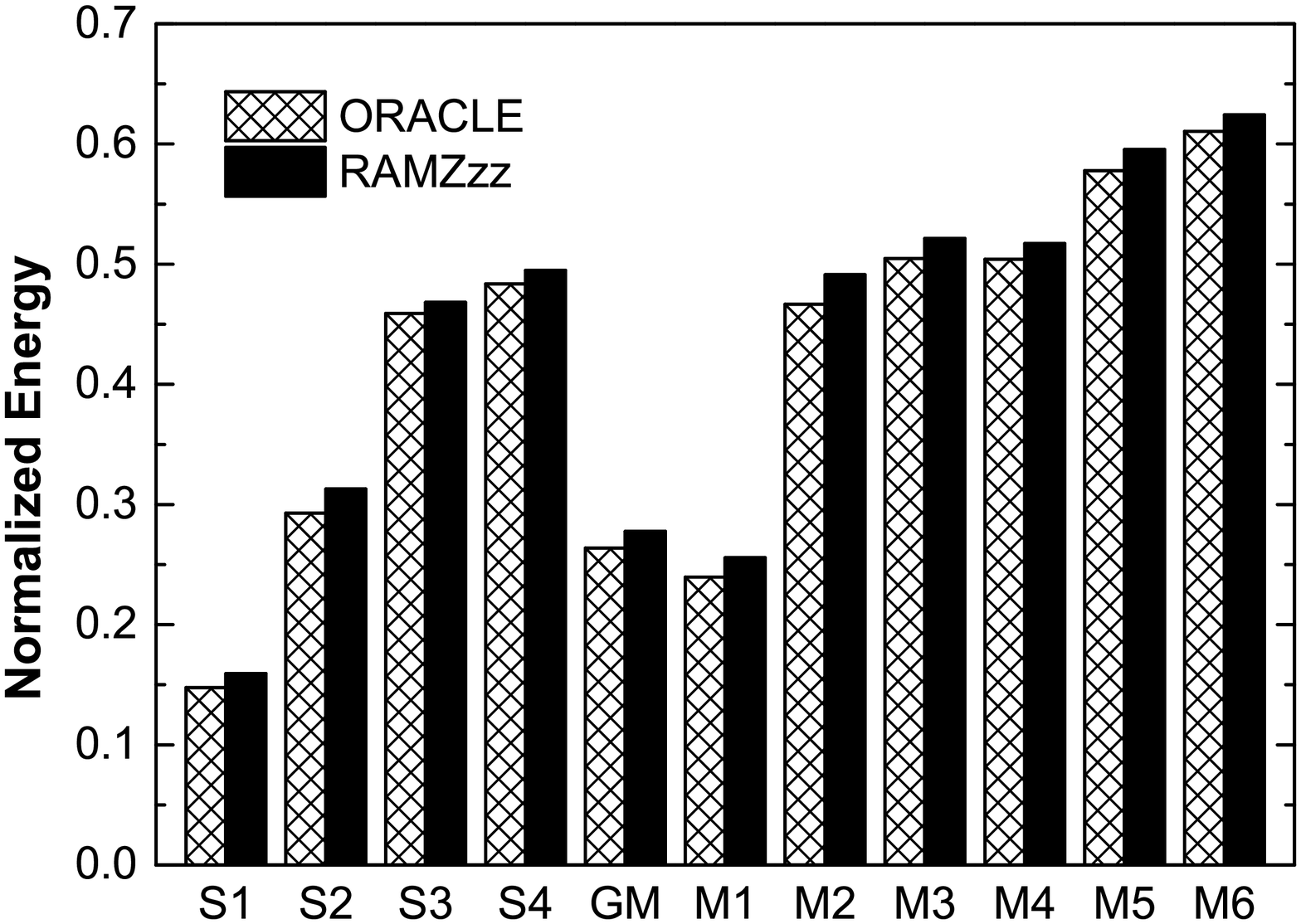}
\end{minipage}}
\hspace{0.1cm}
\subfigure[Results on energy consumption for DDR2]{
\label{fig:ddr2_energy_overall}
\begin{minipage}[b]{0.31\linewidth}
\centering
\includegraphics[width=0.90\linewidth]{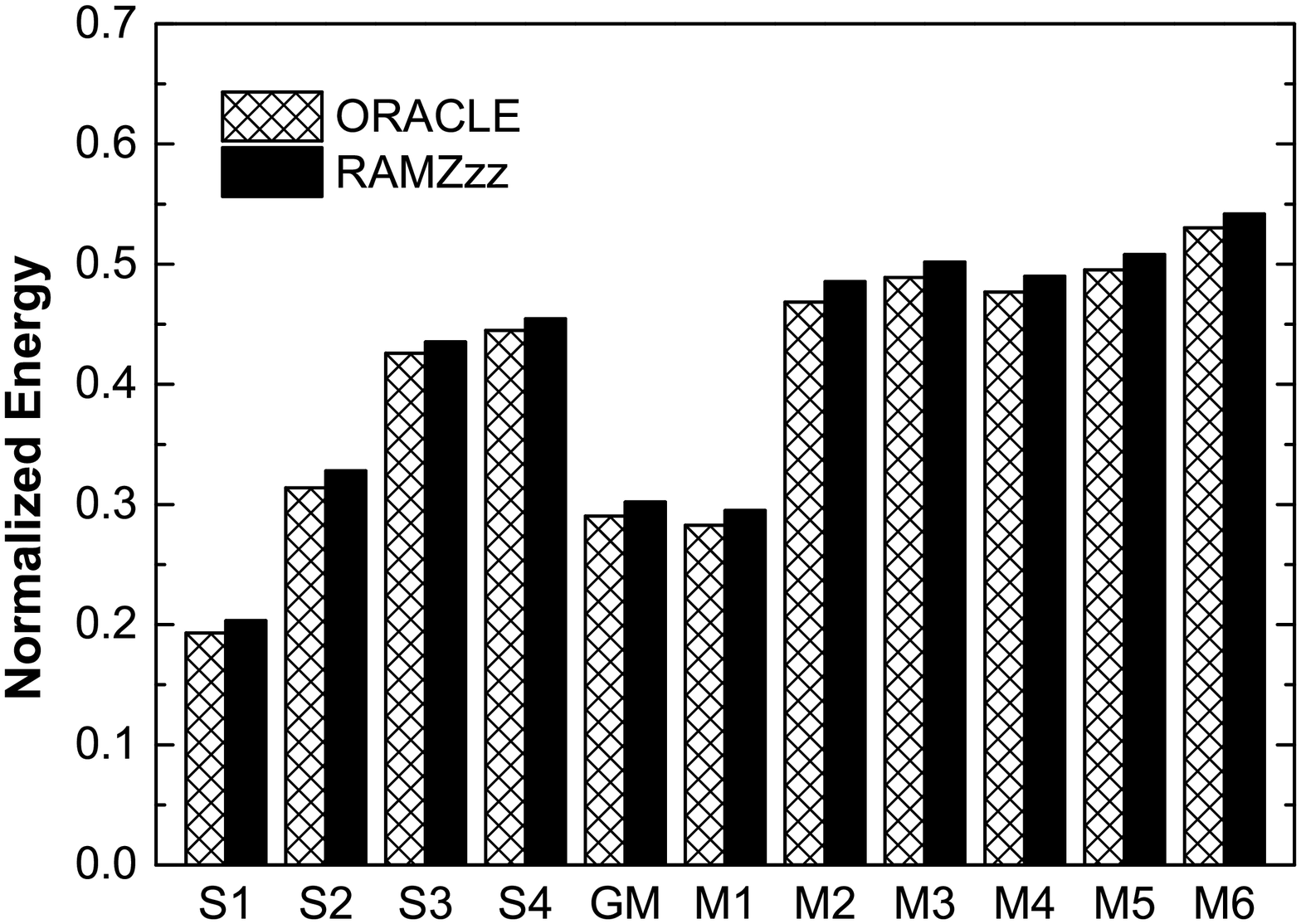}
\end{minipage}}
\hspace{0.1cm}
\subfigure[Results on energy consumption for LPDDR2]{
\label{fig:lpddr2_energy_overall}
\begin{minipage}[b]{0.31\linewidth}
\centering
\includegraphics[width=0.90\linewidth]{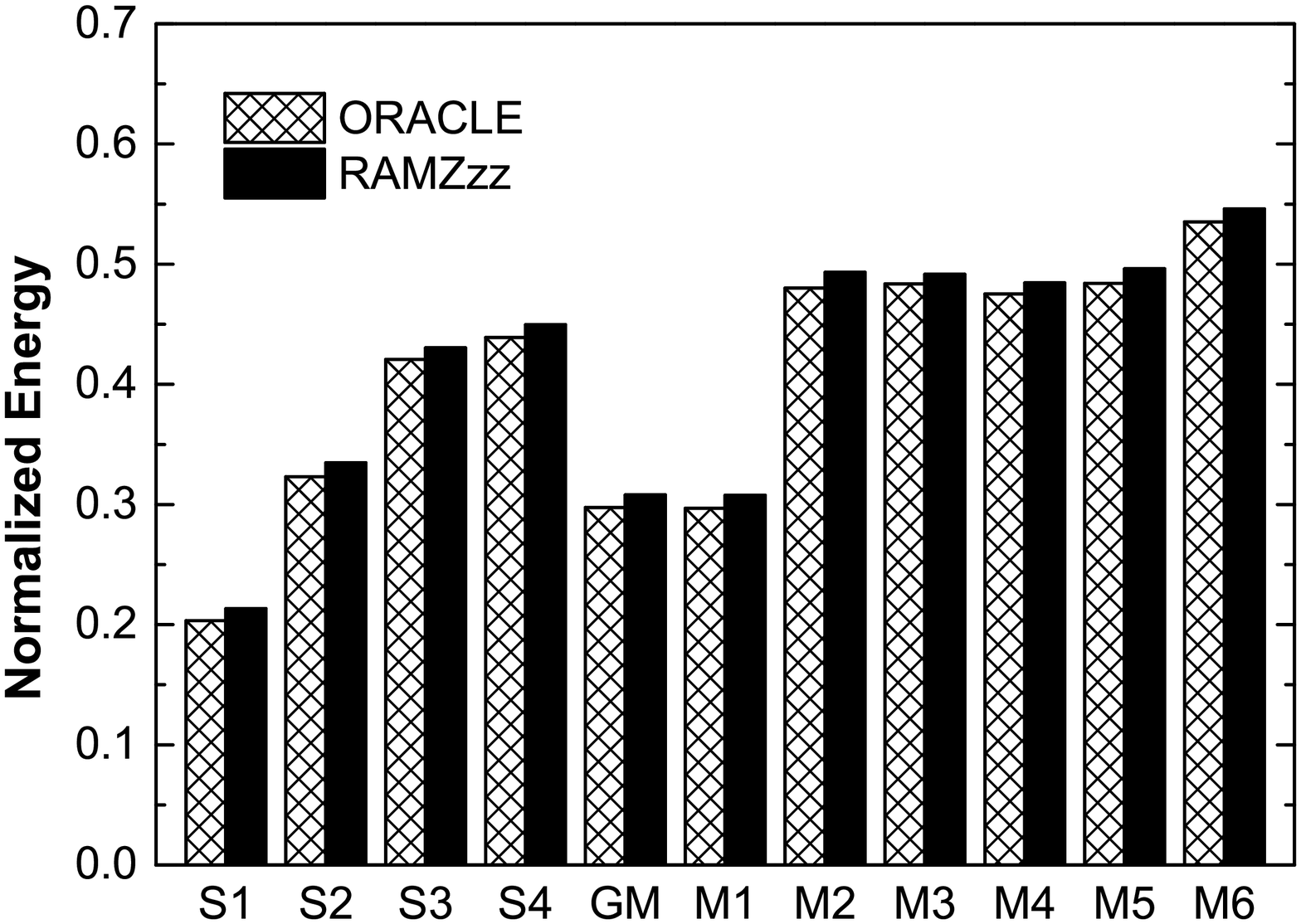}
\end{minipage}}
\caption{Comparing energy consumption of RAMZzz and ORACLE with the optimization goal of
energy consumption on three memory architectures.}
\label{fig:dram_energy_overall}
\vspace{-1ex}
\end{figure*}

\begin{figure*}[htb]
\centering
\subfigure[The breakdown of time for DDR3]{
\label{fig:ddr3_energy_ramzzz_bd}
\begin{minipage}[b]{0.31\linewidth}
\centering
\includegraphics[width=1.0\linewidth]{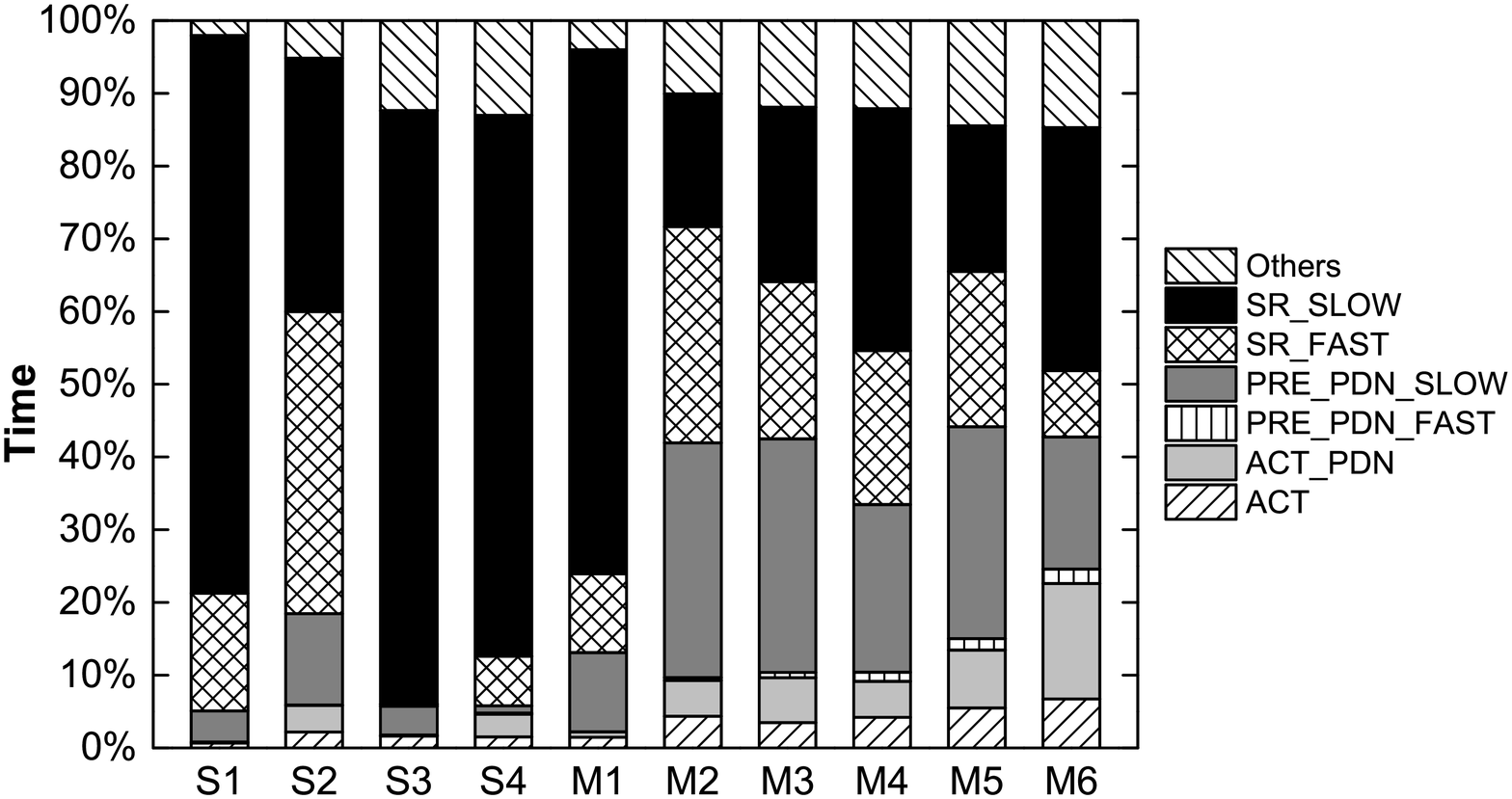}
\end{minipage}}
\hspace{0.1cm}
\subfigure[The breakdown of time for DDR2]{
\label{fig:ddr2_energy_ramzzz_bd}
\begin{minipage}[b]{0.31\linewidth}
\centering
\includegraphics[width=1.0\linewidth]{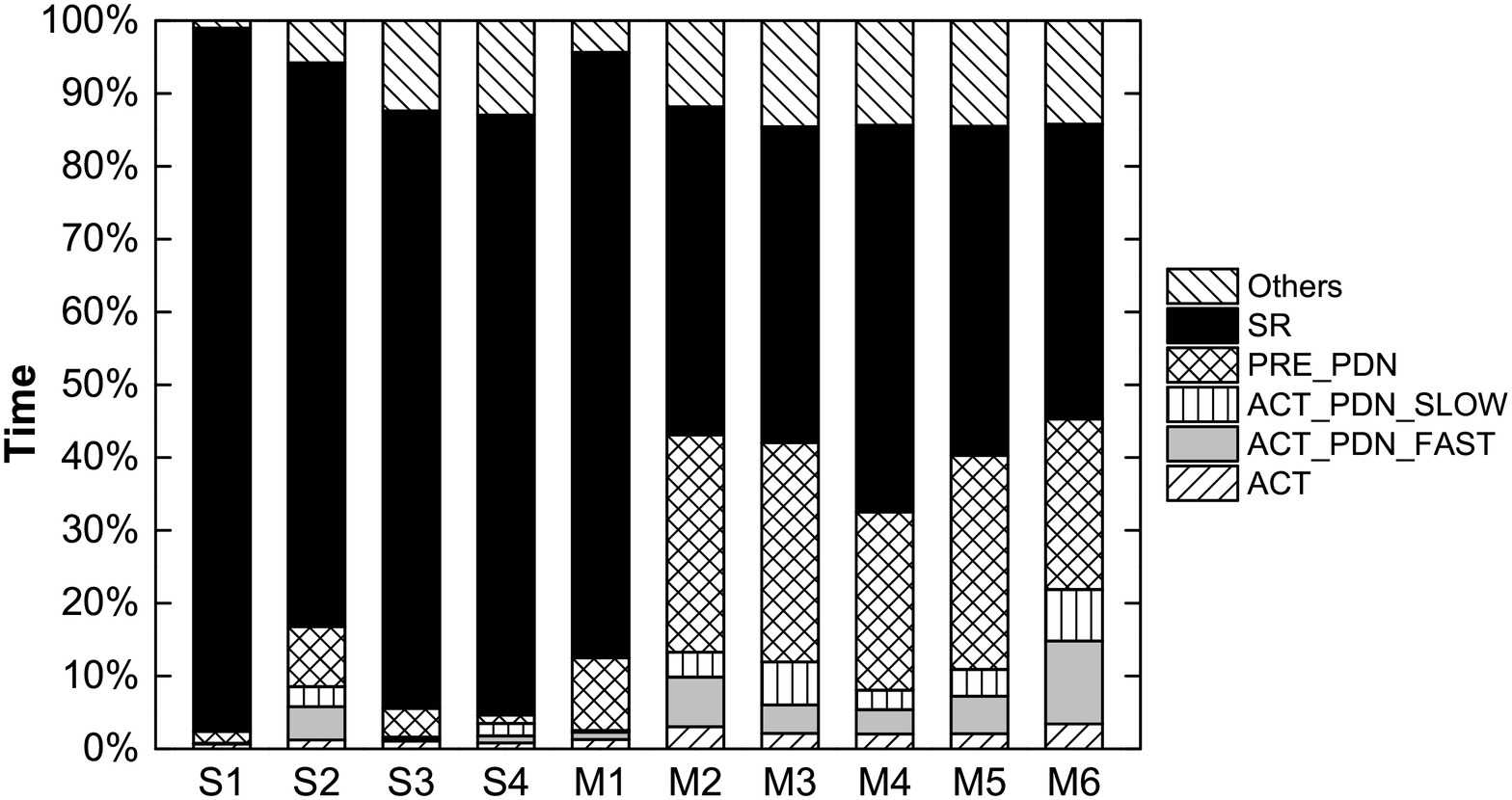}
\end{minipage}}
\hspace{0.1cm}
\subfigure[The breakdown of time for LPDDR2]{
\label{fig:lpddr2_energy_ramzzz_bd}
\begin{minipage}[b]{0.31\linewidth}
\centering
\includegraphics[width=0.93\linewidth]{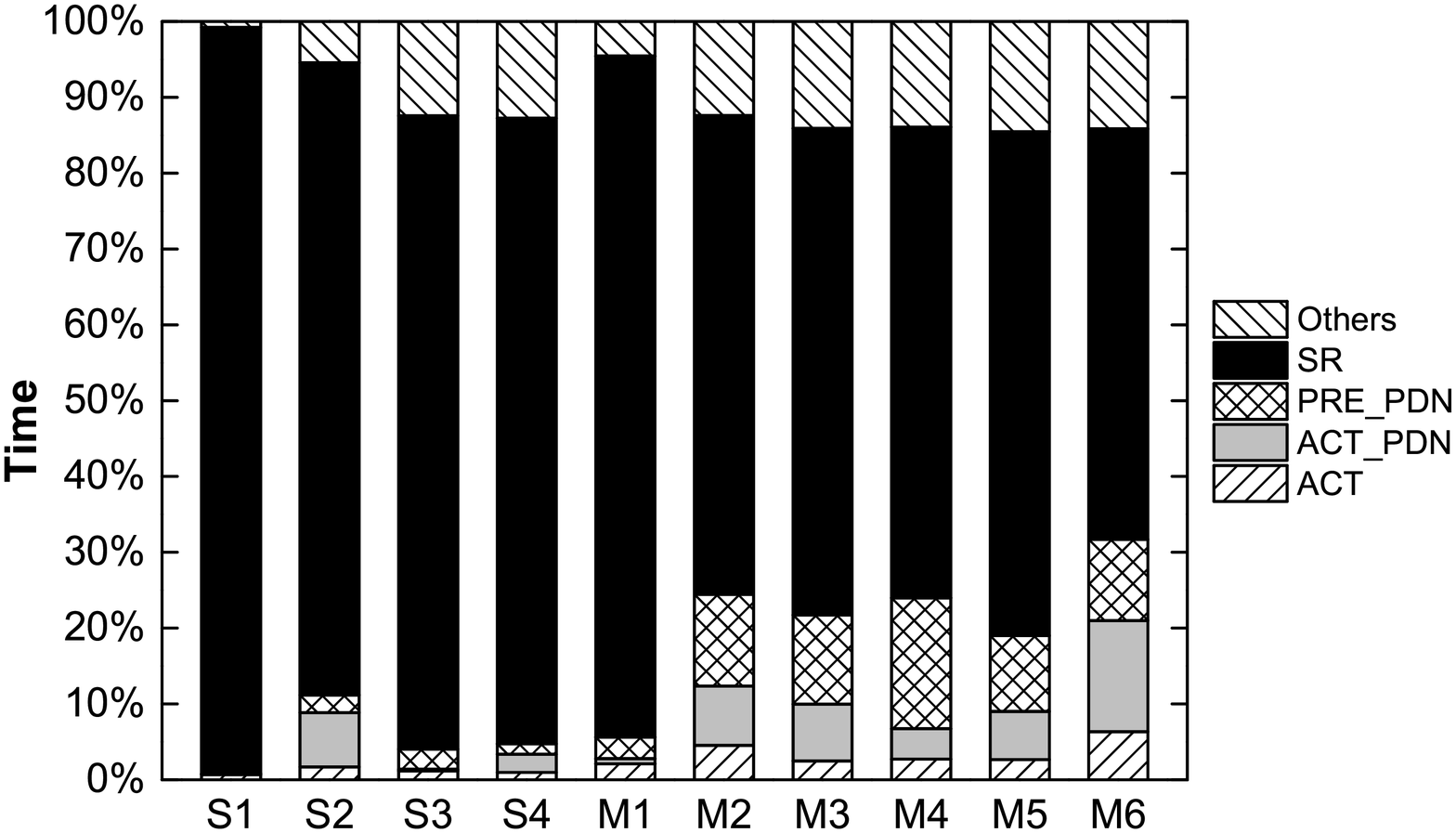}
\end{minipage}}
\caption{The breakdown of time stayed in different power states for RAMZzz with the optimization goal of
energy consumption on three memory architectures.}
\label{fig:dram_energy_ramzzz_bd}
\vspace{-1ex}
\end{figure*}

In this section, we present results of SPEC 2006 workloads when RAMZzz's
optimization metric is set to energy consumption. We set a relatively high delay
budget (10\%, 2.5 times of that used in the ED$^2$ experiment) to unleash the
potential of energy saving. Other system settings of DRAM system and
RAMZzz are the same as those used in Section 5.2 (e.g., 2GB DRAM with 8 ranks).
Figure~\ref{fig:dram_energy_overall} compares the energy consumption of RAMZzz,
ORACLE and BASE on different memory architectures.
Figure~\ref{fig:dram_energy_ramzzz_bd} shows the breakdown of time stayed in
different power states for RAMZzz on DDR3, DDR2 and LPDDR2. We make the following observations.
Firstly, RAMZzz is still more energy-efficient than BASE. The reduction on
energy consumption is 66.9\%, 65.8\% and 65.3\% on average on DDR3, DDR2 and
LPDDR2, respectively. The reduction is also more significant on the workloads of
single applications than the mixed workloads. This is consistent with our
observations in the ED$^2$ experiment.

RAMZzz consumes only 5.8\%, 4.1\% and 3.5\% on average more energy than ORACLE
on all workloads on DDR3, DDR2 and LPDDR2, respectively.
Our study on the power-down timeout shows that our prediction model is very
close to ORACLE. RAMZzz achieves the effectiveness and flexibility in different
optimization goals. While the optimization metric is set to energy consumption,
the total delay of RAMZzz, including remapping delay, migration delay and
resynchronization delay, is less than 7.5\% for all three DRAM architectures as
well as all workloads. Recall that our delay budget for energy-oriented
optimizations is 10\%.

We briefly present results of the individual impact of dynamic migrations and
adaptive demotions on RAMZzz with the optimization goal of energy consumption.
For dynamic page migrations, we observe RAMZzz has an average reduction of
18.4\%, 11.7\% and 11.1\% over RZ--SP, and with a range of 5.6--38.0\%,
3.4--29.8\% and 2.1--30.0\% on DDR3, DDR2 and LPDDR2, respectively. For adaptive
demotions, we observe RAMZzz has the reduction of 29.7--53.8\% (39.5\% on
average), 11.6--51.0\% (25.6\% on average) and 5.2--43.7\% (23.6\% on average)
over RZ--SD on DDR3, DDR2 and LPDDR2, respectively.

\section{Studies on Full System ED$^2$ and Energy Consumption}
\label{sec:full_system}

\begin{figure*}[htb]
\begin{center}
\begin{minipage}[b]{0.31\linewidth}
\centering
\includegraphics[width=0.90\linewidth]{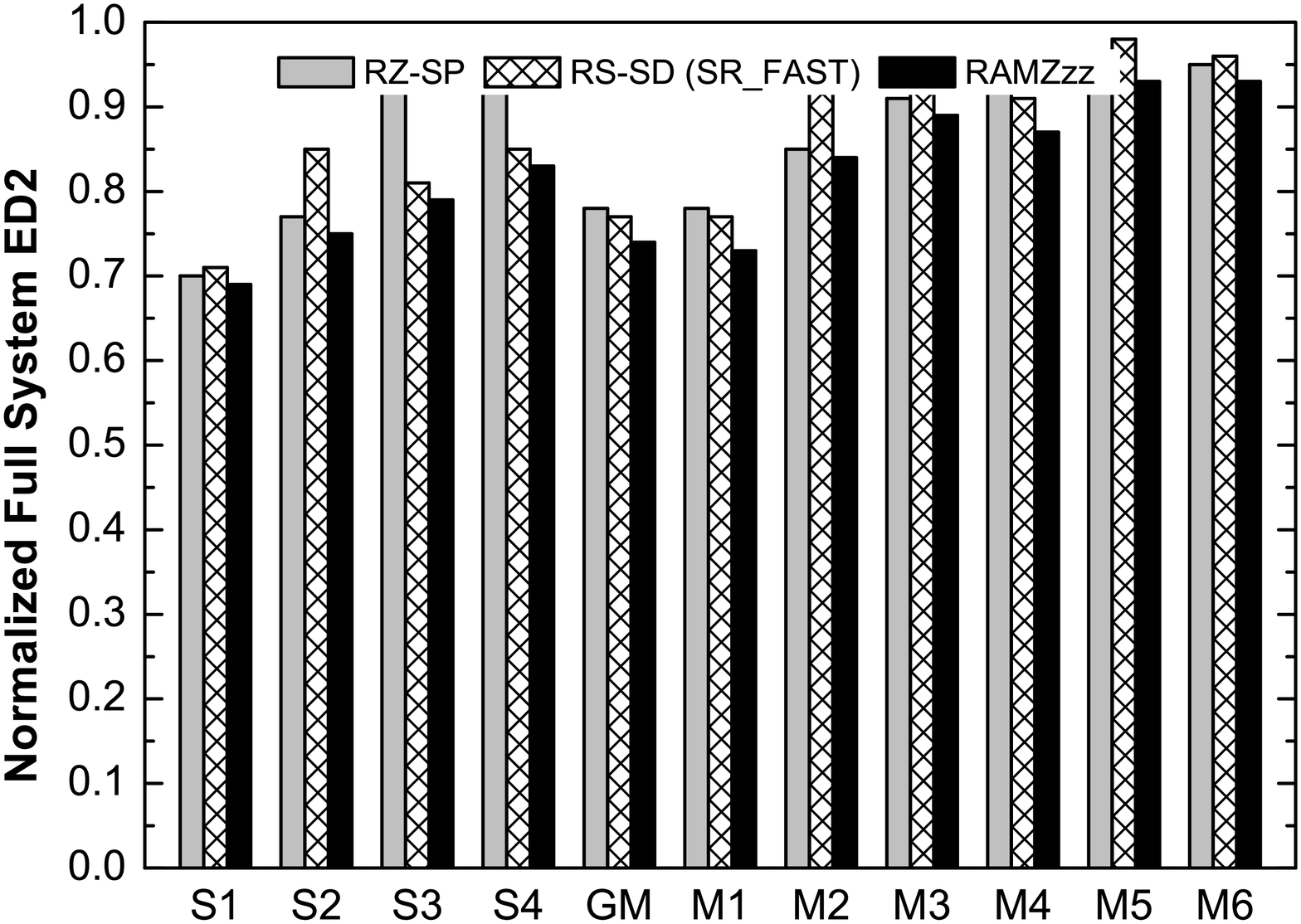}
\caption{Comparing full-system ED$^2$ with the optimization goal of
ED$^2$ on DDR3.} \label{fig:sys_ed2}
\end{minipage}
\hspace{0.1cm}
\begin{minipage}[b]{0.31\linewidth}
\centering
\includegraphics[width=0.90\linewidth]{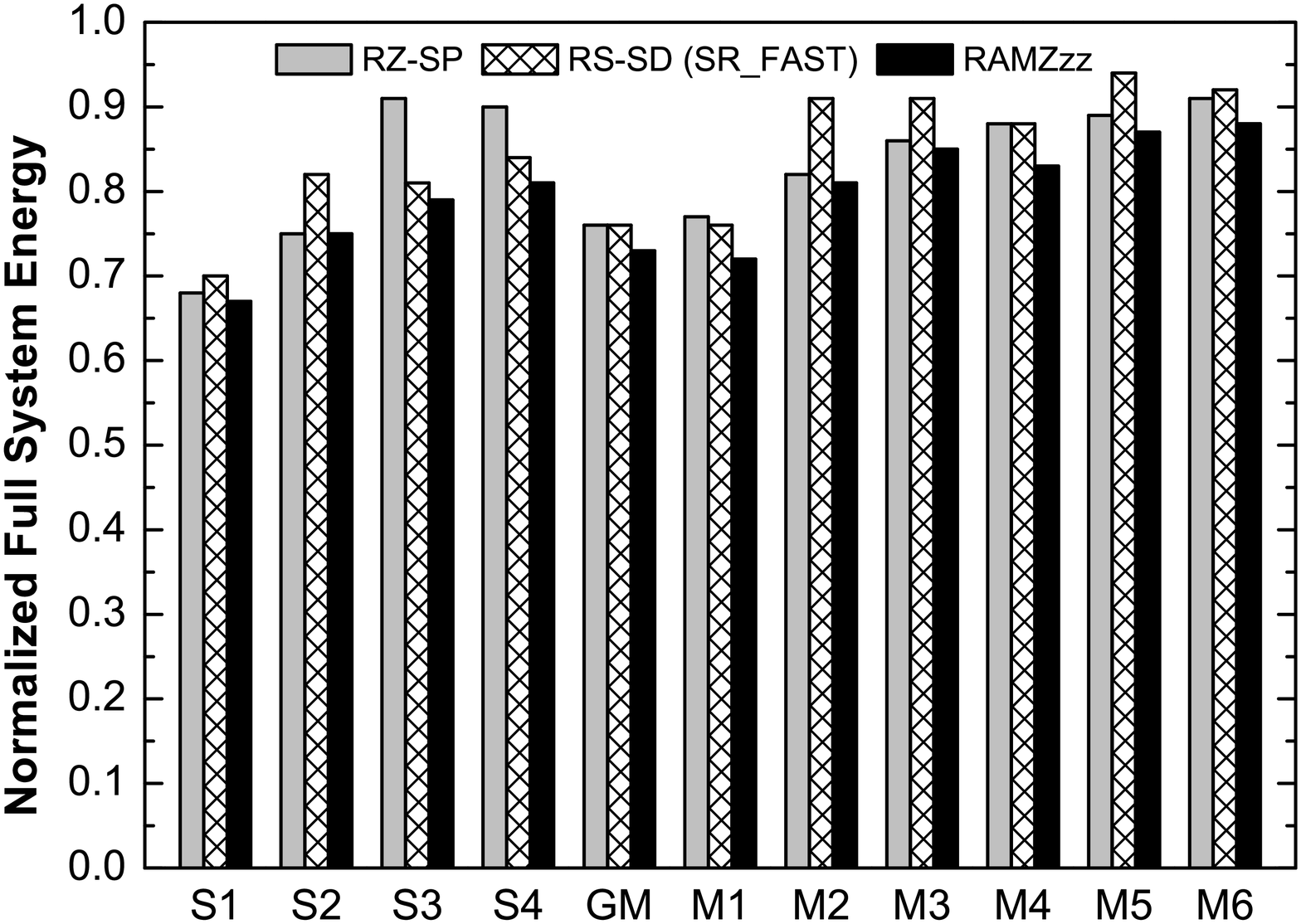}
\caption{Comparing full-system energy with the optimization goal of
energy consumption on DDR3.} \label{fig:sys_energy}
\end{minipage}
\hspace{0.1cm}
\begin{minipage}[b]{0.31\linewidth}
\centering
\includegraphics[width=0.90\linewidth]{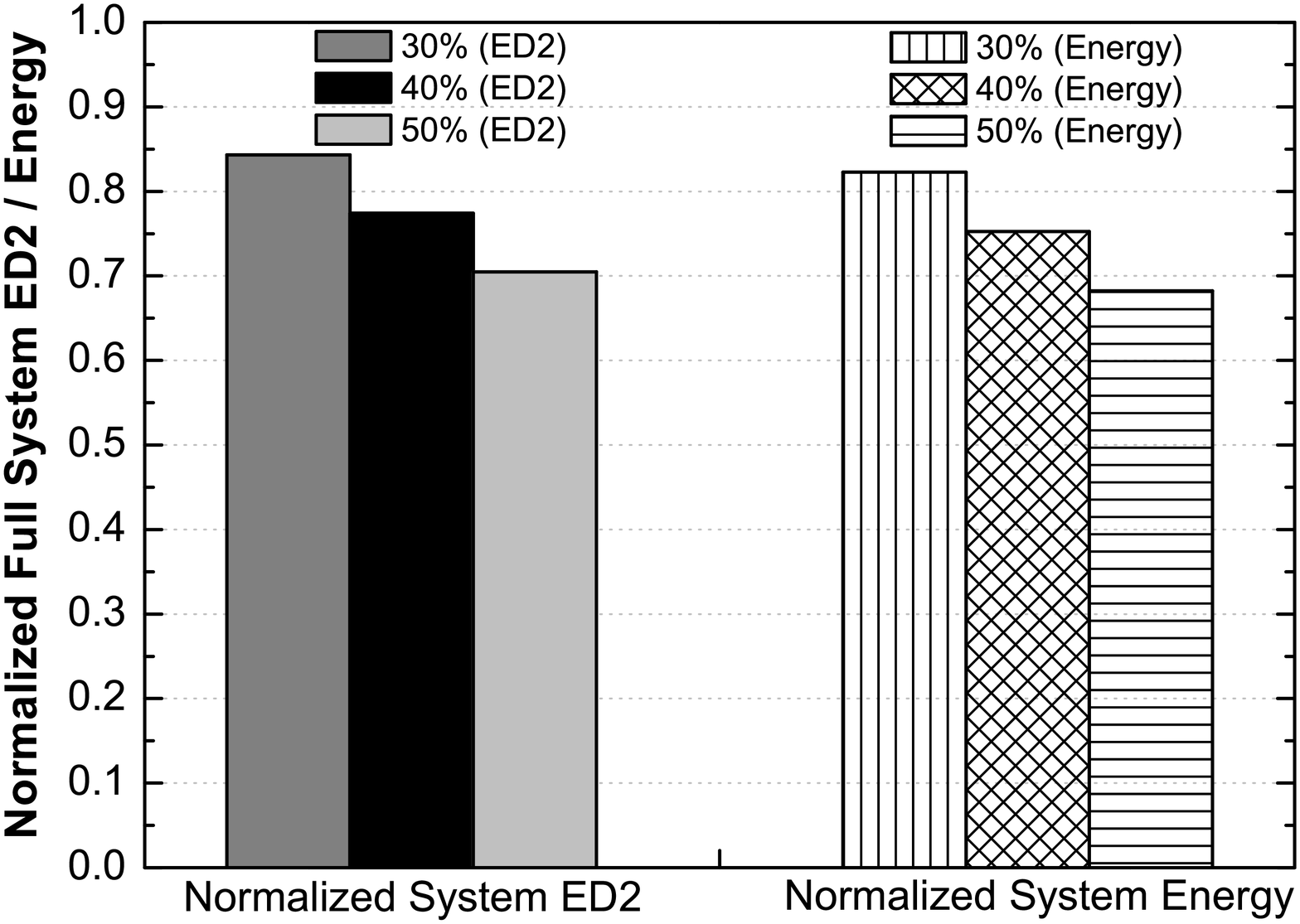}
\caption{The impact of the ratio of memory power to total system power.}
\label{fig:mem_power_ratio}
\end{minipage}
\vspace{-3ex}
\end{center}
\end{figure*}

In this section, we evaluate the impact of RAMZzz on
full-system energy consumption and performance with SPEC 2006
workloads. We start by performing back-of-envelop calculations,
following previous
studies~\cite{Deng:2011:MAL:1950365.1950392,Chatterjee:2012:LHD:2457472.2457482}.
We assume that the average power consumption of the DRAM system
accounts for 40\% of the total system power in the baseline policy
(i.e., BASE), and compute a fixed average power estimate (i.e., the
remaining 60\%) for all other components. Thus, the energy
consumption of all other components (excluding DRAM) scales with the
program execution time, which is usually consistent with the
real-world
case~\cite{Deng:2011:MAL:1950365.1950392,Chatterjee:2012:LHD:2457472.2457482}.
This ratio has been identified as the current contribution of DRAM
system to entire system power
consumption~\cite{Hoelzle:2009:DCI:1643608,Ousterhout:2010:CRS:1713254.1713276,Tsirogiannis:2010:AEE:1807167.1807194}.
We also study the impact of varying this ratio in this evaluation.
Architectural characteristics and experimental parameters for
ED$^2$-oriented and energy-oriented optimizations are the same as
those used in Section 5.2 and Appendix~\ref{sec:energy_overall},
respectively.

Figure~\ref{fig:sys_ed2} presents full system ED$^2$ of RAMZzz, RZ--SP and
RZ--SD (SR\_FAST is used as the pre-selected low-power state) when the
optimization metric is set to ED$^2$ on DDR3. All three approaches still
outperform BASE on all workloads in terms of full system ED$^2$. Compared with
BASE, the reduction in full system ED$^2$ is 23.0\%, 18.0\% and 17.8\% on
average for RAMZzz, RZ--SP and RZ--SD, respectively. RAMZzz outperforms both
RZ--SP and RZ--SD in full-system ED$^2$, but leads to slightly higher
performance degradations. We observe that RAMZzz has an average reduction of
4.8\% (from 1.6\% to 17.9\%) and 5.3\% (from 1.7\% to 8.6\%) over RZ--SP and
RZ--SD in full system ED$^2$, respectively. When the optimization metric is set
to energy consumption on DDR3, all three approaches are also energy-efficient
than BASE in full system energy as shown in Figure~\ref{fig:sys_energy}. RAMZzz
outperforms both RZ--SP and RZ--SD, with the average reduction of 4.3\% (from
1.4\% to 13.8\%) and 5.9\% (from 1.8\% to 9.0\%) in full system energy,
respectively. We observe similar results on other DRAM architectures.

We further study the ratio of power consumption of the memory
subsystem to the overall power consumption of the full system.
Particularly, we vary the ratio from 30\% to 50\%.
Figure~\ref{fig:mem_power_ratio} shows that the fraction of memory power
has a significant effect on both full system ED$^2$ and energy
consumption. Increasing the ratio from 30\% to 50\% (i.e., the power
contribution of other components are reduced from 70\% to 50\%), the normalized
full-system ED$^2$ and energy consumption of RAMZzz decrease from 0.84 to 0.70
and 0.83 to 0.68, respectively.

\section{Sensitivity Studies}\label{subsec:sensitivity}

\begin{figure*}[htb]
\begin{center}
\begin{minipage}[b]{0.31\linewidth}
\centering
\includegraphics[width=0.90\linewidth]{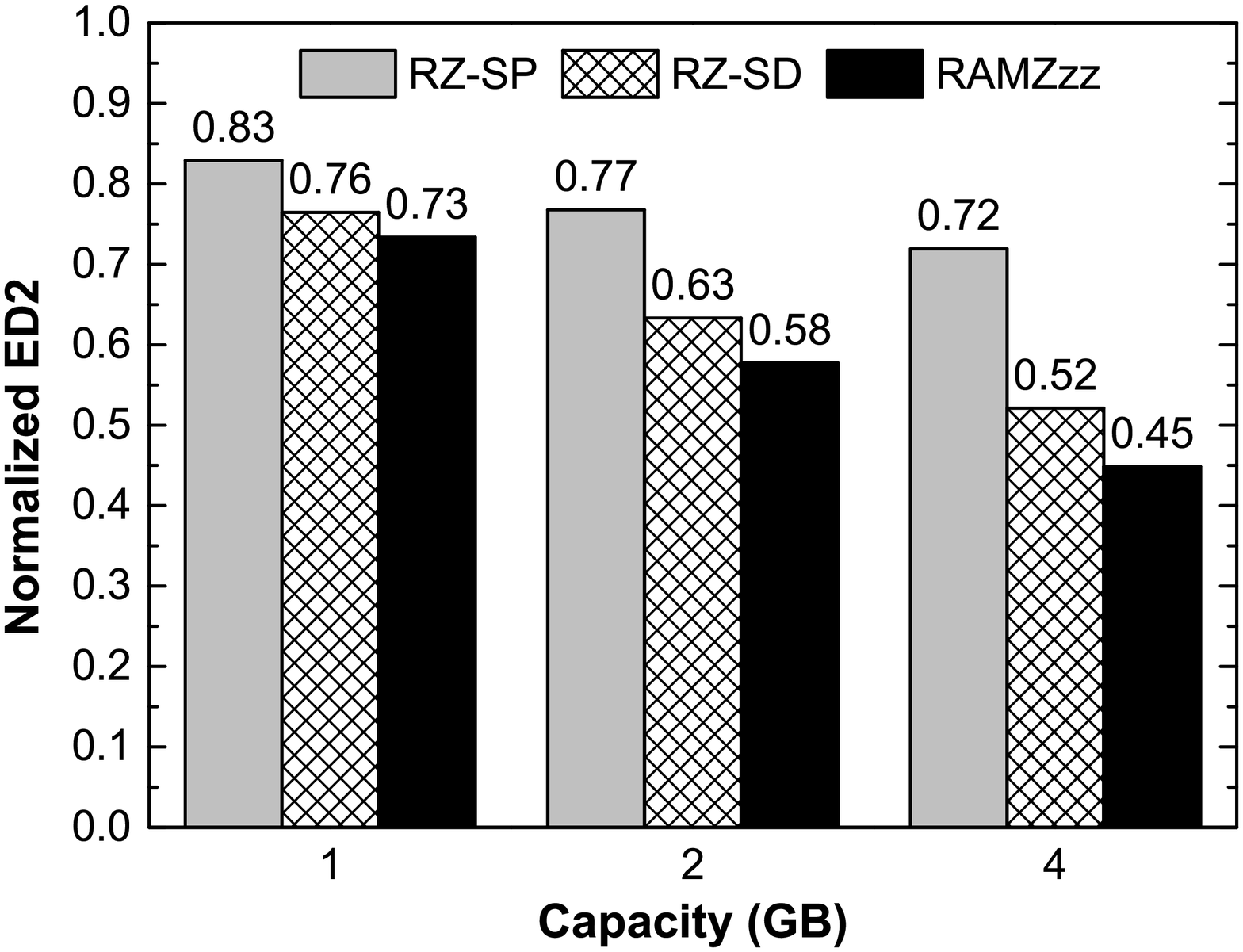}
\caption{Comparing ED$^2$ with varying DRAM capacity on M4 on
DDR3.}
\label{fig:capacity}
\end{minipage}
\hspace{0.1cm}
\begin{minipage}[b]{0.31\linewidth}
\centering
\includegraphics[width=0.90\linewidth]{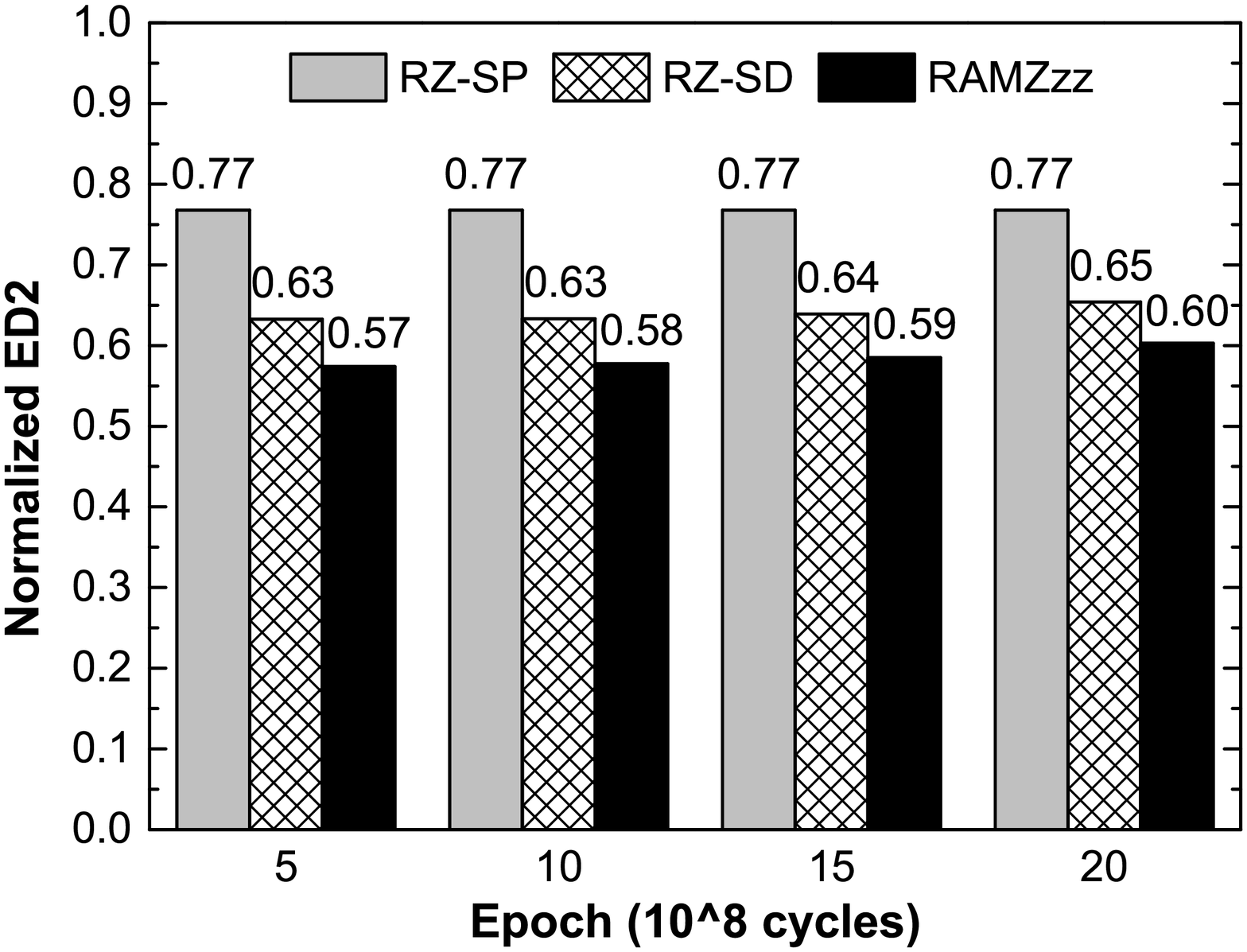}
\caption{Comparing ED$^2$ with varying epoch size on M4 on
DDR3.}
\label{fig:epoch}
\end{minipage}
\hspace{0.1cm}
\begin{minipage}[b]{0.31\linewidth}
\centering
\includegraphics[width=0.90\linewidth]{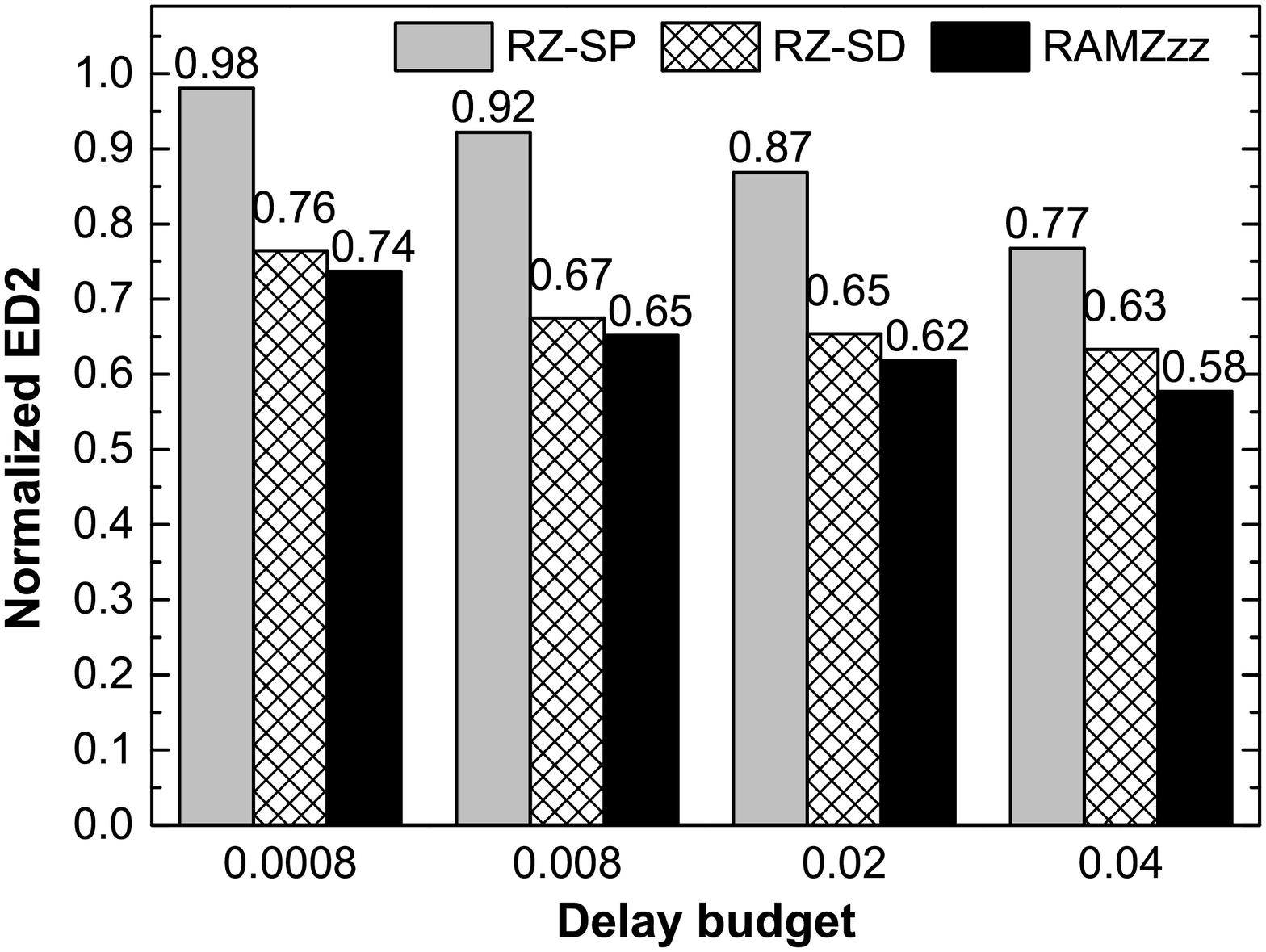}
\caption{Comparing ED$^2$ with varying delay budget on M4 on
DDR3.}
\label{fig:delaybudget}
\end{minipage}
\vspace{-3ex}
\end{center}
\end{figure*}

We use ED$^2$ as the optimization metric, DDR3 as the target memory
architecture and SPEC 2006 workloads to conduct the sensitivity analysis. Since
RAMZzz is very close to ORACLE, we present results for RAMZzz, RZ--SP and RZ--SD
(choosing PRE\_PDN\_SLOW as the target low-power state) only. In those studies,
we vary one parameter at a time and keep other parameters in their default
settings. Due to the space limitation, we present the figures for M4 (a modest
case among all those workloads) and comment on other workloads without figures
when appropriate.

{\bf DRAM parameters.} We study the impact of different numbers of ranks and
memory capacities of DRAM. As the number of ranks increases, we observe a
rather stable ED$^2$ for RAMZzz, RZ--SD and RZ--SP. For all three approaches,
when the number of ranks increases from 2 to 4, ED$^2$ drops less than 1\%,
because of a finer grained power control on ranks. When the number of ranks
increases from 4, 8 to 16, ED$^2$ increases less than 3\%. The major reason for
increasing ED$^2$ is the increased amount of page migrations caused by
increasing number of ranks. Figure~\ref{fig:capacity} shows the results for
varying the memory capacity. As memory capacities increase, all three methods achieve a
lower ED$^2$, and the ED$^2$ improvement of RAMZzz over RZ--SD and RZ--SP both
becomes larger. That indicates the effectiveness of our approach on
larger-memory systems.

{\bf RAMZzz parameters.} We study the impact of different epoch/slot sizes and
delay budgets of RAMZzz.

Figure~\ref{fig:epoch} shows the results of varying epoch size. RZ--SP is not
sensitive to the epoch size, whereas the ED$^2$ of RAMZzz and RZ--SD both
increases slightly. That is because, for a longer epoch, the rank hotness does
not affect the changes in page access locality in time, and the ED$^2$
improvement brought by page migrations is slightly reduced. We observed a
similar result when varying the slot size in $(0.125\times 10^8\times 2^i)$
cycles ($i=$0, 1, ..., 3). The ED$^2$ of RAMZzz varies by less than 2\%. In
practice, we set the slot size to be $10^8$ cycles, and the epoch size to be
$10^9$ cycles as a compromise on the prediction overhead and the accuracy.

Figure~\ref{fig:delaybudget} compares ED$^2$ for varying delay
budget. A small delay budget limits the potential for energy saving,
whereas a large delay budget leads to too aggressive energy saving
and exaggerates the delay incurred by miss-predictions. In practice,
we set the delay budget within 1--4\% for optimizing ED$^2$.

\begin{figure*}[htb]
\begin{center}
\begin{minipage}[b]{0.31\linewidth}
\centering
\includegraphics[width=0.95\linewidth]{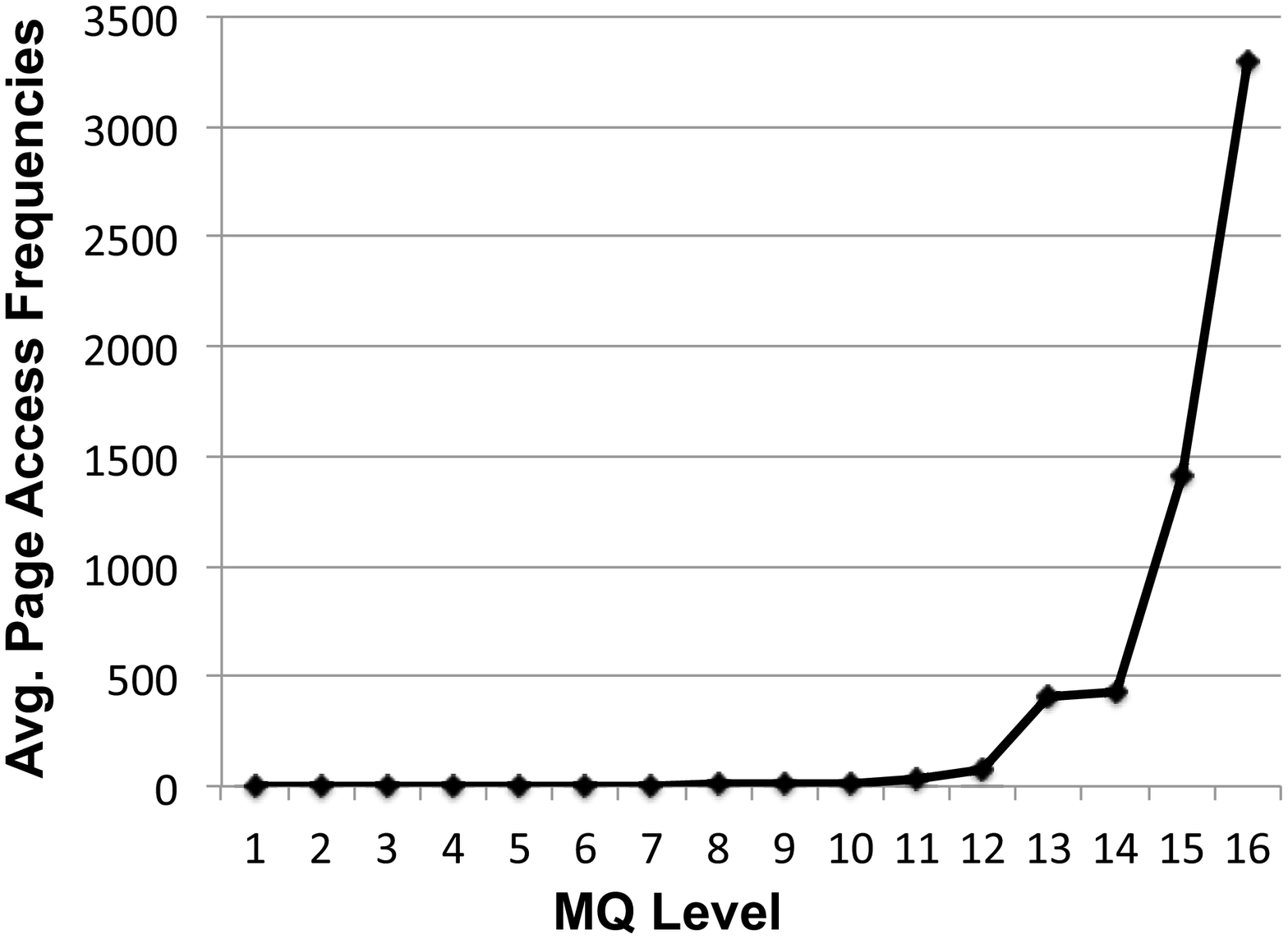}
\caption{The average of page access frequencies for different MQ
levels.}
\label{fig:mq_study}
\end{minipage}
\hspace{0.1cm}
\begin{minipage}[b]{0.31\linewidth}
\centering
\includegraphics[width=0.95\linewidth]{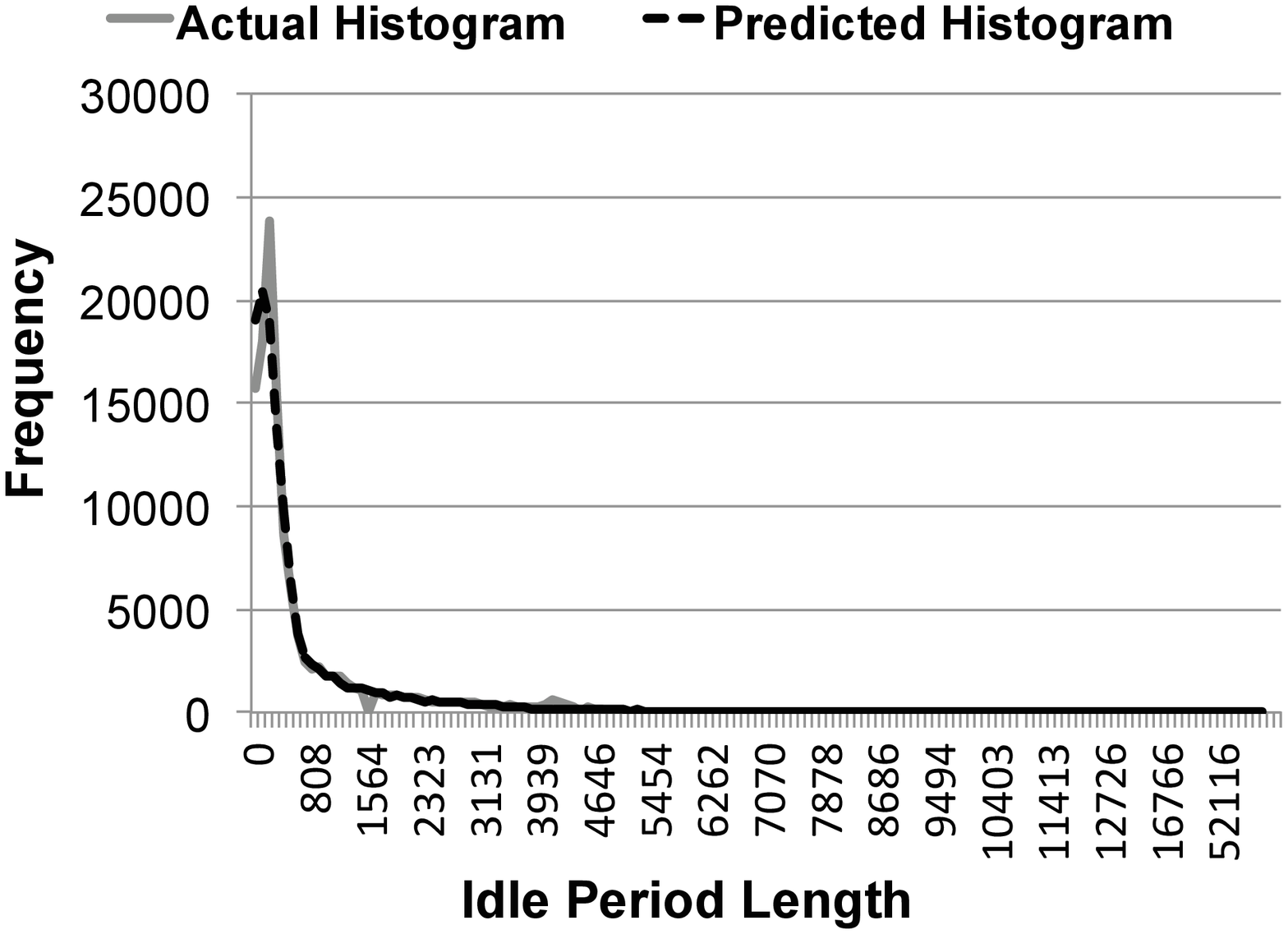}
\caption{Comparing the actual and predicted idle histogram of a slot
that is not the start of an epoch.}
\label{fig:hist_in_an_epoch}
\end{minipage}
\hspace{0.1cm}
\begin{minipage}[b]{0.31\linewidth}
\centering
\includegraphics[width=0.95\linewidth]{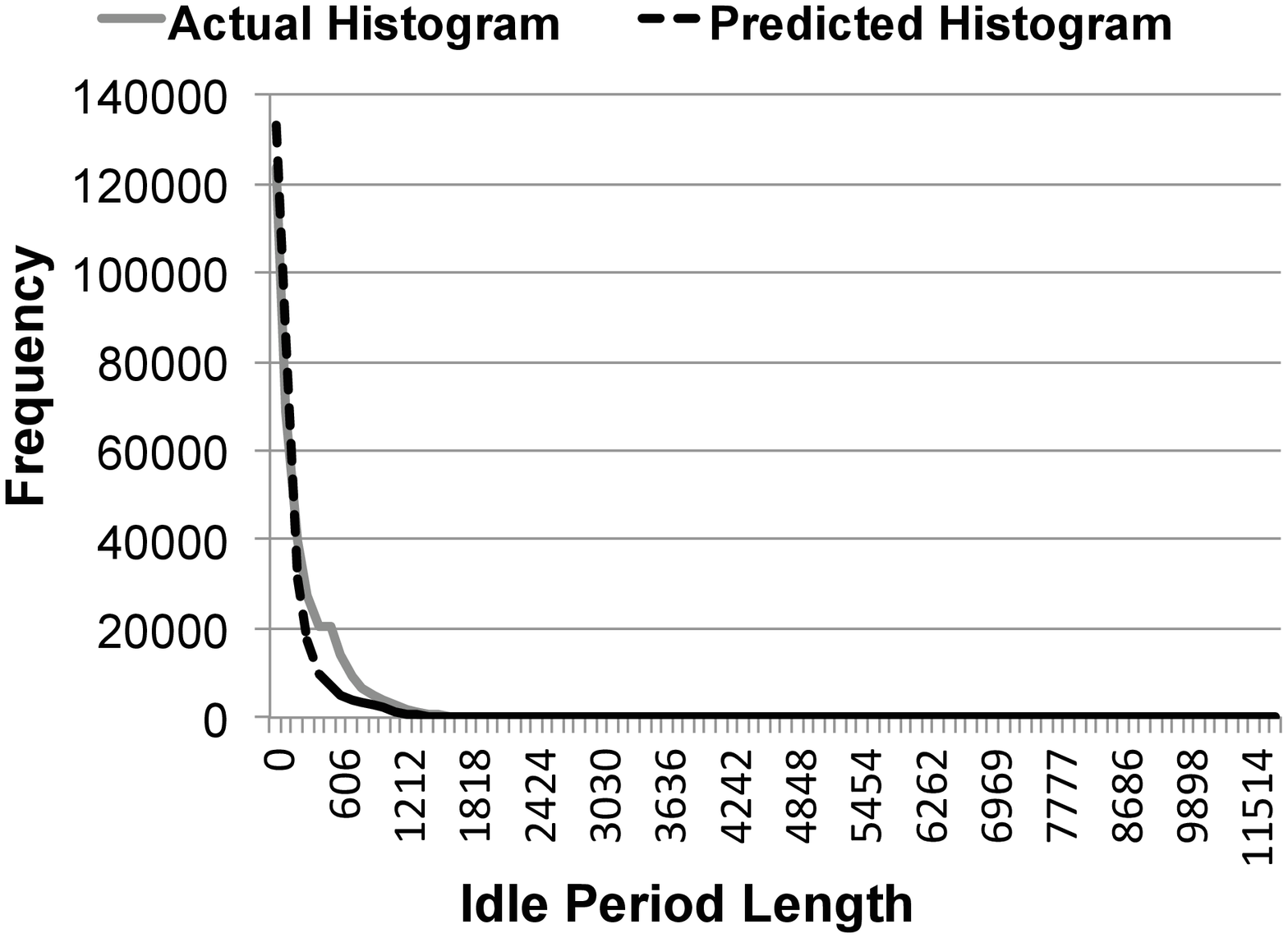}
\caption{Comparing the actual and predicted idle histogram of a slot that
is the start of an epoch.}
\label{fig:hist_out_an_epoch}
\end{minipage}
\vspace{-3ex}
\end{center}
\end{figure*}

\section{Studies on MQ Structure}

Figure~\ref{fig:mq_study} shows physical page access frequencies for
pages stored in different levels of the MQ structure during an epoch after page
migrations of RAMZzz on DDR3 architecture and M4 workload. We observe that pages
stored in high MQ levels have higher access frequencies, while pages stored in
low MQ levels have lower access frequencies. This shows that the MQ structure
actually correlates with page access distribution. We have consistent
observations on other workloads and memory architectures.

\section{Studies on Prediction of Idle Period Distribution}

We compare the predicted idle histogram to the actual idle histogram of
RAMZzz on Rank 0 on DDR3 architecture for a selected workload (M4) in this
section. The predicted histogram is close to the actual histogram in our
evaluation in both cases: 1) the slot is not the beginning of an epoch as shown in
Figure~\ref{fig:hist_in_an_epoch} (in this case, the actual histogram in the
previous slot is used as the prediction of the current slot); 2) the slot is the
beginning of an epoch as shown in Figure~\ref{fig:hist_out_an_epoch} (in this
case, the algorithm developed in Section 4.2 is used to estimate the predicted
histogram). We have observed similar results on other workloads and memory
architectures.

We also find that the confidence of the regression curve being an
exponential function (i.e., $R^2$) on the actual histogram of idle periods is high (0.94 and
0.92 in Figures~\ref{fig:hist_in_an_epoch} and~\ref{fig:hist_out_an_epoch},
respectively). Thus, our assumption that the inter-arrival times of memory
requests follow a Poisson distribution is valid.

\bibliographystyle{IEEEtran}
\bibliography{DRAM}

\begin{thebibliography}{10}
\providecommand{\url}[1]{#1}
\csname url@samestyle\endcsname
\providecommand{\newblock}{\relax}
\providecommand{\bibinfo}[2]{#2}
\providecommand{\BIBentrySTDinterwordspacing}{\spaceskip=0pt\relax}
\providecommand{\BIBentryALTinterwordstretchfactor}{4}
\providecommand{\BIBentryALTinterwordspacing}{\spaceskip=\fontdimen2\font plus
\BIBentryALTinterwordstretchfactor\fontdimen3\font minus
  \fontdimen4\font\relax}
\providecommand{\BIBforeignlanguage}[2]{{%
\expandafter\ifx\csname l@#1\endcsname\relax
\typeout{** WARNING: IEEEtran.bst: No hyphenation pattern has been}%
\typeout{** loaded for the language `#1'. Using the pattern for}%
\typeout{** the default language instead.}%
\else
\language=\csname l@#1\endcsname
\fi
#2}}
\providecommand{\BIBdecl}{\relax}
\BIBdecl

\bibitem{Hoelzle:2009:DCI:1643608}
U.~Hoelzle and L.~A. Barroso, \emph{The Datacenter as a Computer: An
  Introduction to the Design of Warehouse-Scale Machines}, 1st~ed.\hskip 1em
  plus 0.5em minus 0.4em\relax Morgan and Claypool Publishers, 2009.

\bibitem{Lefurgy:2003:EMC:957964.957972}
C.~Lefurgy, K.~Rajamani, F.~Rawson, W.~Felter, M.~Kistler, and T.~W. Keller,
  ``Energy management for commercial servers,'' \emph{IEEE Computer}, vol.~36,
  no.~12, 2003.

\bibitem{Meisner:2009:PES:1508244.1508269}
D.~Meisner, B.~T. Gold, and T.~F. Wenisch, ``Powernap: eliminating server idle
  power,'' in \emph{ASPLOS '09}, 2009.

\bibitem{s.ware:architecting}
M.~S. Ware, K.~Rajamani, M.~S. Floyd, B.~Brock, J.~C. Rubio, F.~L.~R. III, and
  J.~B. Carter, ``Architecting for power management: The ibm power7 approach,''
  in \emph{HPCA '10}, 2010.

\bibitem{Huang:2003:DIP:1247340.1247345}
H.~Huang, P.~Pillai, and K.~G. Shin, ``Design and implementation of power-aware
  virtual memory,'' in \emph{USENIX ATC '03}, 2003.

\bibitem{Huang:2005:IEE:1077603.1077696}
H.~Huang, K.~G. Shin, C.~Lefurgy, and T.~Keller, ``Improving energy efficiency
  by making dram less randomly accessed,'' in \emph{ISLPED '05}, 2005.

\bibitem{Delaluz:2002:SDE:513918.514095}
V.~Delaluz, A.~Sivasubramaniam, M.~Kandemir, N.~Vijaykrishnan, and M.~J. Irwin,
  ``Scheduler-based dram energy management,'' in \emph{DAC '02}, 2002.

\bibitem{DBEnergy}
C.~Bae and T.~Jamel, ``Energy-aware memory management through database buffer
  management,'' in \emph{WEED '11}, 2011.

\bibitem{Fan:2001:MCP:383082.383118}
X.~Fan, C.~Ellis, and A.~Lebeck, ``Memory controller policies for dram power
  management,'' in \emph{ISLPED '01}, 2001.

\bibitem{Diniz:2007:LPC:1250662.1250699}
B.~Diniz, D.~Guedes, W.~Meira, Jr., and R.~Bianchini, ``Limiting the power
  consumption of main memory,'' in \emph{ISCA '07}, 2007.

\bibitem{bienia11benchmarking}
C.~Bienia, ``Benchmarking modern multiprocessors,'' Ph.D. dissertation,
  Princeton University, January 2011.

\bibitem{Lebeck:2000:PAP:378993.379007}
A.~R. Lebeck, X.~Fan, H.~Zeng, and C.~Ellis, ``Power aware page allocation,''
  in \emph{ASPLOS '00}, 2000.

\bibitem{Zheng:2010:PPT:1850266.1850273}
H.~Zheng and Z.~Zhu, ``Power and performance trade-offs in contemporary dram
  system designs for multicore processors,'' \emph{IEEE Trans. on Comput.},
  vol.~59, no.~8, 2010.

\bibitem{Malladi:2012:TED:2337159.2337164}
K.~T. Malladi, B.~C. Lee, F.~A. Nothaft, C.~Kozyrakis, K.~Periyathambi, and
  M.~Horowitz, ``Towards energy-proportional datacenter memory with mobile
  dram,'' in \emph{ISCA '12}, 2012.

\bibitem{ddr3spec}
{Micron Tech., Inc.}, \emph{MT41J256M4JP-15E Datasheet}, 2010.

\bibitem{ddr2spec}
{Micron Tech., Inc.}, \emph{MT47H128M8CF-25E Datasheet}, 2007.

\bibitem{lpddr2spec}
{Micron Tech., Inc.}, \emph{MT42L128M32D1LF-25WT Datasheet}, 2011.

\bibitem{calc}
{Micron Tech., Inc.}, \emph{System Power Calculator},
  http://www.micron.com/products/support/power-calc, 2012.

\bibitem{hur:a}
I.~Hur and C.~Lin, ``A comprehensive approach to dram power management,'' in
  \emph{HPCA '08}, 2008.

\bibitem{Sudan:6212453}
K.~Sudan, K.~Rajamani, W.~Huang, and J.~Carter, ``Tiered memory: An iso-power
  memory architecture to address the memory power wall,'' \emph{IEEE Trans. on
  Comput.}, vol.~61, no.~12, 2012.

\bibitem{Delaluz:2000:ECO:354880.354900}
V.~Delaluz, M.~Kandemir, N.~Vijaykrishnan, and M.~J. Irwin, ``Energy-oriented
  compiler optimizations for partitioned memory architectures,'' in \emph{CASES
  '00}, 2000.

\bibitem{Delaluz:2001:DEM:580550.876438}
V.~Delaluz, M.~Kandemir, N.~Vijaykrishnan, A.~Sivasubramaniam, and M.~J. Irwin,
  ``Dram energy management using software and hardware directed power mode
  control,'' in \emph{HPCA '01}, 2001.

\bibitem{Zheng:2008:MAD:1521747.1521797}
H.~Zheng, J.~Lin, Z.~Zhang, E.~Gorbatov, H.~David, and Z.~Zhu, ``Mini-rank:
  Adaptive dram architecture for improving memory power efficiency,'' in
  \emph{MICRO '41}, 2008.

\bibitem{Ahn:2009:FSP:1654059.1654102}
J.~H. Ahn, N.~P. Jouppi, C.~Kozyrakis, J.~Leverich, and R.~S. Schreiber,
  ``Future scaling of processor-memory interfaces,'' in \emph{SC '09}, 2009.

\bibitem{Zheng:2009:DDB:1555754.1555788}
H.~Zheng, J.~Lin, Z.~Zhang, and Z.~Zhu, ``Decoupled dimm: Building
  high-bandwidth memory system using low-speed dram devices,'' in \emph{ISCA
  '09}, 2009.

\bibitem{bi:delay-hiding}
M.~Bi, R.~Duan, and C.~Gniady, ``Delay-hiding energy management mechanisms for
  dram,'' in \emph{HPCA '10}, 2010.

\bibitem{Cooper-Balis:2010:FAP:1849300.1849320}
E.~Cooper-Balis and B.~Jacob, ``Fine-grained activation for power reduction in
  dram,'' \emph{IEEE Micro}, vol.~30, 2010.

\bibitem{David:2011:MPM:1998582.1998590}
H.~David, C.~Fallin, E.~Gorbatov, U.~R. Hanebutte, and O.~Mutlu, ``Memory power
  management via dynamic voltage/frequency scaling,'' in \emph{ICAC '11}, 2011.

\bibitem{Deng:2011:MAL:1950365.1950392}
Q.~Deng, D.~Meisner, L.~Ramos, T.~F. Wenisch, and R.~Bianchini, ``Memscale:
  active low-power modes for main memory,'' in \emph{ASPLOS '11}, 2011.

\bibitem{kim:HPCA2010}
Y.~Kim, D.~Han, O.~Mutlu, and M.~Harchol-Balter, ``Atlas: A scalable and
  high-performance scheduling algorithm for multiple memory controllers,'' in
  \emph{HPCA '10}, 2010.

\bibitem{Udipi:2010:RDD:1815961.1815983}
A.~N. Udipi, N.~Muralimanohar, N.~Chatterjee, R.~Balasubramonian, A.~Davis, and
  N.~P. Jouppi, ``Rethinking dram design and organization for
  energy-constrained multi-cores,'' in \emph{ISCA '10}, 2010.

\bibitem{He:2006:CAX:1175893.1176184}
B.~He, Q.~Luo, and B.~Choi, ``Cache-conscious automata for xml filtering,''
  \emph{IEEE Trans. on Knowl. and Data Eng.}, vol.~18, no.~12, 2006.

\bibitem{He:2008:CDL:1366102.1366105}
B.~He and Q.~Luo, ``Cache-oblivious databases: Limitations and opportunities,''
  \emph{ACM Trans. Database Syst.}, vol.~33, no.~2, 2008.

\bibitem{DBLP:conf/cidr/HeL07}
B.~He and Q.~Luo, ``Cache-oblivious query processing,'' in \emph{CIDR '07},
  2007.

\bibitem{Kumar:2011:MEM:2016802.2016864}
K.~Kumar, K.~Doshi, M.~Dimitrov, and Y.-H. Lu, ``Memory energy management for
  an enterprise decision support system,'' in \emph{ISLPED '11}, 2011.

\bibitem{Wu:2012:RRD:2388996.2389040}
D.~Wu, B.~He, X.~Tang, J.~Xu, and M.~Guo, ``Ramzzz: rank-aware dram power
  management with dynamic migrations and demotions,'' in \emph{SC '12}, 2012.

\bibitem{Zhou:2001:MRA:647055.715773}
Y.~Zhou, J.~Philbin, and K.~Li, ``The multi-queue replacement algorithm for
  second level buffer caches,'' in \emph{USENIX ATC '01}, 2001.

\bibitem{Ramos:2011:PPH:1995896.1995911}
L.~E. Ramos, E.~Gorbatov, and R.~Bianchini, ``Page placement in hybrid memory
  systems,'' in \emph{ICS '11}, 2011.

\bibitem{Xilinx}
{Xilinx, Inc.}, \emph{Spartan-6 FPGA Memory Controller User Guide}, 2010.

\bibitem{Sudan:2010:MID:1736020.1736045}
K.~Sudan, N.~Chatterjee, D.~Nellans, M.~Awasthi, R.~Balasubramonian, and
  A.~Davis, ``Micro-pages: increasing dram efficiency with locality-aware data
  placement,'' in \emph{ASPLOS '10}, 2010.

\bibitem{ptlsim}
M.~T. Yourst, ``Ptlsim: A cycle accurate full system x86-64 microarchitectural
  simulator,'' in \emph{ISPASS '07}, 2007.

\bibitem{carlson2011etloafsaapms}
T.~E. Carlson, W.~Heirman, and L.~Eeckhout, ``Sniper: Exploring the level of
  abstraction for scalable and accurate parallel multi-core simulations,'' in
  \emph{SC '11)}, 2011.

\bibitem{Guo:2010:RCA:1815961.1816012}
X.~Guo, E.~Ipek, and T.~Soyata, ``Resistive computation: Avoiding the power
  wall with low-leakage, stt-mram based computing,'' in \emph{ISCA '10}, 2010.

\bibitem{Zhang:2000:PPI:360128.360134}
Z.~Zhang, Z.~Zhu, and X.~Zhang, ``A permutation-based page interleaving scheme
  to reduce row-buffer conflicts and exploit data locality,'' in \emph{MICRO
  '00}, 2000.

\bibitem{Wang:2013:EHM:2523721.2523737}
B.~Wang, B.~Wu, D.~Li, X.~Shen, W.~Yu, Y.~Jiao, and J.~S. Vetter, ``Exploring
  hybrid memory for gpu energy efficiency through software-hardware
  co-design,'' in \emph{PACT '13}, 2013.

\bibitem{Chatterjee:2012:LHD:2457472.2457482}
N.~Chatterjee, M.~Shevgoor, R.~Balasubramonian, A.~Davis, Z.~Fang, R.~Illikkal,
  and R.~Iyer, ``Leveraging heterogeneity in dram main memories to accelerate
  critical word access,'' in \emph{MICRO '45}, 2012.

\bibitem{Ousterhout:2010:CRS:1713254.1713276}
J.~Ousterhout, P.~Agrawal, D.~Erickson, C.~Kozyrakis, J.~Leverich,
  D.~Mazi\`{e}res, S.~Mitra, A.~Narayanan, G.~Parulkar, M.~Rosenblum, S.~M.
  Rumble, E.~Stratmann, and R.~Stutsman, ``The case for ramclouds: scalable
  high-performance storage entirely in dram,'' \emph{SIGOPS Oper. Syst. Rev.},
  vol.~43, no.~4, 2010.

\bibitem{Tsirogiannis:2010:AEE:1807167.1807194}
D.~Tsirogiannis, S.~Harizopoulos, and M.~A. Shah, ``Analyzing the energy
  efficiency of a database server,'' in \emph{SIGMOD '10}, 2010.

\end{thebibliography}

\end{document}